\documentclass[
amsmath, amssymb, floatfix, aps, prd,prev,twocolumn,preprintnumbers,floatfix,nofootinbib
]{revtex4-2}

\usepackage{graphicx}
\usepackage{dcolumn}
\usepackage{bm}
\usepackage{float}
\usepackage{hyperref}
\usepackage{xcolor}

\begin{document}

\title{Distributions and Collision Rates of ALP Stars in the Milky Way}

\author{Dennis Maseizik}
 \email{dennis.maseizik@desy.de}
\author{Günter Sigl}%
 \email{guenter.sigl@desy.de}
\affiliation{%
	II. Institute for theoretical Physics, Hamburg University, Luruper Chaussee 149, 22761 Hamburg, Germany
}%

\date{\today}

\begin{abstract}
We apply current analytical knowledge on the characteristic mass and linear growth of miniclusters down to redshift $z=0$ to the hypothetical minicluster distribution of the Milky Way.
Using the mass-radius relation and a core-halo relation for stable soliton solutions composed of axion-like particles (ALPs), we connect the galactic minicluster mass distribution to that of their ALP star cores.
We consider different temperature evolutions of the ALP field with masses in the range $10^{-12}\,\mathrm{eV} \leq m_a \leq 10^{-3}\,$eV and infer the abundance and properties of QCD axion- and ALP stars in our galaxy.
We re-evaluate detection prospects for collisions of neutron stars with both ALP stars and miniclusters as well as relativistic ALP bursts, so-called Bosenovae.
Our analysis shows that the collision rates between miniclusters and neutron stars can become as large as $\sim 10^5\,$yr$^{-1}$ galaxy$^{-1}$, but that the fraction of encounters that can lead to resonance between ALP mass and magnetosphere plasma frequency is generally well below $\sim 1\,$yr$^{-1}$ galaxy$^{-1}$, depending on the ALP model.
We confirm previous results that merger rates of ALP stars are extremely small $< 10^{-12}\,$yr$^{-1}$ galaxy$^{-1}$, while their host miniclusters can merge much more frequently, up to $\sim 10^3\,$yr$^{-1}$ galaxy$^{-1}$ for the QCD axion.
We find that Bosenovae and parametric resonance are much more likely to lead to observable signatures than neutron star encounters.
We also suggest that a combination of accretion and parametric resonance can lead to observable radio lines for a wide range of ALP masses $m_a$ and photon-couplings $g_{a\gamma\gamma}$.
\end{abstract}

\maketitle


\section{Introduction} \label{sec:intro}
\begin{figure*}[t]
\centering
\includegraphics[width=.8\textwidth]{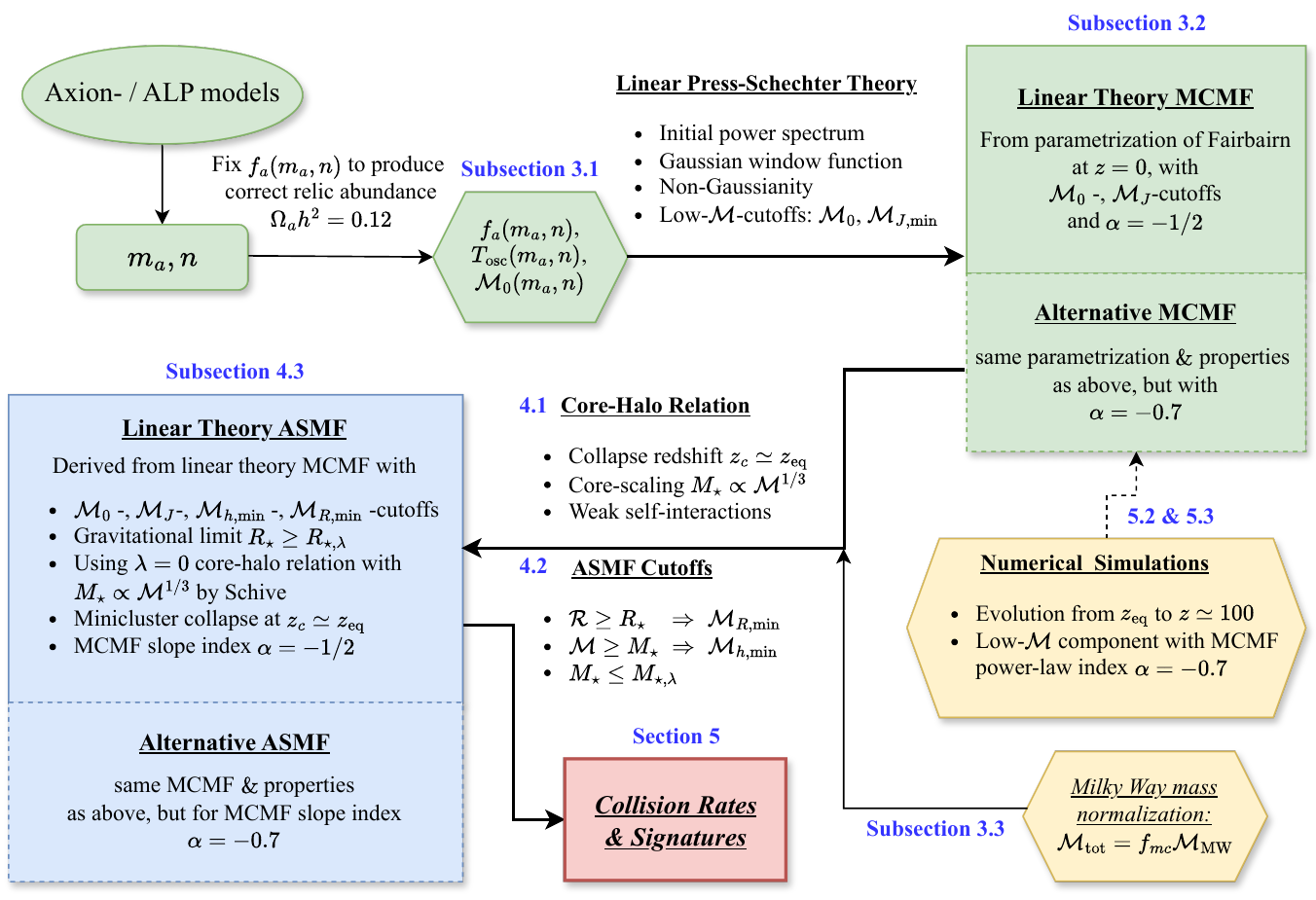}
\caption{Schematic Representation of the structure of this paper.
Green panels indicate methods derived from \cite{fairbairn_structure_2018}, yellow panels relate to other literature \cite{eggemeier_first_2020, mcmillan_mass_2011} and blue and red elements indicate the results obtained from the approach in this paper.
Underlying assumptions are explained by text items and the corresponding sections are shown in blue.} \label{fig:Scheme}
\end{figure*}

The current standard model of cosmology predicts that the majority of the matter content in the universe is present in the form of some collection of unknown, weakly interacting non-relativistic particles, generally referred to as cold dark matter. 
While the observational hints for the existence of such particles have consolidated over time, such as from X-ray emission from the Bullet Cluster, from galaxy rotation curves and from simulations of large-scale structure formation, they have not yet been detected  experimentally and their nature is still unclear. 
For recent reviews see, e.g., \citep{Bertone:2004pz, Klasen:2015uma}\\
The axion is one of several proposed candidates to constitute the dark matter in the universe and it has received increasing attention by the scientific community in the past years, see, e.g., \citep{Marsh:2015xka, di_Cortona_2016} for reviews. 
Originally proposed as a solution to the strong CP-problem of QCD \cite{PecceiCP1977, PecceiConstraints1977, Peccei_2008, WeinbergCP1978, Abbott_1982, Preskill_1982, Dine_1982, KimAxions2010}, the name-giving QCD axion is a pseudo-scalar particle with masses in the range $10^{-12}\,$eV to $10^{-3}\,$eV.
Further motivation for axion-like particles, generally referred to as ALPs, comes from string theory, where they arise naturally from reduction of higher gauge fields \cite{Svrcek_2006}.
In contrast to QCD axions, for which the self-coupling is fixed by the relation $m_af_a \simeq (78\,{\rm MeV})^2 \sim \Lambda_\mathrm{QCD}^2$ between the symmetry breaking scale $f_a$ and the mass $m_a$, with the QCD scale $\Lambda_\mathrm{QCD}$, the generic axion-like particles can be considered to have arbitrary self-interaction and photon-coupling e.g. ultra-light axions with masses in the range of $10^{-19}$ - $10^{-22}\,$eV.\\
We consider ALP models with domain wall number $N_\mathrm{DW}=1$ and limit our analysis to the post-inflationary scenario where miniclusters can be produced generically from the large initial fluctuations of the ALP field.
Our study of the ALP star distribution is further constrained to models with weak attractive self-interactions and a QCD-like (truncated) cosine potential yielding soliton solutions with a maximum stable mass \cite{chavanis_mass-radius_2011}.
\\
These solitons can lead to a range of observable signatures such as relativistic ALP bursts during collapse of critical ALP star configurations \cite{levkov_relativistic_2017, eby_probing_2022, arakawa2024bosenovae, Fox_2023} and radio emission through resonant conversion of ALP dark matter \cite{levkov_radio-emission_2020, di2023stimulated, bai_diluted_2022, chungjukko_2024_multimessenger, witte_transient_2023, chungjukko2023electromagnetic, kouvaris2022radio, Amin_2021}.
Both of these scenarios can be triggered by external interactions.
Neutron star encounters \cite{bai_diluted_2022, witte_transient_2023, edwards_transient_2021, kouvaris2022radio}, soliton mergers \cite{chungjukko_2024_multimessenger, hertzberg_merger_2020, amin_dipole_2021, Du_2024, Schwabe_2016} and accretion \cite{dmitriev_self-similar_2024, chen_new_2021, eggemeier_formation_2019, Eggemeier_2022} can drive the solitons to reach a critical configuration, where either parametric resonance into photons or the self-interaction instability and relativistic ALP emission develop.
The observation of these signatures essentially depends on two quantities:
the single event signal strength and the rate with which these events occur.
The modulation and strength of single ALP star events have been calculated and estimated in the literature for both radio \cite{witte_transient_2023, di2023stimulated, arakawa2024bosenovae, bai_diluted_2022, amin_dipole_2021, chungjukko2023electromagnetic, Amin_2021} and ALP emission \cite{arakawa2024bosenovae, levkov_radio-emission_2020, eby_probing_2022, chungjukko2023electromagnetic} before.
There have also been several studies calculating the event rates \cite{eby_collisions_2017, bai_diluted_2022, edwards_transient_2021, hertzberg_merger_2020} of the corresponding signals, yet all of them are inherently limited by large uncertainties in the determination of the mass distribution and properties of ALP stars.
Our paper extends these studies by constraining ALP star properties from current knowledge of miniclusters and their soliton cores.
We further apply our results to the Milky Way and discuss the observational implications for different detection mechanisms.\\ 
An overview on the structure of this paper is given in figure \ref{fig:Scheme}.
\noindent
In section \ref{sec:intro_Mass-Radius} we review the properties of the Gross-Pitaevskii-Poisson system and derive the resulting Mass-Radius relation of ALP stars (ASs) with weak attractive self-interactions. 
Section \ref{sec:MCMF} introduces the basic properties of ALP miniclusters (MCs) and the halo mass function of ALP miniclusters, hereafter termed the minicluster mass function (MCMF) to denote the difference to the fuzzy dark matter halo scenario.
In section \ref{sec:ASMF} we use the mass-radius relation of stable ALP stars together with the relation for solitonic cores of wavelike dark matter in order to derive different estimates for the present-day ALP star mass function (ASMF) from the corresponding MCMF.
The resulting ASMFs provide predictions for the mass and radius of solitonic ALP cores in the Milky Way, which we use to calculate the collision rates of ALP stars with various astrophysical objects in section \ref{sec:detection}.
In the case of AS-NS and MC-NS collisions, we update previous estimates on the resulting event rates in subsections \ref{subsec:NS-AS} and \ref{subsec:NS-MC}, respectively. 
We find that collision rates with neutron stars with sufficiently strong magnetic fields and sufficiently high plasma densities to enable resonances between ALP mass and plasma frequency are in general strongly suppressed compared to total collision rates rendering them unlikely to be detectable. 
On the other hand, in subsection \ref{subsec:bosenova} we find that mergers of individual MCs can produce ALP star cores that are unstable with respect to ALP self-interactions leading to emission of relativistic ALPs in a so-called Bosenova. 
The MC merger rates in the Milky Way are estimated to be sufficiently large to render them observable if individual Bosenovae are detectable. 
In subsection \ref{subsec:Glowing_AS} we have estimated the number of galactic ASs that are above the threshold for parametric ALP conversion into photons and find it to be substantial. 
This can lead to line-like radio emission that could be detectable for ALP masses above $\sim10^{-8}\,$eV. 
Finally, we estimate the impact of our results to the cosmological context in subsection \ref{subsec:extra-galactic} and summarize our work in section \ref{sec:conclusions}.\\
Throughout this work we use natural units $\hbar=c=1$ and Planck \cite{planck_2016} cosmological parameters $h=0.67$, $\Omega_m=0.32$ $\Omega_a h^2 = 0.12$, $z_\mathrm{eq}=3402$.
We label the minicluster parameters with calligraphic letters, namely $\mathcal{M},\,\mathcal{E},\,\mathcal{R}$ for the MC mass, energy and radius respectively.
ALP star properties are indicated by '$\star$' indices and italic letters, as shown in the overview in table \ref{tab:Params}.
For the minicluster masses $\mathcal{M}$ in the top part of table \ref{tab:Params}, the sub-indices '$i,\min$' refer to different low-mass cutoffs $i$ applied to the galactic AS-MC systems.

\section{Mass-radius relation with attractive self-interactions} \label{sec:intro_Mass-Radius}
\begin{figure}[t]
\centering
\includegraphics[width=\columnwidth]{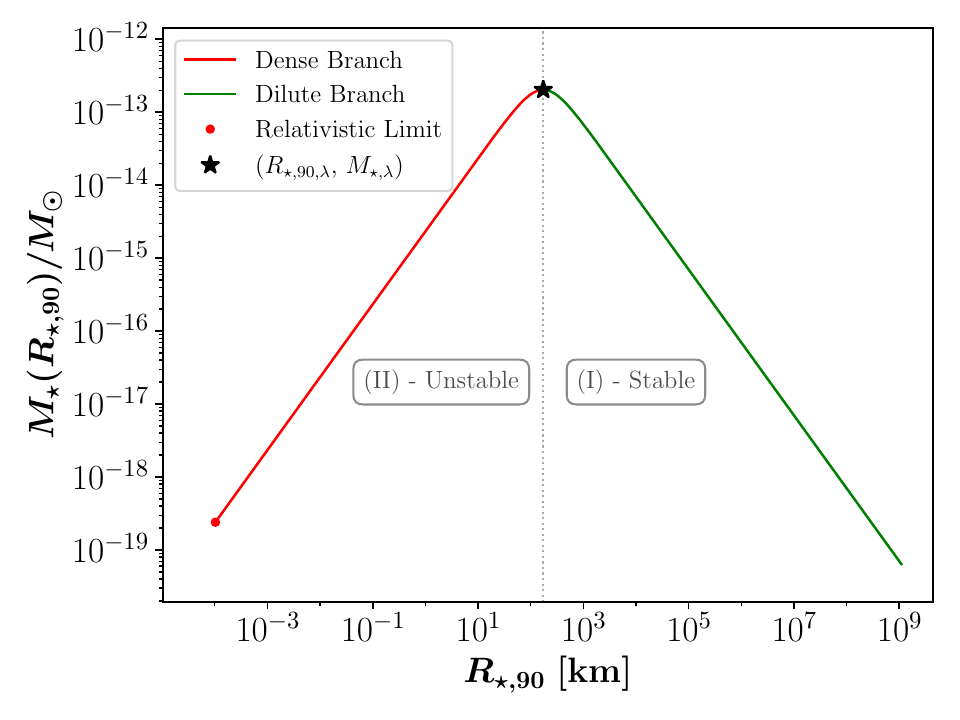}
\caption{Mass-radius relation \eqref{eq:Mass-Radius-Rel_Physical} for QCD axions with $m_a=50\,\mu$eV and $f_a \simeq 10^{11}\,$GeV (see also figure \ref{fig:f_a}) using $R_{\star,90}$ as characteristic ALP star radius. 
The dense branch of unstable solutions (II) is given in red, together with the relativistic limit \eqref{eq:non-relativistic_Cond2} indicated by the red dot.
Stable solutions of dilute axion stars (I) are shown in green while the critical solution with maximum mass $M_{\star,\lambda}$ and minimum radius $R_{\star,90,\lambda}$, which separates (I) and (II), is labelled with a black star.}\label{fig:Mass-Radius_QCD_minimal}
\end{figure}

We start by introducing the Gross-Pitaevskii-Poisson equations which govern the evolution of the ALP field in the non-relativistic regime and use them to derive the mass-radius relation of axion stars with attractive self-interactions.
We follow the standard derivation of the GPP equations for the QCD axion similar to \citep{hertzberg_merger_2020, chavanis_phase_2018, chavanis_mass-radius_2011} and extend this approach to ALPs by also considering light scalar particles with arbitrary combinations of $m_a$ and $f_a$, in addition to the QCD axion.\\
The Lagrangian density of the axion field can be written in the canonical form as
\begin{align}
\mathcal{L}=\sqrt{-g}\left[\frac{\mathrm{R}}{2 \kappa}+\frac{g^{\mu \nu}}{2} \nabla_\nu \phi \nabla_\mu \phi-V(\phi)\right], \label{eq:Lagrangian}
\end{align}
where $g=\operatorname{det}\left(g_{\mu \nu}\right)$ is the determinant of the metric tensor with signature $(+--\,-)$, $\mathrm{R}$ is the Ricci scalar and $\kappa=8 \pi G$ is the gravitational coupling. 
In the non-relativistic regime, the field values $\phi$ are small, so that we can expand the potential $V(\phi)= m_a^2 f_a^2 [1 - \cos(\phi/f_a)]$ around the CP conserving minimum $\phi=0$ and keep only the two leading-order terms
\begin{align}
V(\phi)=\frac{m_a^2}{2 }  \phi^2 + \frac{\lambda}{4!} \phi^4+\mathcal{O}\left(\lambda^2 \phi^6 / m_a^2\right), \label{eq:Potential_expansion}
\end{align}
where $m_a$ is the ALP mass and $\lambda$ is the quartic coupling constant.
To obtain the non-relativistic limit, it is useful to express the real field $\phi(\vec{x},t)$ in terms of a slowly varying complex Schrödinger field $\psi(\vec{x},t)$ using the transformation
\begin{align}
\phi =  \frac{1}{\sqrt{2m_a}}\left[\psi(\vec{x}, t) e^{-i m_a t}+\psi^*(\vec{x}, t) e^{i m_a t}\right] \,. \label{eq:psi_definition}
\end{align}
Inserting equation \eqref{eq:psi_definition} into the Lagrangian \eqref{eq:Lagrangian}, the rapidly oscillating terms proportional to $e^{\pm im_at}$ may be neglected since they average to zero over time.
Additionally taking $|\dot{\psi}|/m_a \ll |\psi|$ and using the Newtonian metric $g_{00}= 1 + 2\Phi$, the non-relativistic evolution of the complex field $\psi$ can be shown to follow the \textit{Gross-Pitaevskii-Poisson system} (GPP):
\begin{align}
i  \frac{\partial \psi}{\partial t}
&=-\frac{1}{2 m_a} \Delta \psi + m_a \Phi \psi - \frac{|\lambda|}{8m_a^2}|\psi|^2\psi \,, \label{eq:GP}\\
\Delta \Phi
&=4 \pi G m_a |\psi|^{2} \,,\label{eq:Poisson}
\end{align}
where $\Phi$ is the Newtonian potential and  $\lambda = - m_a^2/f_a^2$ is the self-interaction parameter of the ALP or axion \cite{chavanis_phase_2018}.
More precisely, the QCD axion self-coupling $\lambda=-c_\lambda m_a^2/f_a^2$ depends on the up- and down quark masses $m_u$, $m_d$ with $c_\lambda = 1- 3 m_u m_d/(m_u^2+m_d^2) \approx0.3$ according to more accurate calculations using chiral perturbation theory and lattice QCD \cite{di_Cortona_2016}.
For simplicity we will assume $c_\lambda=1$ for different ALP models in this paper \cite{Marsh_2016, chavanis_phase_2018}, which coincides with the standard dilute instanton gas approximation for the QCD axion.\\
The stationary solutions to the GPP system \eqref{eq:GP}, \eqref{eq:Poisson} are generally termed solitons, ALP/axion stars or boson stars, depending on the self-interaction $\lambda$.
For the QCD axion, the axion mass is related to the decay constant $f_a$ by \cite{Marsh_2016}
\begin{align}
m_a & \approx 50\, \mu \mathrm{eV}\left(\frac{1.2 \cdot 10^{11} \mathrm{GeV}}{f_a}\right) \,. \label{eq:f_a-m_a-QCD}
\end{align}
In the more general case of ALPs, arbitrary combinations of $m_a$ and $f_a$ may be considered, however we will show in subsection \ref{subsec:MCMF_Parametr} that this choice can be constrained by requiring the correct relic abundance of dark matter.\\
The analytic expression for the mass-radius relation of ALP stars can be derived from equations \eqref{eq:GP}, \eqref{eq:Poisson} using a \textit{Gaussian ansatz} for the wave function \citep{chavanis_mass-radius_2011} with respect to the radial coordinate $r$
\begin{align}
\rho(r) &\equiv m_a|\psi(r)|^2 = \left(\frac{M_\star}{\pi^{3 / 2} R_\star^3}\right) e^{-\frac{r^2}{R_\star^2}} \,, \label{eq:GaussianProfile}
\end{align}
where $M_\star$ and $R_\star$ are the mass and radius of the star respectively.
It should be noted that different approaches similar to equation \eqref{eq:GaussianProfile} have been suggested in the literature (see also \citep{schiappacasse_analysis_2018} for a detailed comparison).
We choose the Gaussian profile for simplicity but keep our approach general by tracking the corresponding ansatz-specific coefficients $\alpha_\mathrm{kin}, \alpha_\mathrm{grav}$ and $\alpha_\mathrm{int}$, which will be introduced in the following.\\
Independent of the specific profile, we can express the Newtonian potential $\Phi(\vec{x},t)$ in equation \eqref{eq:E_grav} through the Green's function for the Poisson equation \eqref{eq:Poisson} \citep{schiappacasse_analysis_2018}, and write the different energy contributions of the non-relativistic ALP star as
\begin{align}
E_{\text {kin }} & =\frac{1}{2 m_a} \int d^3 x\, |\nabla \psi(\vec{x})|^2 
= \alpha_\mathrm{kin} \frac{ M_*}{m_a^2 R_*^2} \,, \label{eq:E_kin}\\
E_{\text {grav }} & = -\frac{m_a}{2} \int d^3 x\, \Phi |\psi(\vec{x})|^2 
= -\alpha_\mathrm{grav} \frac{G M_*^2}{R_\star} \label{eq:E_grav}\,,\\
E_{\text {int }} & =\frac{\lambda}{16 m_a^2} \int d^3 x\, |\psi(\vec{x})|^4
= - \alpha_\mathrm{int} \frac{|\lambda| M_*^2}{m_a^4 R_*^3}  \label{eq:E_int}\,.
\end{align}
where the ansatz-specific coefficients 
\begin{align}
    \alpha_\mathrm{kin} &=\frac{3}{4}\,, \quad \alpha_\mathrm{grav} =\frac{1}{\sqrt{2 \pi}}\,, \quad \alpha_\mathrm{int} =\frac{1}{32 \pi \sqrt{2 \pi}}
\end{align}
are obtained for the Gaussian profile \eqref{eq:GaussianProfile}.
Under this assumption, the total energy of the soliton solution with mass $M_\star$ and radius $R_\star$ may be written as
\begin{align}
E_{\star,\mathrm{tot}} &=  
 \frac{3 M_\star}{4m_a^{2} R_\star^{2}} - \frac{G M_\star^{2}}{\sqrt{2\pi}R_\star} 
 - \frac{|\lambda| M_\star^{2}}{32\pi \sqrt{2\pi} m_a^4 R_\star^{3}} \,.  \label{eq:E_total}
\end{align}
In order to obtain the mass-radius relation from the energy \eqref{eq:E_total}, it is useful to transform the physical variables of the GPP system onto dimensionless quantities of order unity by means of the rescaling
\begin{align}
x & = \tilde{x} / (\sqrt{G} m_a f_a)\,, & t & = \tilde{t} / (G m_a f_a^2)\,, \\
\psi & =\sqrt{Gm_a} f_a^2 \, \tilde{\psi}\,, & \Phi & = G f_a^2\, \tilde{\Phi}\,, \label{eq:Rescaling_Coordinates}
\end{align}
where the rescaled variables are labelled with a tilde \citep{hertzberg_merger_2020}.
The transformation \eqref{eq:Rescaling_Coordinates} excludes all factors $G$, $m_a$ and $\lambda=-m_a^2/f_a^2$ from the equations \eqref{eq:GP} and \eqref{eq:Poisson}.
Writing the total energy of the ALP star in equation \eqref{eq:E_total} in its rescaled and profile-independent form yields the energy relation
\begin{align}
\tilde{E}_{\star,\mathrm{tot}}(\tilde{R}_\star)= \alpha_\mathrm{kin} \frac{\tilde{M}_{\star}}{\tilde{R}_\star^2}- \alpha_\mathrm{grav} \frac{\tilde{M}_{\star}^2}{\tilde{R}_\star} - \alpha_\mathrm{int} \frac{\tilde{M}_{\star}^2}{\tilde{R}_\star^3} \,, \label{eq:E_total_Rescaled}
\end{align}
which we extremize with respect to the star radius $\tilde{R}_\star$ to obtain the rescaled version of the mass-radius relation of ALP stars
\begin{align}
 & \tilde{R}_\star = \frac{\alpha_\mathrm{kin} \pm \sqrt{\alpha_\mathrm{kin}^2-3 \alpha_\mathrm{int} \alpha_\mathrm{grav} \tilde{M}_\star^2}}{\alpha_\mathrm{grav} \tilde{M}_\star} \label{eq:Radius-Mass-Rel}
 \end{align}
The plus and minus sign in equation \eqref{eq:Radius-Mass-Rel} divide the stationary solutions to the GPP system into two branches: the stable dilute branch, given by the plus sign, and the unstable dense branch of ALP stars indicated by the minus sign (see also figure \ref{fig:Mass-Radius_QCD_minimal}).
The critical point between the two branches constitutes what is commonly referred to as the \textit{maximum mass} $M_{\star,\lambda}$ and \textit{minimum radius} $R_{\star,\lambda}$ of stable ALP stars.
For the Gaussian ansatz we obtain
\begin{align}
M_{\star,\lambda } &= \sqrt{\frac{3 }{G}} \frac{2 \pi  f_a}{m_a }  \quad \,, \quad
R_{\star,\lambda} = \sqrt{\frac{3}{32 \pi G}} \frac{1}{m_a f_a} \label{eq:M_Star_Max_R_Star_min} \,.
\end{align}
in dimensionful units.
For later calculations related to the radius cutoff of ALP stars \cite{kavanagh_stellar_2021} we also express the mass-radius relations given by \eqref{eq:Radius-Mass-Rel} in physical units
\begin{align}
M_\star &= \frac{\sqrt{2 \pi} R_\star}{\frac{2 m_a^2 G}{3 } R_\star^2 + \frac{1}{16 \pi f_a^2}} \label{eq:Mass-Radius-Rel_Physical}\,,\\
R_\star =&\,\, \frac{\alpha_\mathrm{kin}  }{\alpha_\mathrm{grav} G m_a^2 M_\star} \nonumber \\
&\pm \sqrt{ \left( \frac{\alpha_\mathrm{kin}  }{\alpha_\mathrm{grav} G m_a^2 M_\star} \right)^2 - \frac{3 \alpha_\mathrm{int}}{\alpha_\mathrm{grav} G m_a^2 f_a^2}} \,. \label{eq:Radius-Mass-Rel_Physical}
\end{align}
An important constraint that applies to the dense branch of ALP stars with $R_\star \leq R_{\star, \min}$ is the validity of the non-relativistic approximation inherent to the GPP system \eqref{eq:GP}, \eqref{eq:Poisson}.
With decreasing $\tilde{R}_\star \ll \tilde{R}_{\star,\lambda} = \sqrt{3 \alpha_\mathrm{int} / \alpha_\mathrm{grav}}$ in equation \eqref{eq:Radius-Mass-Rel}, the density of the soliton with $M_\star \propto R_\star$ increases as $\rho_\star \propto M_\star / R_\star^3 \propto 1 / R_\star^2$, eventually reaching a point where the Taylor expansion \eqref{eq:Potential_expansion} breaks down and higher order terms would have to be taken into account.\\
In this limit, the invariance of the Lagrangian density under the transformation $\phi \longrightarrow \phi + 2 \pi f_a$ is broken, leading to the non-relativistic condition
\begin{align}
\frac{\phi_0}{2 \pi f_a}
&=\frac{\psi_0}{\pi f_a \sqrt{2 m_a}}
= \sqrt{\frac{ M_\star}{2 \pi^{7/2} m_a^2 f_a^2 R_\star^3}} \nonumber \\
&=\sqrt{\frac{G f_a^2 \tilde{M}_\star}{2 \pi^{7/2}  \tilde{R}_\star^3}} \ll 1 \label{eq:non-relativistic_Cond1} \,
\end{align}
for the mass $M_\star$ and radius $R_\star$ of the dense ALP star and where we have used the Gaussian profile \eqref{eq:GaussianProfile} to express $\psi_0 \equiv \psi(\Vec{x}=0)$ \citep{schiappacasse_analysis_2018}.
Equation \eqref{eq:non-relativistic_Cond1} can be interpreted as a lower bound on the dense-branch radius
\begin{align}
R_\star & \gg \left(\frac{M_\star}{2 \pi^{7/2} m_a^2 f_a^2}   \right)^{1/3}  \label{eq:non-relativistic_Cond2} \,,
\end{align}
which we implement in any consideration of the mass-radius relation in the following.
Lastly, we convert the scale radius $R_\star$ to the radius $R_{\star,90}$ containing 90\% of the total star mass, where $R_{\star,90}=1.76796\,R_\star$ for the Gaussian profile.
Note that in the following sections, we will take $R_{\star,90}$ as the physical AS radius and drop the index '$90$' for simplicity.\\
Figure \ref{fig:Mass-Radius_QCD_minimal} shows the mass-radius relation \eqref{eq:Radius-Mass-Rel_Physical} of QCD axion stars with $m_a = 50\,\mu$eV and $f_a \simeq 10^{11}\,$GeV, where the approximate value of the decay constant is related to the QCD axion properties in section \ref{sec:MCMF} and figure \ref{fig:f_a} that we use in this paper.
The dilute branch solutions (I) in green are dominated by Newtonian gravity and have been shown to be stable against perturbations, which is why we expect the present-day ALP star distribution to be mainly composed of dilute solitons.
In contrast, the stars with $R<R_{\star,\lambda}$ on the red curve make up the dense branch of solutions (II).
These solitons are dominated by the attractive ALP self-interactions and have been shown to be unstable against perturbations in numerous studies, with their evolution resulting in a relativistic collapse or Bosenova \citep{levkov_relativistic_2017, eby_probing_2022}.\\
Using the mass-radius relation \eqref{eq:Mass-Radius-Rel_Physical}, we can fix the ALP star radius, profile and energy as a function of its mass $M_\star$ for specific values of $m_a$ and $f_a$.
In the next sections we will use the present-day minicluster distribution to determine the remaining free ALP star parameter - the star mass $M_\star$ - from the minicluster mass $\mathcal{M}$.

\section{ALP Miniclusters and the Minicluster Mass Function} \label{sec:MCMF}
Numerical simulations of axion dark matter in the post-inflationary scnario show the formation of $\mathcal{O}(1)$ density fluctuations, so-called minicluster seeds, which collapse to form gravitationally bound miniclusters around matter-radiation equality \cite{Hogan_1988, eggemeier_first_2020, Xiao_2021, kolb_large-amplitude_1994, kolb_axion_1993, kolb_femtolensing_1996}. 
Kolb and Tkachev \cite{kolb_axion_1993} were the first to predict the abundance of gravitationally bound axions to be roughly equal to 70\%, suggesting that the majority of dark matter particles will be contained in miniclusters.
Similar results were confirmed numerically in \cite{eggemeier_first_2020} revealing a rich substructure in intermediate- and high-mass miniclusters.\\
In this section, we will follow the analytical Press-Schechter approach introduced in \cite{fairbairn_structure_2018} to estimate the halo mass function of ALP miniclusters at $z=0$ for arbitrary ALP mass and -temperature evolution.
The basic procedure is outlined by the green elements in figure \ref{fig:Scheme}.
We assume that a fraction $f_\mathrm{mc}=0.75$ \citep{eggemeier_first_2020} of the total dark matter in the Milky Way is composed of ALP miniclusters with a mass distribution similar to the $z=0$ prediction obtained from \citep{fairbairn_structure_2018} and that each minicluster contains at most a single ALP star.\\
We start by summarizing the basic MC properties in subsection \ref{subsec:M0} and introduce the parametrization of the MCMF obtained from \cite{fairbairn_structure_2018} in subsection \ref{subsec:MCMF_Parametr}.
The latter can subsequently be used to estimate the galactic MC distribution by normalization to the total mass of the galactic dark matter halo in subsection \ref{subsec:MCMF_Norm}.
We also discuss the relevance of low-mass cutoffs of the MCMF in subsection \ref{subsec:MCMF_Parametr}.

\subsection{Characteristic Minicluster Mass} \label{subsec:M0}
\begin{figure}[t]
\centering
\includegraphics[width=\columnwidth]{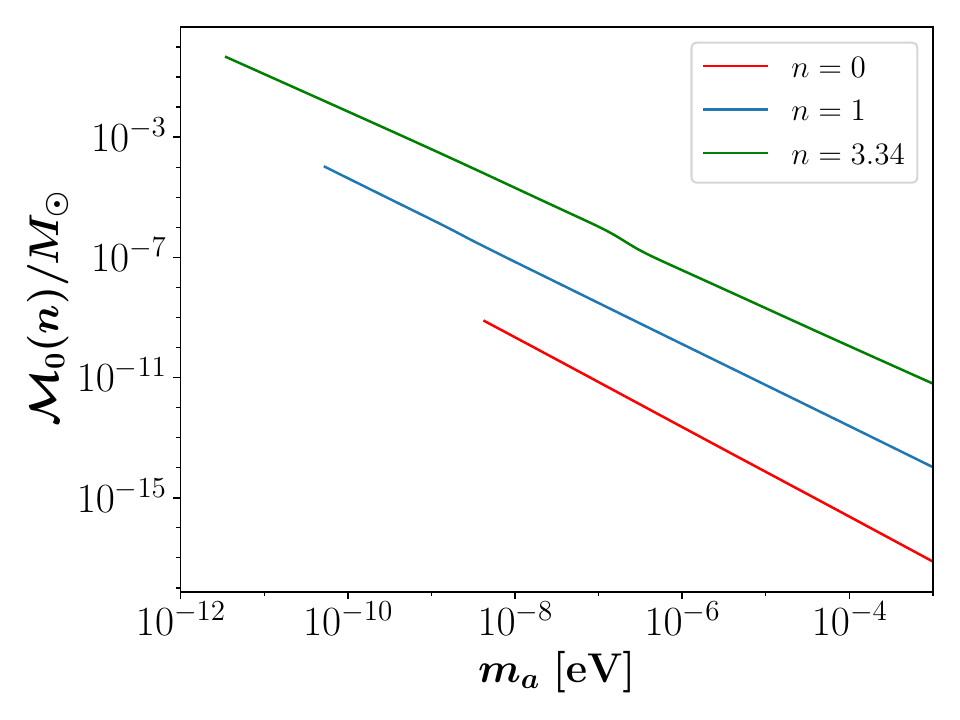}
\caption{Characteristic MC mass $\mathcal{M}_0(m_a,n)$ as a function of ALP mass $m_a$ and its (colored) temperature dependence $n$, reproduced from the procedure in \cite{fairbairn_structure_2018} and subsection \ref{subsec:M0}.
\label{fig:M0}}
\end{figure}

\begin{figure}[t]
\centering
\includegraphics[width=\columnwidth]{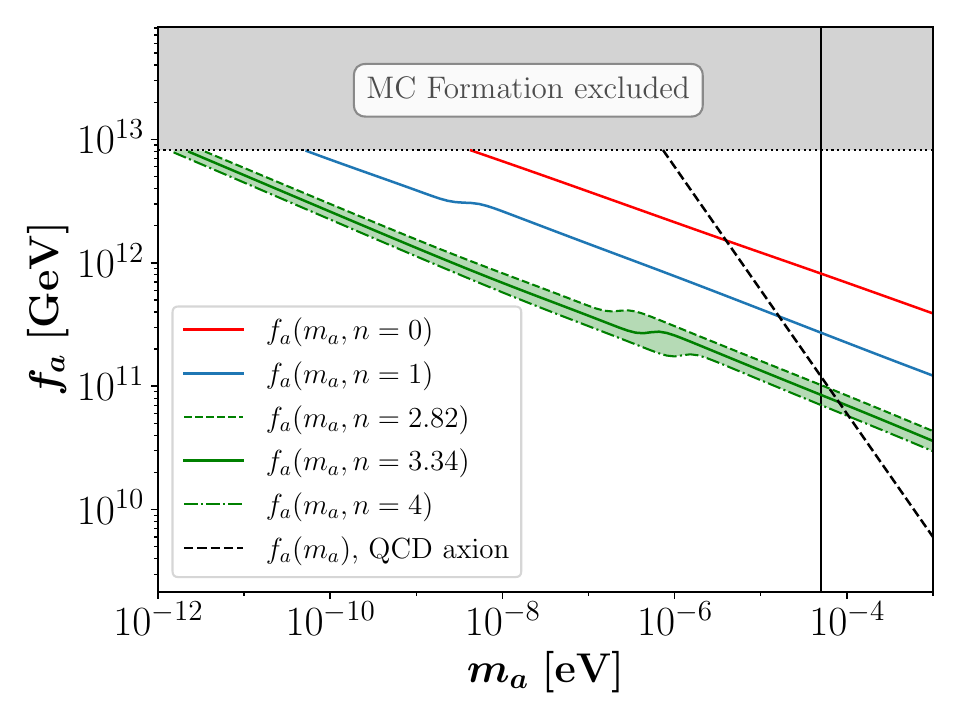}
\caption{ALP decay constant $f_a(m_a,n)$ fixed by matching the relic abundance \eqref{eq:relic_abundance} at different $n$ in colored lines, compared to the black dashed relation \eqref{eq:f_a-m_a-QCD} for the QCD axion.
The green band and lines correspond to different predictions for the temperature index $n$ of the QCD axion \cite{Berkowitz_2015, wantz_axion_2010, gross_1981}.
Post-inflationary MC formation is excluded by CMB constraints on the tensor-to-scalar ratio for $f_a \geq 8.2\cdot 10^{12}\,$GeV in grey shades \cite{array_bicep2_2016, fairbairn_structure_2018}.
\label{fig:f_a}}
\end{figure}

Kolb and Tkachev\cite{kolb_large-amplitude_1994} used a spherical collapse model to predict the characteristic minicluster density
\begin{align}
\rho_{mc} &\simeq 7\cdot10^6 \, \delta^3 (1+\delta)\left(\frac{\Omega_a h^2}{0.12}\right)^4 \frac{\mathrm{GeV}}{\mathrm{cm}^3} \,
\label{eq:rho_mc}
\end{align}
from the initial overdensity parameter $\delta = \delta \rho_a / \Bar{\rho}_a$, where $\Bar{\rho}_a$ is the background density of the ALP field today and we assume the typical value of $\delta \simeq 1$ if not stated otherwise \cite{kolb_axion_1993}.
Approximating the minicluster as a homogeneous sphere with total mass $\mathcal{M}$, we can define a characteristic radius
\begin{align}
\mathcal{R} &\simeq \frac{3.4\cdot 10^7}{\delta(1 + \delta)^{1/3}} \left(\frac{\mathcal{M}}{10^{-12}M_\odot}\right)^{1/3}\,\mathrm{km} \, \label{eq:R_mc}
\end{align}
for the broad range of minicluster masses $\mathcal{M}$ similar to $\mathcal{M}_0$.
The characteristic mass $\mathcal{M}_0$ is determined by the total mass of ALP dark matter contained within the horizon at the oscillation temperature $T_\mathrm{osc}$, when $3H(T_\mathrm{osc})\approx m_a(T_\mathrm{osc})$ and the ALP mass becomes relevant.
Using a spherical geometry for the collapsing minicluster and writing the horizon size in terms of the comoving wavenumber $k_\mathrm{osc} = aH(T_\mathrm{osc})$, one finds
\begin{align}
\mathcal{M}_0 &= \Bar{\rho}_a \frac{4 \pi}{3}\left(\frac{\pi}{k_\mathrm{osc}}\right)^3 \label{eq:M0} \,,
\end{align}
which is equivalent to other definitions of $\mathcal{M}_0$, e.g. in \cite{kolb_femtolensing_1996}, up to a factor of $4\pi^4/3\simeq 130$ \citep{fairbairn_structure_2018}.
In order to calculate $\mathcal{M}_0$ from equation \eqref{eq:M0}, we have to find the oscillation temperature $T_\mathrm{osc}$, given by the equality $3H(T) = m_a(T)$.
Starting with the left-hand side, we can express $H(T)$ using the second Friedmann equation 
\begin{align}
3 H(T)^2 M_p^2 &= \frac{\pi^2}{30} g_{\star,R}(T) T^4 \label{eq:Friedmann}
\end{align}
where the number of relativistic degrees of freedom $g_{\star,R}(T)$ is obtained from the fit in \cite{wantz_axion_2010} and $M_p$ is the reduced Planck mass.
The ALP mass $m_a(T)$ on the right-hand side, depends on the index $n$ describing its temperature evolution according to
\begin{align}
m_a(T)= m_{a,0} \left(\frac{T}{\mu} \right)^{-n} \label{eq:m_T}
\end{align}
and on the ALP decay constant $f_a$, which sets $\mu = \sqrt{m_a f_a}$ as in \citep{fairbairn_structure_2018}.
Unless the temperature dependence is explicitly written out, we refer to the low-temperature value of the axion mass, i.e. $m_a \equiv m_{a,0}$.
Instead of keeping the decay constant as a free parameter, we fix $f_a$ and $\mu$ in equation \eqref{eq:m_T} by requiring the correct relic abundance $\Omega_a h^2=0.12$ of ALP dark matter with particle mass $m_a$ and temperature index $n$ using the relation \cite{fairbairn_structure_2018}
\begin{widetext}
\begin{align}
\Omega_a(f_a) &= \frac{1}{6 H_0^2 M_{p l}^2}\left(1+\beta_{\mathrm{dec}}\right) \frac{c_n \pi^2}{3} m_a\left(T_{\mathrm{CMB}}\right) m_a\left(T_{\mathrm{osc}}\right) f_a^2\left[\frac{a\left(T_{\mathrm{osc}}\right)}{a\left(T_{\mathrm{CMB}}\right)}\right]^3 \label{eq:relic_abundance}
\end{align}
\end{widetext}
where $\beta_\mathrm{dec}=2.48$ is computed from the decay of the axion string-wall network \cite{Kawasaki_2015, Harari:1987ht, fairbairn_structure_2018}, $a(T_\mathrm{CMB})=1$ and $T_\mathrm{CMB} = 2.725\,$K is the CMB temperature \cite{planck_2016}.
The coefficient $c_n$ in equation \eqref{eq:relic_abundance} accounts for anharmonicities in the ALP cosine potential $V(\phi)$ and can be approximated by the relation
\begin{align}
c_n
&= \frac{3}{2\pi^3} \int _{-\pi}^\pi d\theta \, \theta^2 \left[ \ln \left( \frac{e}{1- \left( \frac{\theta}{\pi}\right)^4} \right) \right]^{\frac{3}{2}-\frac{n}{2n+4}}
\end{align}
under the assumption that the relativistic degrees of freedom can be treated as constant over the timescale in which anharmonic corrections act (see \cite{fairbairn_structure_2018} for details).
From this, the value of $f_a$ obtained by fixing $\Omega_a h^2 = 0.12$ with equation \eqref{eq:relic_abundance} can be used to calculate $k_\mathrm{osc}$ from $T_\mathrm{osc}$ by writing
\begin{align}
k_\mathrm{osc} &= a(T_\mathrm{osc}) H(T_\mathrm{osc}) \nonumber \\
&= \left[ \frac{g_{\star,S}(T_\mathrm{CMB}) }{g_{\star,S}(T_\mathrm{osc})} \right]^{1/3} \frac{T_\mathrm{CMB}}{T_\mathrm{osc}}  H(T_\mathrm{osc})  \,, \label{eq:k_osc}
\end{align}
where $g_{\star,S}(T)$ are the entropic degrees of freedom from \cite{wantz_axion_2010}.
Combining equation \eqref{eq:k_osc} with equation \eqref{eq:M0}
directly yields the characteristic minicluster mass $\mathcal{M}_0(m_a,n)$.\\
Note that according to equation \eqref{eq:m_T}, we have to repeat this process for every given value of $m_a=m_{a,0}$ and $n$.
For simplicity, we choose three different representative values for $n=\{0, 1, 3.34\}$, the latter of which coincides with the numerical results in \cite{wantz_axion_2010} for the QCD axion using an  interacting instanton liquid model.\\
For every value of $n$, we determine $\mathcal{M}_0(m_a,n)$ with ALP masses in the range $10^{-12}\,\mathrm{eV} \leq m_a \leq 10^{-3}\,\mathrm{eV}$.
The resulting characteristic minicluster masses from equation \eqref{eq:M0} and ALP decay constants fixed by \eqref{eq:relic_abundance} are plotted in figures \ref{fig:M0} and \ref{fig:f_a}.  
\noindent
Figure \ref{fig:M0} shows $\mathcal{M}_0$ for temperature-independent ALPs in red and for $n=1$, $n=3.34$ in blue and green.
The colored lines in figure \ref{fig:M0} are truncated in the low-$m_a$ region by the condition $f_a < 8.2\cdot 10^{12}\,$GeV \cite{fairbairn_structure_2018, array_bicep2_2016}, shown in figure \ref{fig:f_a} in grey, which is derived from constraints on the tensor-to-scalar ratio $r < 0.07$ of the CMB.
\noindent
For $f_a \geq 8.2\cdot10^{12}\,$GeV, the CMB constraint on the temperature of the universe during inflation lies below the expected temperature $T=f_a$ of the PQ phase transition, thus violating the post-inflationary symmetry breaking assumption in our approach.\\
The scaling of $\mathcal{M}_0$ in figure \ref{fig:M0} shows both an increase with decreasing ALP mass $m_a$ and an increase with larger $n$.
The first of the two is due to the fact, that for smaller $m_a$ the ALP mass becomes relevant later $3H(T_\mathrm{osc}) \approx m_a $, yielding a smaller $T_\mathrm{osc}$ and hence smaller $k_\mathrm{osc}$ in equation \eqref{eq:M0}, while the latter can be explained by larger $n$ with $m_a(T)\propto T^{-n}$ yielding smaller $T_\mathrm{osc}$ similarly.
The precise scaling and shape of $\mathcal{M}_0(m_a,n)$ in figure \ref{fig:M0} is caused by the temperature dependence of the relativistic degrees of freedom $g_{\star,R}(T)$ in equation \eqref{eq:Friedmann} and the temperature evolution of the ALP mass in equation \eqref{eq:m_T}.
In the case of temperature-independent ALPs with $n=0$, the characteristic mass scales as $\mathcal{M}_0\propto m_a^{-3/2}$ as shown by the original authors in \cite{fairbairn_structure_2018}.
For $n=1,3.34$ the scaling of $\mathcal{M}_0$ is slightly changed due to the different temperature evolution of the ALP mass.
We will later use the rough scaling $\mathcal{M}_0 \propto m_a^{-3/2}$ to trace the scaling of our results in subsection \ref{subsec:ASMF-Results} and throughout section \ref{sec:detection} with the ALP mass $m_a$.\\
We also show the ALP decay constants $f_a(m_a,n)$ fixed by $\Omega_a h^2 = 0.12$ with equation \eqref{eq:relic_abundance} in figure \ref{fig:f_a} in colored solid lines together with the QCD-axion $f_a$-$m_a$ scaling from equation \eqref{eq:f_a-m_a-QCD} in black dashed lines.
The green dashed, solid and dash-dotted lines show predictions for the QCD axion temperature dependence $n=2.81,3.34,4$ from different methods in the literature \cite{Berkowitz_2015, wantz_axion_2010, gross_1981}.
As indicated by the green shaded region, the spread in the determination of $f_a(m_a,n)$ arising from the uncertainty in $n$ is large.
In this work, we follow the choice of the original authors of \cite{fairbairn_structure_2018} by taking $n=3.34$ as a representative value.
For this temperature dependence, the QCD axion mass lies in the range $50\,\mu$eV$\lesssim m_a\lesssim 200\,\mu$eV, where uncertainties in the determination of $\Omega_a$ (and more specifically $\beta_\mathrm{dec}$ and $c_n$, see \cite{fairbairn_structure_2018} for details) have been taken into account.
Since we are mainly interested in the observational window of current and next-generation radio-telesopes such as SKA-mid ranging from $350\,$MHz to $14\,$GHz, we choose the lower bound of $m_a\approx 50\,\mu$eV for the QCD axion, which amounts to roughly $12\,$GHz and $f_a\simeq 10^{12}\,$GeV \cite{ska_web}.

\subsection{MCMF Parametrization and -Cutoffs} \label{subsec:MCMF_Parametr}
The parametrization of the MCMF from \cite{fairbairn_structure_2018} is based on analytic Press-Schechter theory describing the evolution of linear density perturbations.
It allows us to infer the mass distribution of ALP miniclusters at $z=0$ and for different $m_a$, $n$ using the characteristic MC mass $\mathcal{M}_0$ from equation \eqref{eq:M0}.
The details can be found in \citep{fairbairn_structure_2018} and references therein;
for the purpose of this paper we will briefly summarize the formalism before applying it in section \ref{sec:ASMF}.
In the standard Press-Schechter formalism, the comoving number density of miniclusters $n(\mathcal{M})$ can be calculated according to
\begin{align}
\frac{\mathrm{d}n}{\mathrm{d}\ln \mathcal{M}} &= \frac{\rho_{a0}}{\mathcal{M}} \left|\frac{\mathrm{d} \ln \sigma}{\mathrm{d} \ln \mathcal{M}}\right| \sqrt{\frac{2}{\pi}} \frac{\delta_c}{\sigma(\mathcal{M})} e^{-\frac{1}{2}\left[\frac{\delta_c}{\sigma(\mathcal{M})}\right]^2}\,, \label{eq:MCMF}
\end{align}
where $\mathrm{d}n/\mathrm{d}\ln \mathcal{M} = \mathcal{M}\mathrm{d}n/\mathrm{d}\mathcal{M}$ is the comoving number density of objects of mass $\mathcal{M}$ per logarithmic mass interval, commonly referred to as the halo mass function, or MCMF in our case.
In this framework, $\sigma^2(\mathcal{M}) \equiv \delta \mathcal{M}^2/\mathcal{M}^2$ denotes the time-dependent mass variance of density fluctuations and $\delta _c = 1.686$ is the (time-independent) overdensity threshold for gravitational collapse.\\
Using a Gaussian window function for the mass variance and a Heaviside initial power spectrum with $P(T_\mathrm{osc}) \propto \Theta(k_\mathrm{osc} - k)$, Fairbairn and Marsh \cite{fairbairn_structure_2018} derived a simplified parametrization of the MCMF as a function of $m_a,n$, which we will use in the following.
In this parametrization, the second out of three characteristic MC masses (the first being $\mathcal{M}_0$) is the minimum mass
\begin{align}
\mathcal{M}_{J,\min}(m_a) \Big |_{z=0} \approx&\,\, 8.3 \cdot 10^{-20}\, M_\odot \left(\frac{m_a}{50\,\mu \mathrm{eV}}\right)^{-3 / 2} \nonumber \\
& \times \left(\frac{\Omega_m}{0.32}\right)^{1 / 4}\left(\frac{h}{0.67}\right)^{1 / 2} \label{eq:M_h_min_J} \,,
\end{align}
related to the Jeans mass and evaluated at $z=0$ \citep{fairbairn_structure_2018}.
Below this mass, miniclusters do not form by gravitational collapse.
The third characteristic mass, the maximum mass of miniclusters in the MCMF, is related to the linear growth of structures with $\mathcal{M}\sim \mathcal{M}_0$ at $z_\mathrm{eq}$ leading to the occurrence of high-mass MCs with $\mathcal{M} \gg \mathcal{M}_0$ at late times.
It is defined in terms of the characteristic MC mass $\mathcal{M}_0$ from equation \eqref{eq:M0} by
\begin{align}
\mathcal{M}_{\max}(m_a,n) \Big |_{z=0} \approx 4.9 \times 10^6 \mathcal{M}_0(m_a,n) \label{eq:M_h_max}
\end{align}
with $z=0$ as before \citep{fairbairn_structure_2018}.
The physical processes driving the formation of heavy MCs with $\mathcal{M} \sim \mathcal{M}_{\max}$ over time are tidal interactions and merger events between miniclusters in close encounters.
Note that the minimum MC mass in equation \eqref{eq:M_h_min_J} is temperature-independent, while the value of $\mathcal{M}_{\max}$ is directly proportional to $\mathcal{M}_0$.
Accordingly, the spread of the MCMF will increase with $\mathcal{M}_0$ and specifically for larger $n$ (c.f. figure \ref{fig:M0}).\\
Fairbairn et al. emphasized that the low-mass end of the MCMF \eqref{eq:MCMF} is subject to large uncertainties related to non-Gaussianity of the field on scales $\mathcal{M} < \mathcal{M}_0$, where the standard Press-Schechter formalism can not be applied anymore.
They argued that due to filter dependence and from dynamical effects, a cut-off in the MCMF is expected for $\mathcal{M}\lesssim \mathcal{M}_0$ (see \cite{fairbairn_structure_2018} for details).
Using the parametrization introduced in subsection \ref{subsec:MCMF_Parametr} based on the Gaussian window function and Heaviside initial power spectrum, Fairbairn and Marsh \cite{fairbairn_structure_2018} found that this cutoff dependence becomes relevant for
\begin{align}
\mathcal{M} &\lesssim \mathcal{M}_0(m_a,n) / 25 \,. \label{eq:M0_cutoff}
\end{align}
To account for the large uncertainties in the low-mass tail of the MCMF \eqref{eq:M0_cutoff}, we will consider two different low-$\mathcal{M}$ cutoffs in the following:
First the cut-off prediction \eqref{eq:M0_cutoff} proportional to $\mathcal{M}_0$ and secondly the Jeans mass cutoff $\mathcal{M}_{J,\min}$ introduced in equation \eqref{eq:M_h_min_J}.\\
In the range where $\mathcal{M}_0/25 \leq \mathcal{M} \leq \mathcal{M}_{\max}$, the MCMF can be parametrized by a power-law
\begin{align}
\frac{\mathrm{d}n}{\mathrm{d}\ln \mathcal{M}} & \propto \mathcal{M}^{-1/2} \,. \label{eq:M_h_scaling_M}
\end{align}
with good precision \cite{fairbairn_structure_2018}.
For simplicity, we also apply the scaling \eqref{eq:M_h_scaling_M} in the range $\mathcal{M}_{J,\min} \leq \mathcal{M} \leq \mathcal{M}_{\max}$ with the Jeans cutoff $\mathcal{M}_{J,\min}$, similar to what was done in \citep{kavanagh_stellar_2021, edwards_transient_2021}.
While we do not address the question of the exact value and shape of the cut-off scale in this paper, we emphasize that our approach can easily be modified once better understanding on the evolution of the MCMF has been made in the future.

\subsection{Galactic MCMF and Normalization} \label{subsec:MCMF_Norm}
\begin{figure}[t]
\centering
\includegraphics[width=\columnwidth]{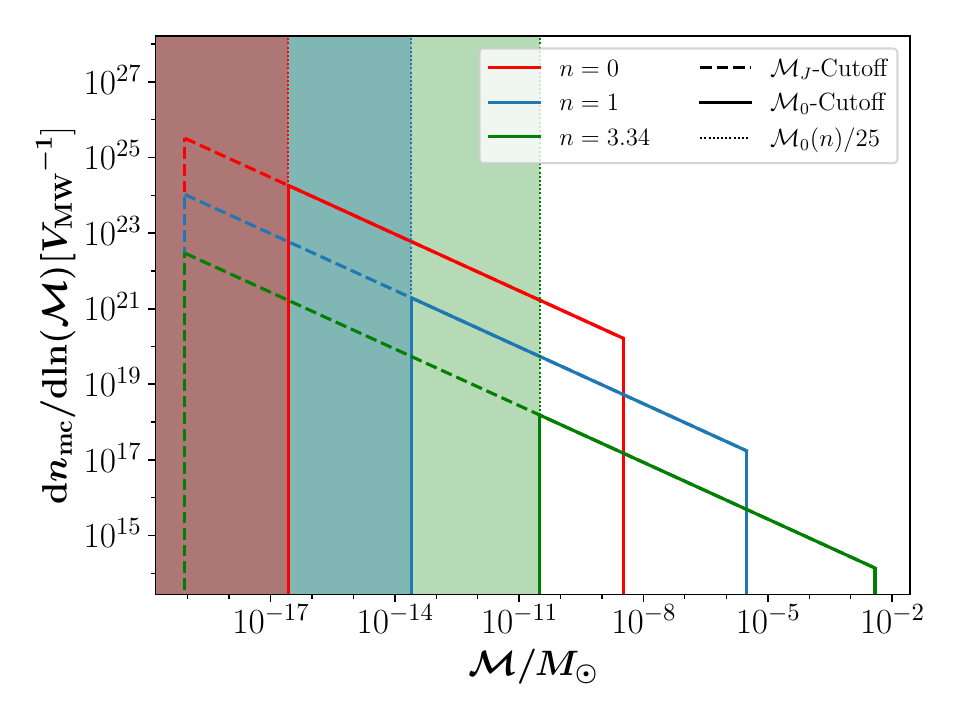}
\caption{MCMF per Milky Way volume obtained using the parametrization by \citep{fairbairn_structure_2018} for $m_a=50\,\mu$eV, $\alpha=-1/2$ and different $n=0,1,3.34$ indicated by colored lines. 
The shaded regions and dotted colored lines denote the low-mass cutoffs given by $\mathcal{M}_0(n)/25$.
Solid lines show the MCMF after applying the $\mathcal{M}_0(n)/25$-cutoffs while dashed lines display the MCMF with the $\mathcal{M}_J$-cutoff from eq. \eqref{eq:M_h_min_J}.
\label{fig:MCMF}}
\end{figure}

Next we will apply the MCMF parametrization of \citep{fairbairn_structure_2018} to the Milky Way dark matter halo using both low-$\mathcal{M}$ cutoffs.
It should be emphasized that the scaling $\mathrm{d}n/\mathrm{d}\ln \mathcal{M}\propto\mathcal{M}^{-1/2}$ in equation \eqref{eq:M_h_scaling_M} is derived from the analytic Press-Schechter approach, and that on the other hand, numerical simulations of minicluster evolution at $z\gtrsim 100$ show a different scaling with $\mathrm{d}n/\mathrm{d}\ln \mathcal{M}\propto\mathcal{M}^{-0.7}$ \citep{eggemeier_first_2020}.
Since the final slope of the MCMF is still subject to open debate, we will assume the corresponding power-law index $\alpha=-1/2$ unless stated otherwise and consider the case $\alpha=-0.7$ separately later.
In a general way, we can define the normalized minicluster mass function
\begin{align}
\frac{\mathrm{d}n}{\mathrm{d}\ln(\mathcal{M})} &= C_\mathrm{ren} \left( \frac{\mathcal{M}}{\mathcal{M}_{\min}} \right)^{\alpha}\, \label{eq:MCMF_Norm}
\end{align}
where $\mathcal{M}_{\min}$ takes the role of a reference MC mass and $C_\mathrm{ren}$ is a normalization constant to be determined in the following.
For simplicity, we will assume that the mass distribution of miniclusters is independent of the galactocentric radius $R$.
The total mass of minclusters can then be calculated from equation \eqref{eq:MCMF_Norm} by integrating over the mass density $\mathrm{d}m/\mathrm{d}\mathcal{M} = \mathcal{M}\mathrm{d}n/\mathrm{d}\mathcal{M}$. 
Assuming a spherically symmetric Milky Way volume $V_\mathrm{MW}=4\pi/(3 R_\mathrm{MW}^3)$ with radius $R_\mathrm{MW}= R_{200} = 237\,$kpc \citep{mcmillan_mass_2011} we get
\begin{align}
\mathcal{M}_{\mathrm{tot}} 
&= V_\mathrm{MW} \int_{\mathcal{M}_{\min}}^{\mathcal{M}_{\max}} d\mathcal{M}\, C_\mathrm{ren}\left( \frac{\mathcal{M}}{\mathcal{M}_{\min}} \right)^{\alpha}\\
&=  V_\mathrm{MW} \mathcal{M}_{\min} \frac{C_\mathrm{ren}}{\alpha + 1} \left[ \left( \frac{\mathcal{M}_{\max}}{\mathcal{M}_{\min}} \right)^{\alpha + 1} - 1 \right] \label{eq:M_tot_mc}
\end{align}
for the total DM mass contained in galactic ALP miniclusters.
We fix $C_\mathrm{ren}$ from equation \eqref{eq:M_tot_mc} by setting $\mathcal{M}_{\mathrm{tot}} \stackrel{!}{=} f_\mathrm{mc} \, \mathcal{M}_\mathrm{MW}$, where $f_\mathrm{mc}\simeq 0.75$ encodes the fraction of dark matter contained in miniclusters \citep{eggemeier_first_2020} and $\mathcal{M}_\mathrm{MW}=1.43\cdot 10^{12}\,M_\odot$ is the DM halo mass taken from the fits of \citep{mcmillan_mass_2017}.
With this normalization, the corresponding total number of galactic MCs is
\begin{align}
\mathcal{N}_{\mathrm{tot}} &= V_\mathrm{MW} \int_{\mathcal{M}_{\min}}^{\mathcal{M}_{\max}} d\mathcal{M} \frac{C_\mathrm{ren}}{\mathcal{M}} \left( \frac{\mathcal{M}}{\mathcal{M}_{\min}} \right)^{\alpha}\\
&= V_\mathrm{MW} \frac{C_\mathrm{ren}}{\alpha} \left[ \left( \frac{\mathcal{M}_{\max}}{\mathcal{M}_{\min}} \right)^{\alpha} - 1 \right] \label{eq:N_tot_mc}\,.
\end{align}
In the following sections, we repeat the above normalization for every $10^{-12}\,\mathrm{eV} \leq m_a \leq 10^{-3}\,\mathrm{eV}$, $n=0,1,3.34$ and for both cutoffs of the MCMF, $\mathcal{M}_{\min} = \mathcal{M}_0/25$ and $\mathcal{M}_{\min} =\mathcal{M}_{J,\min}$.
The results for $m_a=50\,\mu$eV are plotted in figure \ref{fig:MCMF}.
\noindent
The different colors in figure \ref{fig:MCMF} refer to the different values of $n$, which truncate the MCMF through $\mathcal{M}_{\max}$ in equation \eqref{eq:M_h_max} and through $\mathcal{M}_{\min}=\mathcal{M}_0/25, \mathcal{M}_{J,\min}$.
Dotted colored lines and shaded regions indicate the $\mathcal{M}_0$-cutoff.\\
We also highlight some important features of the MCMF in figure \ref{fig:MCMF}:
First, the MCMF peaks around the low-$\mathcal{M}$ cutoff which means that the typical MC mass will be subject to large uncertainty.
Secondly, figure \ref{fig:MCMF} also demonstrates that the intermediate- to high-mass component of the MCMF is essentially insensitive to the low-mass cutoffs and normalization \eqref{eq:M_tot_mc}.
The reason for this is the fact that a large majority of the mass relevant for the normalization of the MCMF is contained in the high-mass tail. 
The total number of miniclusters \eqref{eq:N_tot_mc} on the other hand is very sensitive to the low-mass cutoff as we shall see later.\\
Independent of the low-mass cutoff, large MC masses are predicted in both cases of $\mathcal{M}_{\min}$, especially for larger values of $n$.
This observation has important implications for DM searches, which we discuss in section \ref{sec:detection}.

\section{Galactic ALP Star Mass Distribution} \label{sec:ASMF}

In order to derive the ALP star mass distribution from the galactic MCMF, we need to apply a core-halo relation for AS-MC systems to the MCMF parametrizations from subsection \ref{subsec:MCMF_Parametr} and figure \ref{fig:MCMF}.
We will validate the use of the $\lambda=0$ core-halo relation from \cite{schive_understanding_2014} for weak attractive self-interactions $\lambda <0$ and ALP stars on the dilute branch in subsection \ref{subsec:core-halo} and appendix \ref{app:AS-Params}.
Apart from the $\mathcal{M}_J$- and $\mathcal{M}_0$-cutoffs from subsection \ref{subsec:MCMF_Parametr}, we will need to consider additional cutoffs to the ASMF constraining the formation of composite AS-MC systems.
This will be done in subsection \ref{subsec:ASMF_Cutoff} before applying the core-halo relation together with the corresponding ASMF and MCMF cutoffs for $z=z_\mathrm{eq}$ in subsection \ref{subsec:ASMF-Results}.
We explicitly assume that at most a single ALP star forms inside each minicluster. 
This amounts to neglecting scenarios with strong attractive self-interactions where it is possible for several ALP stars to form locally (see e.g. \cite{chen_new_2021}).

\subsection{Core-Halo Relation of ALP stars} \label{subsec:core-halo}
In the case without self-interactions $\lambda = 0$, numerical simulations of the Schrödinger-Poisson system involving ultra-light dark matter showed the occurrence of cored density profiles inside NFW-like halos \cite{schive_cosmic_2014, schive_understanding_2014, veltmaat_formation_2018}.
A similar study was performed in \cite{eggemeier_formation_2019} for axion miniclusters with $m_a=10^{-8}\,$eV, confirming that the core-halo relation of axion miniclusters coincides with the core-halo relation of FDM halos in the Schrödinger-Poisson regime.
In this paper, we employ the same core-halo relation as in \cite{eggemeier_formation_2019} by following the approach in \cite{schive_understanding_2014}, who used numerical simulations and analytical calculations to derive the core-halo relation
\begin{align}
M_\star(z) & 
=  \mathcal{M}_{h,\min}(z) \left[\frac{\mathcal{M}}{\mathcal{M}_{h,\min}(z)}\right]^{1 / 3}  \,,\label{eq:CoreHalo}
\end{align}
where the redshift-dependent \textit{minimum halo mass}
\begin{align}
\mathcal{M}_{h,\min}(z) =&\,\, 2.36 \cdot 10^{-16}  M_\odot \, 
\left( \frac{1+z}{1+z_\mathrm{eq}} \right)^{3/4} \nonumber \\
& \times \left[\frac{\zeta(z)}{\zeta(z_\mathrm{eq})}\right]^{1 / 4} \left(\frac{m_a}{50\,\mu\text{eV}}\right)^{-3/2} \,, \label{eq:M_h_min_CoreHalo}
\end{align}
is defined by requiring $M_\star=\mathcal{M}$.
The minimum halo mass can be interpreted as the lightest halo or minicluster mass, at which the formation of a composite core-halo system can occur at a given redshift.
Note that the factor $1/4$ from the original definition in \cite{schive_understanding_2014} was dropped, since we use a different definition of the soliton mass than the original authors of \cite{schive_understanding_2014}.\\
While the core-halo relation \eqref{eq:CoreHalo} has been shown to be applicable to MCs for $\lambda=0$ in \cite{eggemeier_formation_2019}, \cite{dmitriev_self-similar_2024}, there is currently little progress on finding a similar relation for self-interacting core-halo systems $|\lambda| > 0$.
The derivation and simulation of an extended core-halo relation for such systems is beyond the scope of this work.
Instead we will argue in the following that, restricting our analysis to the dilute branch (I) of stable ALP stars in figure \ref{fig:Mass-Radius_QCD_minimal}, and for weak attractive self-interactions of the form $\lambda = -m_a^2/f_a^2$, the gravitational core-halo relation \eqref{eq:CoreHalo} for $\lambda=0$ may be used as a reasonable approximation.\\
The authors of \cite{padilla_core-halo_2021} presented an extended redshift-independent formulation of the core-halo relation for arbitrary $|\lambda| > 0$ based on the analytical approaches in \cite{schive_understanding_2014} and \cite{Chavanis_2019}.
They modified the standard assumption $v_\star \simeq v_\mathrm{mc}$ for the virial velocities $v_\star$, $v_\mathrm{mc}$ of the star and minicluster/halo system by introducing a modified virialization condition of the form
\begin{align}
\frac{G M_\star}{R_\star} \simeq D_h \frac{G \mathcal{M}}{\mathcal{R}}\,, \label{eq:virial_Ansatz_v2}
\end{align}
where the perturbative coefficient $D_h$ was determined by matching the $\lambda=0$ results to those of Schive et al. \cite{schive_understanding_2014}.\\
Inserting equation \eqref{eq:virial_Ansatz_v2} into the mass-radius relation of self-interacting ALP stars, Padilla et al. \cite{padilla_core-halo_2021} showed that the extended core-halo relation scales as $M_\star \propto \sqrt{1+\Delta \lambda(\mathcal{M})}$ at $z=0$, where the corresponding perturbation term
\begin{align}
\Delta \lambda
&= 6.87 \cdot 10^{-9} \left( \frac{f_a}{ 10^{11}\, \mathrm{GeV} } \right)^{-2}  \left(\frac{\mathcal{M}}{10^{-12} M_\odot}\right)^{2 / 3}  \,\label{eq:DeltaLambda}
\end{align}
quantifies the expected modification of $M_\star$ compared to \eqref{eq:CoreHalo} and \eqref{eq:M_h_min_CoreHalo}.
We apply the above extension to $z=z_\mathrm{eq}$ and find that the predicted perturbation from self-interactions remains negligble $\Delta \lambda \leq 3\cdot 10^{-5}$ for every ALP configuration $(m_a,n)$ considered in this work (see also figure \ref{fig:DeltaLambda}).
This result is not surprising since the dilute stable branch of ALP stars is defined by dominance of gravity over short-range interactions.
Nevertheless, some modifications of the core-halo relation \eqref{eq:CoreHalo} are expected to occur once the star mass approaches the critical point $M_\star=M_{\star,\lambda}$, where both gravitational and self-interacting contributions become important at $\Delta\lambda\approx 3\cdot 10^{-5}$ (see figure \ref{fig:DeltaLambda}).
We also employ a more conservative model for possible modifications to the core-halo relation in appendix \ref{app:AS-Params}, but find for both of our approaches, that the resulting effects should be within uncertainties of the MCMF.\\
In the following, we will thus use the $\lambda=0$ relation \eqref{eq:CoreHalo}, as an order-of-magnitude estimate in the vicinity of $M_{\star,\lambda}$, while keeping in mind the need for an extended core-halo relation for more detailed predictions.
Since the uncertainties contained in the MCMF calculated from the linear theory formalism of \citep{fairbairn_structure_2018} are already large, the precision of the gravitational core-halo relation \eqref{eq:CoreHalo} is more than sufficient for our considerations.\\
We also emphasize that there is an ongoing discussion on the connection between different core-halo relations presented in the literature and that the relation found in \cite{schive_understanding_2014} is not universal.
We refer to \cite{zagorac_soliton_2023, padilla_core-halo_2021, chavanis_core_2020} for reviews and \cite{Liu_2023} for recent results on the topic.
Zagorac et al. \cite{zagorac_soliton_2023} concluded that the canonical relation \eqref{eq:CoreHalo} with power-law index $M_\star \propto \mathcal{M}^{1/3}$ provides the overall best-fit to the (averaged) data.
Based on their results and for simplicity, we will focus on the case $M_\star \propto \mathcal{M}^{1/3}$ in equation \eqref{eq:CoreHalo} throughout this paper.

\subsection{Low-Mass Cutoffs in the ALP Star Distribution} \label{subsec:ASMF_Cutoff}
\begin{figure}[t]
\centering
\includegraphics[width=\columnwidth]{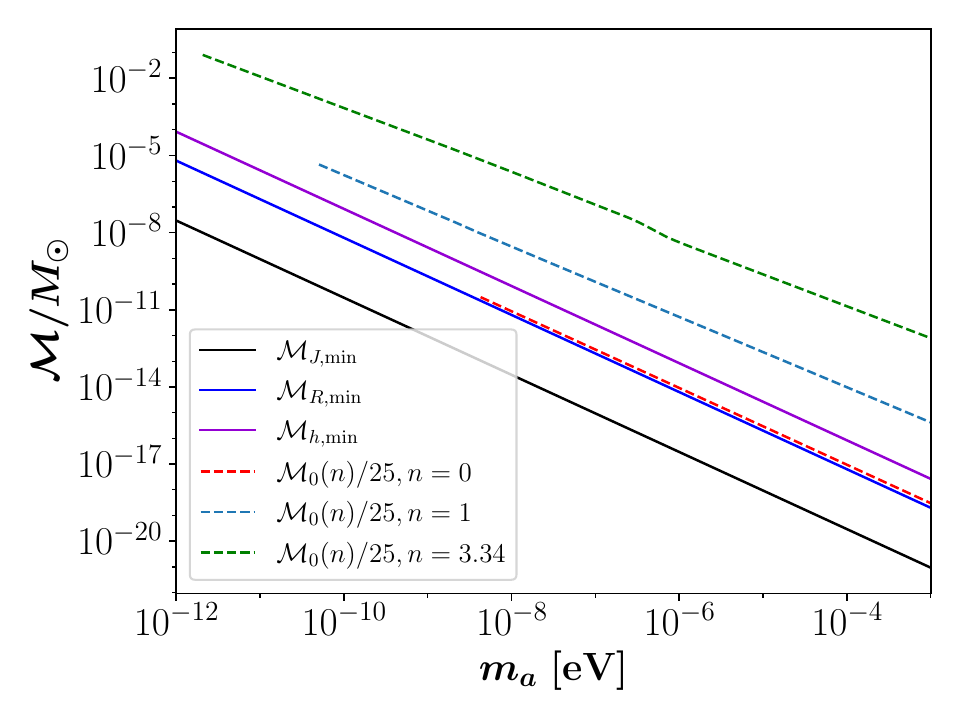}
\caption{Different MC masses setting the low-$M_\star$ cutoffs of the ASMF at $z=z_\mathrm{eq}$ in solar masses and as a function of $m_a$. Solid lines are independent of the ALP mass temperature dependence, the dashed colored lines show the different $n$-dependent $\mathcal{M}_0$-cutoffs according to equation \eqref{eq:M0_cutoff}.
\label{fig:Cutoffs}}
\end{figure}

The low-mass cutoffs to the ALP star distribution are different from the $\mathcal{M}$-cutoffs in subsection \ref{subsec:MCMF_Parametr} in the sense that they do only apply to the ASMF and not to the MCMF.
Accordingly, the total number of miniclusters in the Milky Way is less constrained compared to the total number of ALP stars, i.e. $\mathcal{N}_{\mathrm{tot}} \geq N_{\star,\mathrm{tot}}$.
Physically, this can be understood by demanding two conditions for the existence of a composite AS-MC system:
First, that the total mass of the MC predicted by the core-halo relation \eqref{eq:CoreHalo} is larger or equal to the mass of its core and secondly that the radius of the ALP star should not exceed that of its host minicluster:
\begin{align}
M_\star( \mathcal{M} ) ~&\stackrel{!}{\leq} ~\mathcal{M} \label{eq:ASMF_Cond1}\\
R_\star (M_\star) ~&\stackrel{!}{\leq}~ \mathcal{R} \label{eq:ASMF_Cond2} \,,
\end{align}
where we use the mass-radius relation \eqref{eq:Radius-Mass-Rel_Physical} in the second condition.
Note that the equality in \eqref{eq:ASMF_Cond1} is equivalent to the definition of the redshift-dependent minimum halo mass \eqref{eq:M_h_min_CoreHalo}.
Inserting the redshift of MC formation $z=z_\mathrm{eq}$ into equation \eqref{eq:M_h_min_CoreHalo}, we directly obtain the low-$M_\star$ cutoff $M_{\star,h}(z_\mathrm{eq}) \equiv \mathcal{M}_{h,\min}(z_\mathrm{eq}) = 2.36 \cdot 10^{-16}  M_\odot$.\\
The second condition, equation \eqref{eq:ASMF_Cond2}, is derived from the mass-radius relation \eqref{eq:Radius-Mass-Rel_Physical} and from the characteristic MC radius in equation \eqref{eq:R_mc}.
Setting $\mathcal{R} \stackrel{!}{=} R_\star$ and using equations \eqref{eq:Radius-Mass-Rel_Physical}, \eqref{eq:R_mc} and \eqref{eq:CoreHalo}, we find the critical minimum ALP star mass
\begin{widetext}
\begin{align}
M_{\star,R} &= 4.87 \cdot 10^{-17}\,M_\odot\,
\sqrt{\delta} (1 + \delta)^{1/6}\, 
\left( \frac{\alpha_\mathrm{kin} R_{\star,90}}{\alpha_\mathrm{grav} R_\star} \right)^{1/2}
\left( \frac{1 + z}{1 + z_\mathrm{eq}} \right)^{1/4}
\left[ \frac{\zeta(z)}{\zeta(z_\mathrm{eq})} \right]^{1/12}  \left(\frac{m_a}{50\,\mu\text{eV}}\right)^{-3/2} \,,\label{eq:M_h_min_RadiusCutoff_AS} 
\end{align}
\end{widetext}
where we dropped one term which can be neglected as long as the condition
\begin{align}
f_a \gg&\,\, 18\, \mathrm{GeV} \,\sqrt{\delta} (1+\delta)^{1/6}  \left( \frac{m}{50\,\mu\mathrm{eV}} \right)^{1/2} \nonumber \\
& \times \left( \frac{1+z}{1+z_\mathrm{eq}} \right)^{1/4} \left[ \frac{\zeta(z)}{\zeta(z_\mathrm{eq})} \right]^{1/12} 
\end{align}
is fulfilled.
In our framework with $f_a \gtrsim 10^{10}\,$GeV (see figure \ref{fig:f_a}) and for $10^{-12}\,\mathrm{eV} \leq m_a \leq 10^{-3}\,\mathrm{eV}$, this condition remains valid even for the densest miniclusters with $\delta \sim 10^4$.
It should be noted that our predictions for the radius cutoff are different from the ones in \cite{kavanagh_stellar_2021} for the simple fact that we evaluate the core-halo relation \eqref{eq:CoreHalo} at $z=z_\mathrm{eq}$ compared to $z=0$ taken by the previous authors.
Kavanagh et al. \cite{kavanagh_stellar_2021} reported that none of the MCs with $\mathcal{M} \leq 5 \cdot 10^{-16}\,M_\odot$ passed the AS cutoff at $m_a=20\,\mu$eV, $\delta \sim 0.1$ and $z=0$ using the spherical radius \eqref{eq:R_mc}.
Expressing equation \eqref{eq:M_h_min_RadiusCutoff_AS} in terms of the core-halo relation, we find that the corresponding critical minicluster mass at the radius cutoff is
\begin{widetext}
\begin{align}
\mathcal{M}_{R,\min}(z) 
&= 2.07 \cdot 10^{-18}\,M_\odot \left( \frac{\alpha_\mathrm{kin} R_{\star,90}}{\alpha_\mathrm{grav} R_\star} \right)^{3/2} \sqrt{\delta^3 (1 + \delta)}\, \left( \frac{1 + z}{1 + z_\mathrm{eq}} \right)^{-3/4} 
\left[ \frac{\zeta(z)}{\zeta(z_\mathrm{eq})} \right]^{-1/4} \left(\frac{m_a}{50\,\mu\text{eV}}\right)^{-3/2}
\label{eq:M_h_min_RadiusCutoff}
\end{align}
\end{widetext}
We can compare our prediction \eqref{eq:M_h_min_RadiusCutoff} for $z=0$ and with $m_a,\delta$ as in \citep{kavanagh_stellar_2021} to find $ \mathcal{M}_{R,\min}(0) = 6.55 \cdot 10^{-16} M_\odot$ at $m_a=20\,\mu$eV, $\delta \sim 0.1$, which is in good agreement with $\mathcal{M}_{R,\min}(0)\approx 5\cdot 10^{-16}\,M_\odot$ reported in \cite{kavanagh_stellar_2021}.
We believe that, since the MCs collapse around matter-radiation equality and decouple from the cosmic expansion at this time, thus freezing the redshift-dependence of the collapsed system, taking $z=z_\mathrm{eq}$ is the correct approach to take.
Nevertheless, the redshift dependence of the axion- and ALP minicluster systems is subject to open debate which is why kept track of it in our calculations.\\
For comparison, we have plotted the MC masses predicted from different cutoffs in figure \ref{fig:Cutoffs}.
\noindent
All of the shown cutoffs scale as $\mathcal{M} \propto m_a^{-3/2}$, up to small corrections due to the ALP temperature evolution in the case of the dashed colored $\mathcal{M}_0$-cutoffs with $n>0$.
In the absence of the $\mathcal{M}_0$-cutoff (and for $n=0$) we find that the most stringent requirement for the existence of AS-MC systems is the purple minimum halo mass $\mathcal{M}_{h,\min}$ for every $m_a$ in figure \ref{fig:Cutoffs}, i.e. $\min(M_\star)=M_{\star,h}$.
This prediction is again different from the results in \cite{kavanagh_stellar_2021} due to the different redshift-dependence of $M_{\star,h}(z)$ and $M_{\star,R}(z)$ in equations \eqref{eq:M_h_min_CoreHalo} and \eqref{eq:M_h_min_RadiusCutoff_AS}.\\
Let us mention for completeness, that we also implement an additional cutoff to the ALP parameter space $m_a,n$, where the minimum AS mass in the ASMF becomes comparable to the maximum stable AS mass $\min(M_\star) \approx M_{\star,\lambda}$ and the gravitational limit of the core-halo relation breaks down for the entire AS population.
However this condition only applies to a small region of the low-$m_a$ component of ALPs with $n\geq3.34$ and for the $\mathcal{M}_0$-cutoff.

\subsection{ALP Star Mass Distributions} \label{subsec:ASMF-Results}
\begin{figure}[t]
\centering
\includegraphics[width=\columnwidth]{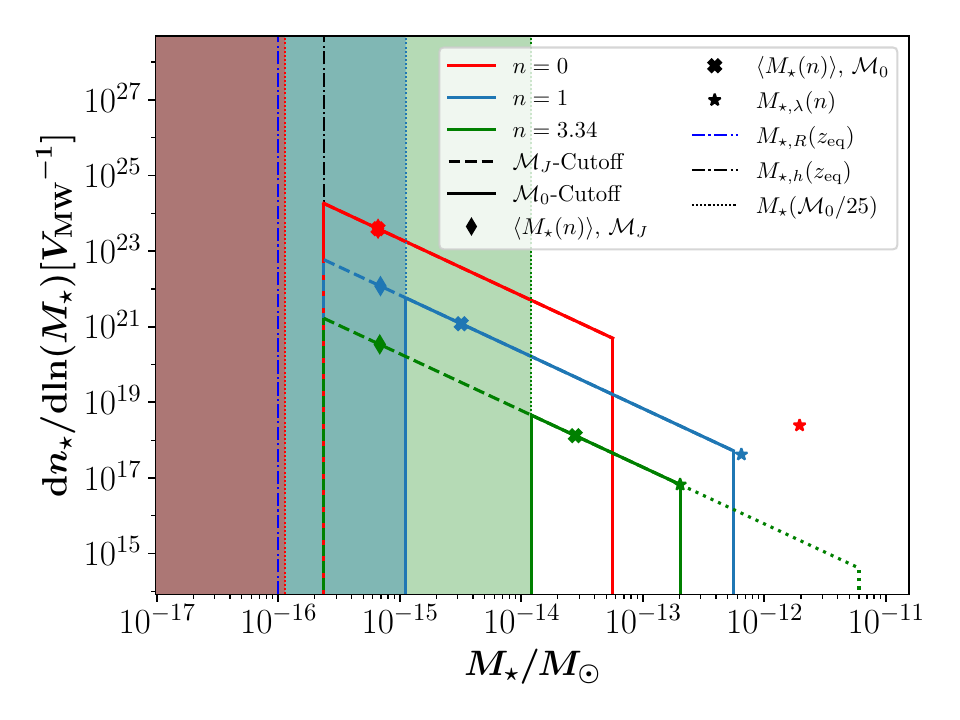}
\caption{ASMF per Milky Way volume obtained from the MCMF for $\alpha=-1/2$ in figure \ref{fig:MCMF}. 
Colored lines and symbols indicate AS masses at different $n=0,1,3.34$; the shaded regions and thin dotted colored lines denote the low-mass cutoffs given by $\mathcal{M}_0(m_a,n)/25$. 
Solid and dashed lines indicate the ASMF with and without applying the $\mathcal{M}_0$-cutoffs; dash-dotted lines represent the radius cutoff \eqref{eq:M_h_min_RadiusCutoff_AS} in purple and the core-halo cutoff \eqref{eq:M_h_min_CoreHalo} in black.
Colored stars refer to the maximum stable AS mass $M_{\star,\lambda}$ from equation \eqref{eq:CoreHalo}, above which the $n=3.34$ component is truncated due to instability (see thick green dotted line).
The average AS masses from equation \eqref{eq:M_average} for the two low-$\mathcal{M}$ cutoffs from subsection \ref{subsec:MCMF_Parametr} are shown in colored diamonds and crosses .
\label{fig:ASMF}}
\end{figure}

\begin{figure}[t]
\centering
\includegraphics[width=\columnwidth]{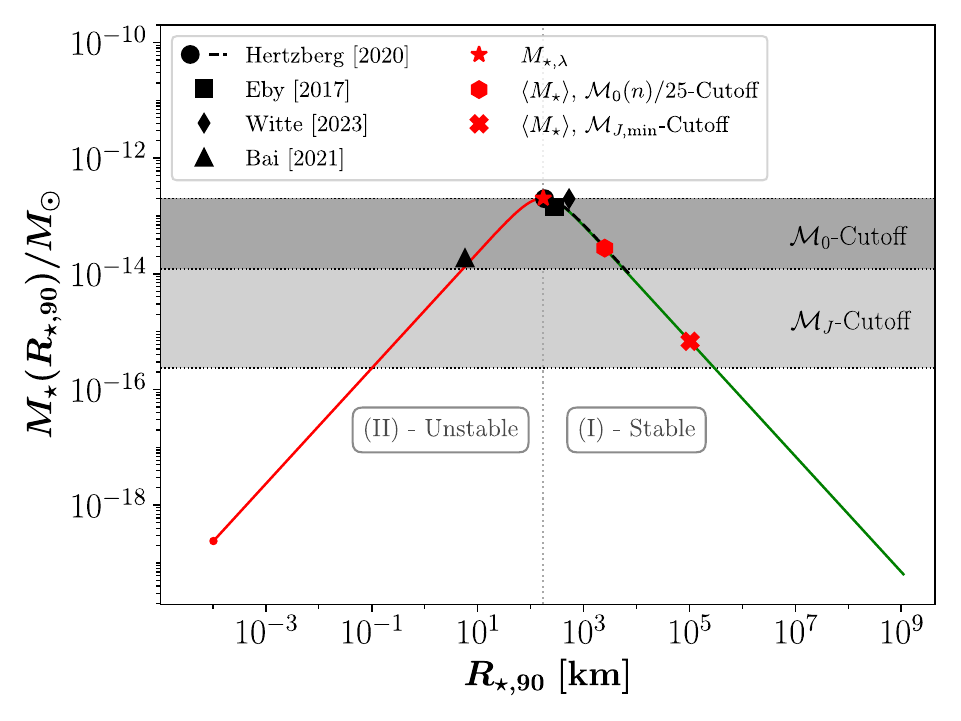}
\caption{Mass-radius relation \eqref{eq:Radius-Mass-Rel} with AS properties inferred from the MCMF in figure \ref{fig:MCMF} using the core-halo relation \eqref{eq:CoreHalo} for QCD axions with $m_a=50\,\mu$eV, $n=3.34$ and $f_a\simeq 10^{11}\,$GeV. 
Green and red lines show the stable and unstable branches, black symbols denote the AS parameters used in the literature.
Light/dark grey shaded areas and dotted lines correspond to the ASMF with the $\mathcal{M}_{J}$-/$\mathcal{M}_0$-cutoff in figure \ref{fig:ASMF}.
The upper (dotted) boundary of the shaded areas corresponds to the upper AS mass limit given by the maximum stable mass $M_{\star,\lambda}$ shown by the red star.
Average AS masses from equation \eqref{eq:M_average} obtained from the two MCMF cutoffs are labelled by red symbols.}\label{fig:Mass-Radius_QCD}
\end{figure}

\begin{figure}[t]
\centering
\includegraphics[width=\columnwidth]{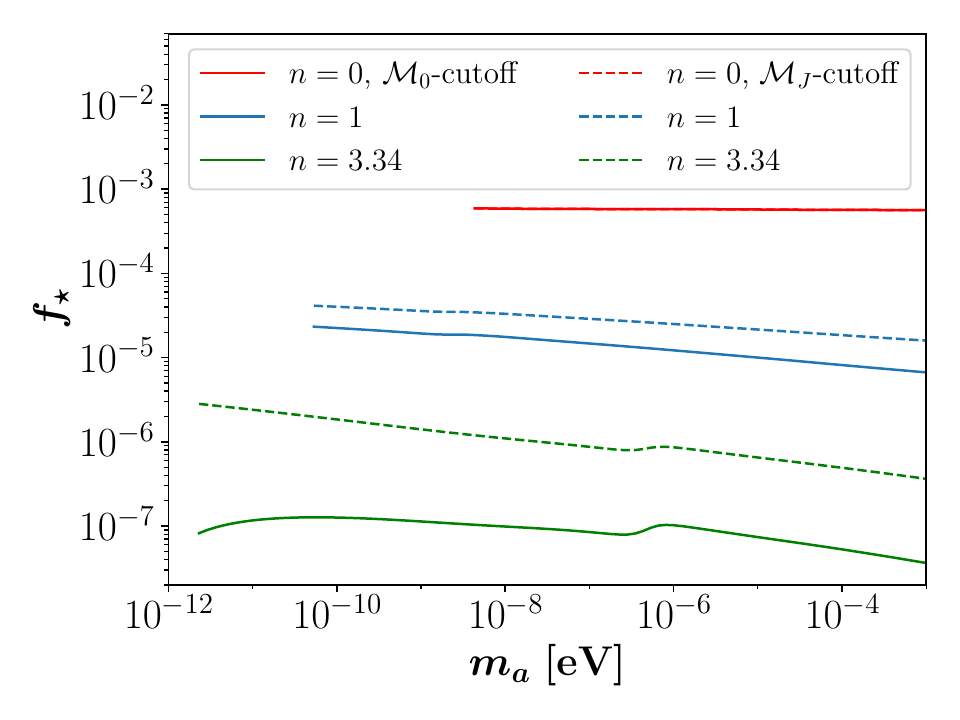}
\caption{ALP star abundance $f_\star$ from equation \eqref{eq:f_AS_alpha} at different ALP masses $m_a$ and for $n=0,1,3.34$ in colored lines.
Dashed/solid lines show the results obtained using the $\mathcal{M}_J$-/$\mathcal{M}_0$-cutoffs of the MCMF from subsection \ref{subsec:MCMF_Parametr}, both for $\alpha=-1/2$.
\label{fig:f_AS}}
\end{figure}

\begin{figure}[t]
\centering
\includegraphics[width=\columnwidth]{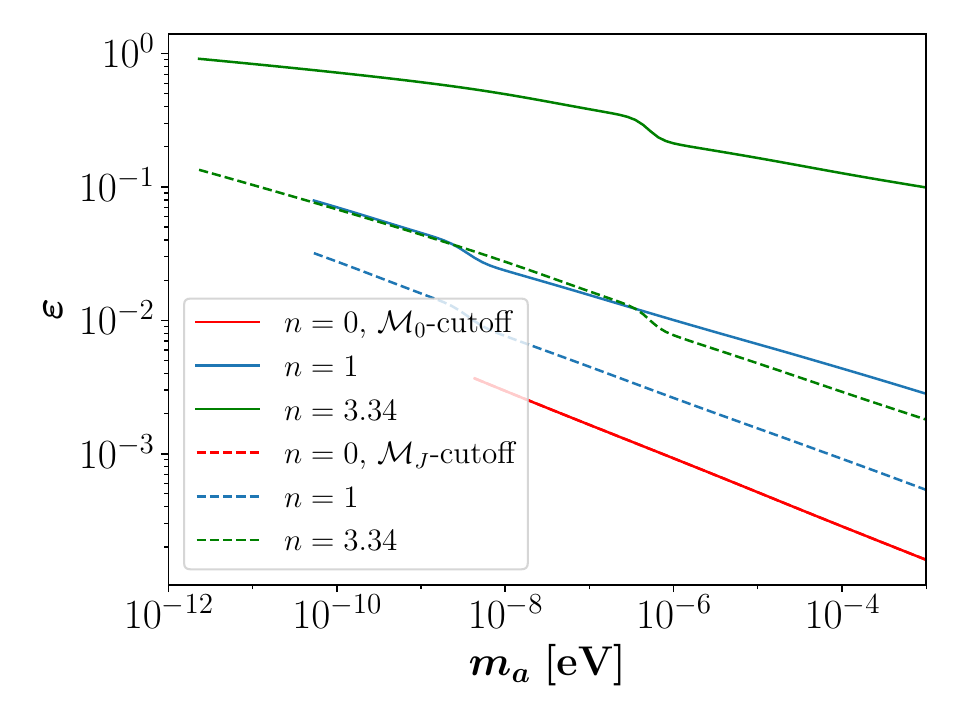}
\caption{ALP star mass parameter $\varepsilon$ from equation \eqref{eq:f_AS_alpha} at different ALP masses $m_a$ and for $n=0,1,3.34$ in colored lines for $\alpha=-1/2$.
\label{fig:epsilon_AS}}
\end{figure}

Considering the restrictions from the core-halo cutoff \eqref{eq:M_h_min_CoreHalo} and from the radius cutoff \eqref{eq:M_h_min_RadiusCutoff_AS}, we can finally infer the ALP star mass distribution from the MCMF for both the $\mathcal{M}_J$- and the $\mathcal{M}_0$-cutoff from subsection \ref{subsec:MCMF_Norm}.
We characterize the ASMF using the AS number density $\mathrm{d}n_\star/\mathrm{d}\ln(M_\star)$ per logarithmic star mass interval and use the identity $\mathrm{d}n = \mathrm{d}n_\star$ together with the core-halo relation \eqref{eq:CoreHalo}.
Figure \ref{fig:ASMF} shows the corresponding ASMF obtained from the representative MCMF in figure \ref{fig:MCMF} with $m_a=50\,\mu$eV and $\alpha=-1/2$.
The maximum stable AS mass $M_\star = M_{\star,\lambda}$ due to self-interactions is indicated with colored stars, while dashed and solid colored lines show the ASMFs obtained from the two different low-$\mathcal{M}$ cutoffs.\\
Note that in the case of $n=3.34$ in green we have additionally applied a high-$M_\star$ cutoff (shown in dotted green lines) introduced by the maximum stable AS mass $M_{\star,\lambda}$.
The relatively small number of such ALP stars could have reached a critical stage resulting in a relativistic \textit{Bosenova} as demonstrated in \cite{levkov_relativistic_2017}.
In theory, the cyclic explosions of this event could introduce repeated mass-loss until the star becomes sub-critical again leading to $M_\star \leq M_{\star,\lambda}$.
However, we will ignore the super-critical AS component in the remainder of this paper due to their small abundance and since the details and long-time evolution of this process are currently unknown.\\
In agreement with figure \ref{fig:Cutoffs}, the $n=0$ population in red in figure \ref{fig:ASMF} is truncated by the core-halo requirement \eqref{eq:ASMF_Cond1} in black dash-dotted lines for both of the MCMF cutoffs.
For $n=1,3.34$ in blue and green, either the core-halo cutoff or the $\mathcal{M}_0$-cutoff truncate the low-$M_\star$ component of the corresponding ASMF.
This means that for the $\mathcal{M}_J$-cutoff, numerous miniclusters will not have an ALP star core due to their mass being below the minimum threshold $\mathcal{M}_{h,\min}$ from equation \eqref{eq:M_h_min_CoreHalo}.\\
We emphasize that one important feature of the ASMF and MCMF in figures \ref{fig:MCMF} and \ref{fig:ASMF} is the approximate independence of the high-mass population from the low-mass cutoffs.
\noindent
This means that even with different $\mathcal{M}_{\min}$ and the corresponding large uncertainties in the low-$\mathcal{M}$-cutoffs, the abundance of high-mass MCs and ASs does not change significantly.
The physical reason for this weak dependence is the fact that we fix the number of AS-MC systems by their total mass, to which the high-mass tail yields the largest contribution.
In contrast, the total number of ASs/MCs depends sensitively on the low-$\mathcal{M}$ cutoffs as mentioned in subsection \ref{subsec:MCMF_Norm} (compare equation \eqref{eq:N_tot_mc} and figures \ref{fig:N_tot_MC}, \ref{fig:N_tot_AS}).\\
Similar to the MCMF in subsection \ref{subsec:MCMF_Norm}, we conclude that the spread of the AS mass distribution is determined by the different low-mass cutoffs and by the temperature dependence $n$ of the ALP.
Using the parametrization from \cite{fairbairn_structure_2018}, the ASMF cutoff $\min(M_\star)$ from subsection \ref{subsec:ASMF_Cutoff} and $\max(M_\star)=\min(M_\star(\mathcal{M}_{\max}), M_{\star,\lambda})$, we can directly calculate the total mass and number of ALP stars in the Milky Way for different $m_a$ and $n$.
A simple integration yields the following expression for the total mass contained in ALP stars
\begin{align}
M_{\star,\mathrm{tot}} 
&= 4 \pi R_{200}^3 \int_{\min(M_\star)}^{\max(M_\star)} dM_\star\,M_\star \frac{C_\mathrm{ren}M_\star^{3\alpha - 1} }{\mathcal{M}_{\min}^\alpha \mathcal{M}_{h,\min}^{2\alpha}(z)} \nonumber \\
&=  \frac{4\pi R_{200}^3 C_\mathrm{ren}}{3\alpha + 1}  \frac{\min(M_\star)^{3\alpha + 1} - \max(M_\star)^{3\alpha + 1}}{\mathcal{M}_{\min}^{\alpha} \mathcal{M}_{h,\min}^{2\alpha}(z)} \,, \label{eq:M_tot_star}
\end{align}
where $C_\mathrm{ren}$ is determined by the normalization \eqref{eq:M_tot_mc} of the MCMF.
Similarly, the total number of ALP stars is given by
\begin{align}
N_{\star,\mathrm{tot}} 
&= 4 \pi R_{200}^3 \int_{\min(M_\star)}^{\max(M_\star)} dM_\star \frac{C_\mathrm{ren} M_\star^{3\alpha - 1}}{\mathcal{M}_{\min}^\alpha \mathcal{M}_{h,\min}^{2\alpha}(z)} \nonumber \\
  &= \frac{4\pi R_{200}^3 C_\mathrm{ren}}{3\alpha}  \frac{\min(M_\star)^{3\alpha} - \max(M_\star)^{3\alpha}}{\mathcal{M}_{\min}^{\alpha} \mathcal{M}_{h,\min}^{2\alpha}(z)}  \label{eq:N_tot_star} \,.
\end{align}
We repeat the corresponding procedure of determining the relevant high- and low-mass cutoffs for both the MCMF and ASMF for $10^{-12}\,\mathrm{eV} \leq m_a \leq 10^{-3}\,$eV and $n=0,1,3.34$.
This way, we can obtain the AS properties $M_\star$, $R_\star $, $N_{\star,\mathrm{tot}}$ as well as their MC equivalents from the MCMF as a function of $m_a,n$.\\
For now we continue to evaluate the exemplary case of the QCD axion with $m=50\,\mu$eV and $n=3.34$ from figures \ref{fig:MCMF} and \ref{fig:ASMF} by updating the corresponding mass-radius relation from figure \ref{fig:Mass-Radius_QCD_minimal} with the results from figure \ref{fig:ASMF}.
The AS distributions derived from the $\mathcal{M}_J$- and $\mathcal{M}_0$-cutoff are indicated by the light and dark grey shaded regions in figure \ref{fig:Mass-Radius_QCD}.
For comparison we have additionally plotted the characteristic ALP star parameters that were previously used in other references on axion/ALP star phenomenology, namely in \cite{hertzberg_merger_2020}, \cite{eby_collisions_2017}, \cite{witte_transient_2023} and \cite{bai_diluted_2022}.
\noindent
Most of the previous authors used the maximum stable mass of ALP stars $M_{\star,\lambda}$ as a representative value for the AS mass, as seen by the clustering of the black symbols in figure \ref{fig:Mass-Radius_QCD}.
We confirm the existence of such critical ALP stars but find that their abundance is generally much lower than previously assumed.
One simple reason for this is the fact that the majority of the galactic dark matter in our approach is contained in miniclusters but not in their often much lighter ALP stars with masses $M_\star \leq M_{\star,\lambda}$.
Especially the intermediate- and high-$\mathcal{M}$ tail of the MCMF in figure \ref{fig:MCMF} has masses $\mathcal{M} \gg M_{\star,\lambda}$ which is why $M_{\star,\mathrm{tot}} \ll \mathcal{M}_\mathrm{tot}$.
An additional factor reducing the AS abundance is the negative slope of the MCMF, which peaks at the lowest MC masses $\mathcal{M}\sim \mathcal{M}_{\min}$ well before the core-halo cutoff mass $\mathcal{M}_{h,\min}$ restricting the presence of ASs (see equations \eqref{eq:M_h_min_J} and \eqref{eq:M_h_min_CoreHalo}).\\
The authors of \cite{hertzberg_merger_2020} and \cite{eby_collisions_2017} used a particularly simple approach for describing this observation, which we will introduce for comparison in the following.
They expressed the number and typical mass of ALP stars in terms of two parameters $f_\star$ and $\varepsilon$ by setting
\begin{align}
f_\star &= \frac{M_{\star,\mathrm{tot}}}{\mathcal{M}_\mathrm{MW}}
\quad \,,\quad
\varepsilon = \frac{\langle M_\star \rangle}{M_{\star,\lambda}} \label{eq:f_AS_alpha} \,,
\end{align}
where $f_\star \in [0,1]$ describes the relative DM abundance of ALP stars and $\varepsilon \in (0,1]$ their typical masses in terms of the maximum mass $M_{\star,\lambda}$.
The typical mass of ALP stars $\langle M_\star \rangle$ is determined from the average \eqref{eq:M_average} over the AS mass distribution and shown in figure \ref{fig:N_tot_AS}.
We can directly calculate our predictions for these two parameters from the galactic AS-MC distributions and show the results for $f_\star$ and $\varepsilon$ in figures \ref{fig:f_AS} and \ref{fig:epsilon_AS}.\\
As before, dashed and solid colored lines show the $\mathcal{M}_J$- and $\mathcal{M}_0$-cutoffs of the MCMF respectively.
\noindent
For $n=0$ in red, both cutoffs coincide since the core-halo cutoff \eqref{eq:M_h_min_CoreHalo} is more stringent than both $\mathcal{M}_{J,\min}$ and $\mathcal{M}_{0,\min}$.
The cases $n=1$ and $n=3.34$ in contrast show significant deviations between the two low-$\mathcal{M}$ cutoffs in both $f_\star$ in figure \ref{fig:f_AS} and $\varepsilon$ in figure \ref{fig:epsilon_AS}, as expected from the MCMFs in figure \ref{fig:MCMF}.
The reason for this is the truncation of the low-$M_\star$ population due to the $\mathcal{M}_0$-cutoff (compare fig. \ref{fig:ASMF}).\\
Apart from the cutoff dependence, the strongest impact on $f_\star$ and $\varepsilon$ is given by the temperature evolution $n$, which sets the characteristic MC mass $\mathcal{M}_0$ determining both the typical MC- and AS parameters.
For $f_\star$ in figure \ref{fig:f_AS}, this effect can be understood in terms of the core-halo scaling $M_\star \propto \mathcal{M}_0^{1/3}$, with temperature index $n$.
Accordingly, the relative fraction of MC mass contained in ALP star cores will decrease with larger $\mathcal{M}_0$ or equivalently with larger $n$ as seen in figure \ref{fig:f_AS}.
Similarly we can trace the scaling of the core-halo relation for $\langle M_\star \rangle$, which implies that $\varepsilon = \langle M_\star \rangle / M_{\star,\lambda}$ will increase with larger $n$ as seen in figure \ref{fig:epsilon_AS}.\\
The weak dependence of $f_\star$ on $m_a$ can be estimated for the exemplary case $n=0$ in red lines in figure \ref{fig:f_AS}.
The normalization condition $\mathcal{M}_\mathrm{tot}\stackrel{!}{=}f_\mathrm{mc}\,\mathcal{M}_\mathrm{MW}$ applied to equation \eqref{eq:M_tot_mc}  implies that the constant $C_\mathrm{ren}$ inherits a scaling $C_\mathrm{ren} \propto \mathcal{M}_0^{-1}$.
Additionally taking $\min(M_\star)^{-1/2} \gg \max(M_\star)^{-1/2}$ in equation \eqref{eq:M_tot_star} and inserting the scalings of $\mathcal{M}_J$, $\mathcal{M}_{h,\min}$, $\mathcal{M}_0(n=0) \propto m_a^{-3/2}$ from equations \eqref{eq:M_h_min_J}, \eqref{eq:M_h_min_CoreHalo}, and figure \ref{fig:M0}, the total galactic ALP star mass scales with $m_a$ as
\begin{align}
M_{\star,\mathrm{tot}}(n=0) & \propto C_\mathrm{ren} \mathcal{M}_{h,\min} \mathcal{M}_{\min}^{1/2} \min(M_\star)^{-1/2}\nonumber \\
& \propto 
\mathcal{M}_0^{-1} \mathcal{M}_{h,\min} \mathcal{M}_{\min}^{1/2} m_a^{-3/4} \nonumber \\
& \propto \mathrm{const}\,, 
 \label{eq:M_tot_star_scaling}
\end{align}
where we have inserted $\mathcal{M}_{\min}\propto m_a^{-3/2}$ for the two low-$\mathcal{M}$ cutoffs of the MCMF and used $\alpha=-1/2$.
The minimum star mass $\min(M_\star)$ for $n=0$ in equation \eqref{eq:M_tot_star_scaling} is derived from the core-halo cutoff $\mathcal{M}_{h,\min}$, which allowed us to rewrite $\min(M_\star)=M_\star(\mathcal{M}_{h,\min})\propto m_a^{-3/2}$ using the core-halo relation \eqref{eq:CoreHalo}.
Combining equation \eqref{eq:M_tot_star_scaling} with $f_\star\propto M_{\star,\mathrm{tot}}$ directly yields $f_\star(m_a) \propto \mathrm{const}$ from this.
For $n>0$ in figure \ref{fig:f_AS}, $f_\star$ is only roughly constant in $m_a$, since the scaling $\mathcal{M}_0\propto m_a^{-3/2}$ is slightly broken by the temperature evolution of the ALP mass (see also figure \ref{fig:M0}).\\
To summarize, our predictions for the ASMF suggest that the assumptions $10^{-4} \leq f_\star \lesssim 1$ and $\varepsilon \lesssim 1$ \cite{hertzberg_merger_2020} taken by previous authors are generally inadequate when dealing with ALP stars inside galactic miniclusters.
While a considerable fraction of ALP stars can reach $\varepsilon \sim 1$ for $n=1,3.34$, their abundance is expected to be strongly suppressed $10^{-7} < f_\star < 10^{-4}$ by the large mass contribution of the MC population.
Conversely, the case $n=0$ yields the largest abundance of ALP stars $f_\star \simeq 10^{-3}$, albeit at drastically smaller star masses $\varepsilon \lesssim 10^{-3}$.

\section{Implications for experimental Detection of ALP Stars} \label{sec:detection}
We will now use the results from section \ref{sec:ASMF} to re-evaluate different detection mechanisms for AS-MC systems.
Some of the most recent and promising scenarios involve the resonant conversion of ALP dark matter in the magnetic field of neutron stars in subsections \ref{subsec:NS-AS} and \ref{subsec:NS-MC} and the collapse of near-critical ALP stars leading to a Bosenovae in subsection \ref{subsec:bosenova}.
We also suggest a new mechanism of radio emission for ALP stars in section \ref{subsec:Glowing_AS} which we analyze in more detail in a follow-up paper.
Before focussing on the specific phenomena in subsections \ref{subsec:NS-AS} - \ref{subsec:Glowing_AS}, we will briefly introduce our calculation of the different collision rates.\\
In the following we calculate the mass-integrated rates of collisions between astrophysical objects and ALP stars using Milky Way parameters (see appendix \ref{app:MW} and \cite{hertzberg_merger_2020}).
We use the indices '$i$' and '$j$' to label different types of objects and introduce the symmetry factor
\begin{align}
S&= \begin{cases}
      \frac{1}{2} & i = j,\\
      1 &  i \neq j
    \end{cases}
\end{align}
to prevent double counting for $i=j$.
The total rate of collisions per year and galaxy can then be obtained by integrating over the galactocentric radius $r$ and over the AS-/MC masss distribution $M_i$
\begin{align}
    \Gamma_{i-j} =&\,\, 4 \pi S \int_{0}^{R}d r\, r^2
    n_i(r)
    n_j(r) \nonumber \\
    & \times \int\,dM_i\, p_i(M_i) \left\langle \sigma_{\text{eff}}(v, M_i) \,v \right\rangle_v \,,\label{eq:Gamma_Integrated_Spherical}
\end{align}
where $n_i(r),n_j(r)$ are the radially symmetric number densities and
\begin{align}
    \sigma_{\mathrm{eff}}(v, M_i, M_j, R_i, R_j)
    & = \pi\left(R_{i} + R_j \right)^{2}(1 + \eta) \,,
\end{align}
where 
\begin{align}
    \eta = \frac{2 G (M_i + M_j)}{(R_i + R_j)\, v^{2}} \label{eq:Cross-Section}
\end{align}
is the scattering cross section with gravitational enhancement $\eta$ as a function of the relative velocity $v$.
The mass distribution of $i$ follows a probability distribution function $p_i(M_i)$  obtained from equation \eqref{eq:MCMF_Norm}.
We set the escape velocity of the Milky Way $v_\mathrm{esc}=622\,$kms$^{-1}$ \citep{mcmillan_mass_2011} as an upper limit on $v$ and define the velocity-averaged cross section indicated by the brackets '$\langle \, \rangle_v$' in equation \eqref{eq:Gamma_Integrated_Spherical} as
\begin{align}
    \langle\sigma_{\mathrm{eff}}(v) v\rangle_v
    &= 4 \pi \int_{0}^{v_{\mathrm{esc}}} p_v(v) \sigma_{\mathrm{eff}}(v) v^{3} d v \label{eq:SigmaVAverage}
\end{align}
with the Gaussian velocity distribution
\begin{align}
    p_v(v) = \frac{1}{(\pi v_0^2)^{3/2}} \,\exp \left(-\frac{v^2}{v_{0}^{2}}\right)\,,\label{eq:Momentum-Distr}
\end{align}
obeying the normalization condition $4\pi \int_0^{v_{\mathrm{esc}}}dv v^2 p_v(v) = 1$.
The reference velocity $v_0=239\,$km\,s$^{-1}$ \citep{mcmillan_mass_2011} is set to the virial velocity of the MW dark matter halo and the normalization constant $p_v(0)\approx 1 / (\pi v_0^2)^{3/2}$ in equation \eqref{eq:Momentum-Distr} was approximated for $v_\mathrm{esc} \gtrsim v_0$ \citep{hertzberg_merger_2020}.\\
While the NFW dark matter halo exhibits a spherical symmetry allowing us to integrate $\Gamma_{i-j}$ according to equation \eqref{eq:Gamma_Integrated_Spherical} in the case of AS- and MC collisions, the baryonic matter distribution of the MW follows a disc and bulge profile.
For collisions involving neutron stars (NS) with $R_{NS}=10\,$km and $M_{NS}=1.4\,M_\odot$ we will thus use cylindric coordinates instead and express the galactocentric radial coordinate $r=\sqrt{\rho^2+z^2}$ in terms of its cylindric counterpart $\rho$.
This way, we can integrate over the galactic NS distribution according to the equivalent form
\begin{align}
    \Gamma_{i-NS}=&\,\,4 \pi S \int_{0}^{R_\rho} d \rho\,\rho \int_0^{R_z} dz\,
    n_{NS}(\rho, z)\, n_j\left(\sqrt{\rho^2+z^2}\right) \nonumber \\
    & \times \int\,dM_i\, p_i(M_i) \left\langle \sigma_{\text{eff}}(v, M_i) \,v \right\rangle_v \,, \label{eq:Gamma_Integrated_Cylindrical}
\end{align}
where the boundaries $R_\rho= 50\,$kpc, $R_z = 25\,$kpc are fixed by the fits in \citep{taani_modeling_2012}.
The number densities $n_i$ used in the following sections are the AS density, MC density and NS density
\begin{align}
n_\star(r) 
&= C_{\star}\,\rho_{NFW}(r),\\
n_\mathrm{mc} (r) 
&= C_\mathrm{mc} \,\rho_{NFW}(r)\,,\\
n_{NS}(\rho,z) 
&= \frac{C_{NS} }{2 \pi \rho} p_\rho(\rho) p_z(\rho,z)
    \label{eq:number-densities} \,,
\end{align}
where the normalization constants $C_\star$, $C_\mathrm{mc}$ with units of inverse mass are set by requiring
\begin{align}
N_{\star,\mathrm{tot}} &= 4\pi \int dr\,r^2 n_\star(r) \\
\mathcal{N}_\mathrm{tot} &= 4\pi \int dr\,r^2 n_\mathrm{mc}(r)\label{eq:n_star_Normalization}
\end{align}
with $N_{\star,\mathrm{tot}}$, $\mathcal{N}_\mathrm{tot}$ according to equations \eqref{eq:N_tot_star} and \eqref{eq:N_tot_mc}.
The neutron star number density and its dimensionless normalization constant $C_{NS}$ are similarly determined by requiring
\begin{align}
N_{NS} 
&= 2 \int_{0}^{R_\rho} d\rho \int_{0}^{R_z} dz\, C_{NS}\, p_\rho(\rho)\, p_z(\rho,z) 
\end{align}
with $p_\rho(\rho), p_z(\rho,z)$ taken from the phenomenological fit to the galactic NS distribution introduced in \citep{taani_modeling_2012} and summarized in appendix \ref{app:MW_NS}.

\subsection{Neutron-Star-ALP-Star Collisions} \label{subsec:NS-AS}
\begin{figure}[t]
\centering
\includegraphics[width=\columnwidth]{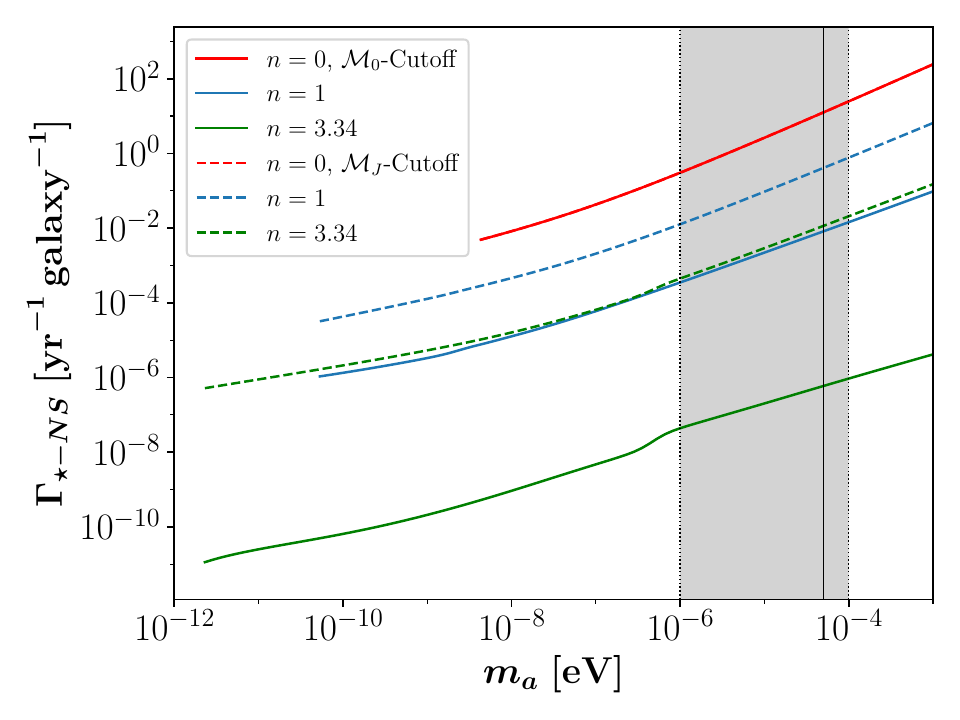}
\caption{Mass-integrated collision rates per year and galaxy between ALP stars and neutron stars in the Milky Way  as a function of $m_a$ with MCMF power-law index $\alpha=-1/2$.
Colored lines indicate the temperature dependence $n$ of the ALP mass, solid and dashed lines represent the two different low-$\mathcal{M}$ cutoffs of the MCMF in subsection \ref{subsec:MCMF_Parametr}. 
The cosmological band $10^{-6}\, \mathrm{eV} \leq m_a \leq 10^{-4}\,\mathrm{eV}$ of the QCD axion is indicated by the grey-shaded region and the black solid line marks $m_a=50\,\mu$eV.
\label{fig:Int_Coll_NS_AS0}}
\end{figure}

\begin{figure}[t]
\centering
\includegraphics[width=\columnwidth]{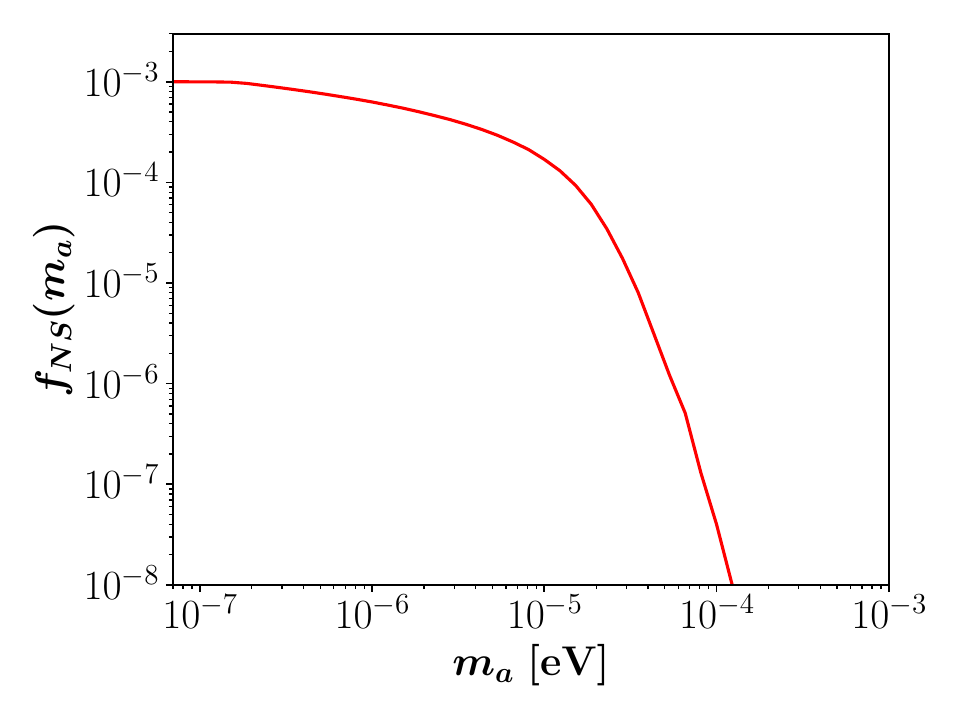}
\caption{Fraction $f_{NS}(m_a)$ of active neutron stars in our mock model which exhibit a plasma frequency $\omega_p \gtrsim m_a$ enabling the radio conversion of axion-like particles with mass $m_a$.
\label{fig:f_NS}}
\end{figure}

\begin{figure}[t]
\centering
\includegraphics[width=\columnwidth]{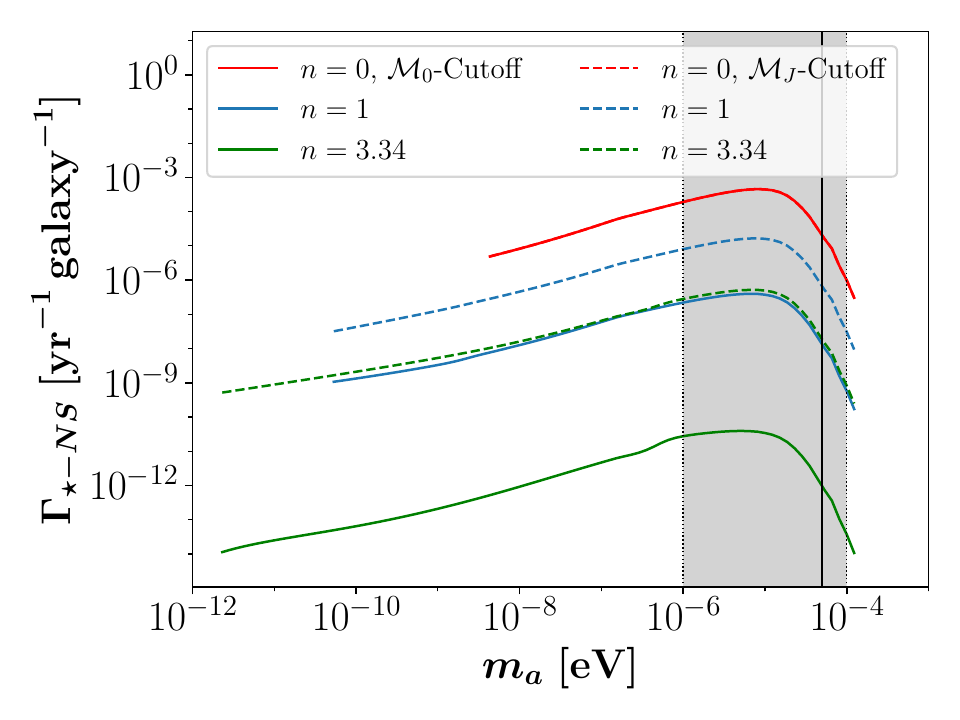}
\caption{Mass-integrated NS-AS signal rates per year and galaxy in the Milky Way from figure \ref{fig:Int_Coll_NS_AS0} after applying the NS fraction $f_{NS}(m_a)$.
Colored lines indicate the temperature dependence of the ALP mass, solid and dashed lines represent the two different low-mass cutoffs of the MCMF. 
\label{fig:Int_Coll_NS-AS}}
\end{figure}

One topic that has received a lot of attention in recent years is the prospect of radio emission from ALP stars by resonant conversion of axion-like dark matter inside the magnetic fields of neutron stars.
We note that the resonance in this scenario amounts to the equality of ALP mass and photon plasma frequency on the conversion surface $R_\mathrm{res}$ of the neutron star and that it is fundamentally different from the parametric resonance occurring inside the ALP stars in subsection \ref{subsec:Glowing_AS}.\\
The expected rate of NS-AS collisions in our galaxy has been estimated by numerous authors before, e.g. \citep{eby_collisions_2017, bai_diluted_2022, edwards_transient_2021} to name some.
Our work improves previous predictions by incorporating the two MCMF cutoffs from subsection \ref{subsec:MCMF_Parametr}, two MCMF slopes $\alpha=-1/2$, $\alpha=-0,7$ and the full ALP star mass distributions from section \ref{sec:ASMF} at different $m_a$ and $n$.
We also extend the results from previous works by calculating the fully mass-integrated collision rates for ALP stars and -miniclusters with other astrophysical objects, similar to what was done for MCs in \citep{edwards_transient_2021} but with a simplified model for MC survival.\\
The resulting event rates for collisions between NSs and ASs are given in figure \ref{fig:Int_Coll_NS_AS0}.
\noindent
As before, solid lines indicate the $\mathcal{M}_0$-cutoff of the MCMF, while dashed lines show the results using the lower Jeans mass cutoff with $\mathcal{M}_{J,\min}$.
The encounter rates in figure \ref{fig:Int_Coll_NS_AS0} suggest that for larger ALP masses $m_a$ and smaller temperature dependence $n$, a considerable range of ALP parameters could be detected.
This is both due to the fact that for smaller $n$, the average AS radius $R_\star \propto 1 / M_\star$ is significantly larger, thus enhancing the cross section, and that equivalently for lighter $M_\star,\mathcal{M}_0$, the total number of ALP stars $N_{\star,\mathrm{tot}} \lesssim \mathcal{N}_\mathrm{tot} \propto 1 / \mathcal{M}_0$ is increased.
Depending on the temperature index $n$, the NS-AS collision rates will have different scalings with $m_a$ and $n$.
In the exemplary case $n=0$, we can use the identities $N_{\star,\mathrm{tot}} \propto \mathcal{M}_0^{-1}\propto m_a^{3/2}$ and $\langle R_\star \rangle \propto m_a^{-2} \langle M_\star \rangle^{-1} \propto m_a^{-2} \mathcal{M}_0^{-1} \propto m^{-1/2}$ from equations \eqref{eq:M_tot_star} and \eqref{eq:Radius-Mass-Rel_Physical} to find that
\begin{align}
\Gamma_{NS-\star}\Big|_{n=0} & \propto
\begin{cases}
N_{\star,\mathrm{tot}} \langle R_\star \rangle^2
\propto m_a^{1/2}, \quad &\eta < 1\\
N_{\star,\mathrm{tot}} \langle R_\star \rangle M_{NS}
\propto m_a, \quad &\eta > 1
\label{eq:NS-AS_Scaling}
\end{cases} \,,
\end{align}
where $M_{NS} \gg M_\star$, $M_{NS}=\text{const}$ and the turnaround is reached when the gravitational enhancement term $\eta$ in equation \eqref{eq:Cross-Section} becomes relevant.
The other cases with $n>0$ roughly follow the same trend, but with different turnarounds for $\eta$ and marginally different scalings with $m_a$ from the different temperature evolution.
Note that for $n=0$ the results in figures \ref{fig:Int_Coll_NS_AS0} and \ref{fig:Int_Coll_NS-AS} are independent of the low-$\mathcal{M}$ cutoff because the minimum MC mass $\mathcal{M}_{h,\min} > \mathcal{M}_0/25 > \mathcal{M}_J$ from equation \eqref{eq:M_h_min_CoreHalo} is the dominant constraint of the ASMF.\\
It is important to note however that the detection of a radio signal from an AS-NS encounter requires the neutron star to have an active magnetic field with suitable photon plasma frequency $\omega_p \gtrsim m_a$ to allow for the conversion of ALPs into photons at the NS conversion surface.
In order to quantify the fraction of suitable NS collisions for both ASs and MCs (in subsection \ref{subsec:NS-MC}), we will introduce the following procedure:
We use a mock population model for the magnetic field strength $B$ and rotation frequency $\Omega_{NS} =2\pi/P$, on which the Goldreich-Julian plasma frequency $\omega_p \approx \sqrt{4\pi n_{GJ} / (137 m_e)}$ with charge density $n_{GJ}=2B \Omega_{NS} / e$ depends, similar to what was done in \citep{noordhuis2023axion}.
Our mock population is composed of a sample of $10^5$ neutron stars with uncorrelated, randomly distributed initial rotational periods $P$, initial magnetic field strengths $B_0$ and misalignment angles $\chi$, each drawn from the distributions
\begin{align}
p_P(P) & =\frac{1}{\sqrt{2 \pi \sigma_p^2}} e^{-\left(P-\mu_p\right)^2 /\left(2 \sigma_p^2\right)}, \\
p_B\left(B_0\right) & =\frac{1}{\sqrt{2 \pi \sigma_B^2}} e^{-\left[\log _{10}\left(B_0\right)-\mu_B\right]^2 /\left(2 \sigma_B^2\right)}, \\
p_\chi(\chi) & =\sin \chi / 2 \,,
\end{align}
with $\mu_p=0.22, \sigma_p=0.42^3, \mu_B=13.2$, and $\sigma_B=0.62$ as in \cite{noordhuis2023axion}.
For each of these stars we assume an average lifetime of $t_{NS} \sim 10 \,\tau_\mathrm{Ohm}$ where we take $\tau_\mathrm{Ohm} = 1\,$Myr \citep{noordhuis2023axion}.
Assuming a constant formation rate over the age of the universe, this yields an overall survival suppression factor $f_{surv} \sim 10 \,\tau_\mathrm{Ohm} / t_H \sim 10\,\mathrm{Myr} / 10\,\mathrm{Gyr} = 10^{-3}$, which has to be combined with an additional resonance factor $f_{res}(m_a)$ accounting for the relative fraction of active neutron stars with a plasma frequency fulfilling the resonance condition $\omega_p \gtrsim m_a$.\\
To obtain $f_{res}(m_a)$, we additionally draw $10^5$ random neutron star ages $t_i \in [0, 10\,\tau_\mathrm{Ohm})$ and evolve each NS $i$ in time until $t_i$ by numerically solving the evolution equations
\begin{align}
\dot{P} & =\beta \frac{B^2}{P}\left(\kappa_0+\kappa_1 \sin ^2 \chi\right), \\
\dot{\chi} & =-\beta \kappa_2 \frac{B^2}{P^2} \sin \chi \cos \chi,
\end{align}
where $\kappa_0 \sim \kappa_1 \sim \kappa_2 \sim 1$ and $\beta= 6 \cdot 10^{-40}\,$s/G$^2$ \cite{noordhuis2023axion}.
The time dependence of the magnetic field strength amounts to an exponential decay
\begin{align}
B(t) = B_0 \exp(- t/\tau_\mathrm{Ohm})
\end{align}
characterized by the Ohmic decay constant $\tau_\mathrm{Ohm}$.\\
Using the above approach we determine the relative NS fraction $f_{res}(m_a)$ numerically by counting the number of neutron stars fulfilling the condition $\omega_p \gtrsim m_a$ for every $m_a$ in the range $10^{-12}\,\mathrm{eV} \leq m_a \leq 10^{-3}\, \mathrm{eV}$.
The resulting effective fraction $f_{NS}(m_a)=f_{surv} f_{res}(m_a)$ is plotted in figure \ref{fig:f_NS}.
\noindent
For $m_a \lesssim 10^{-7}\,$eV, we obtain $f_{res}(m_a)\approx 1$ and the effective NS fraction saturates at $f_{NS}(m_a) \approx f_{surv} \simeq 10^{-3}$.
On the other hand, in the range $m_a > 10^{-7}\,$eV the fraction of NS fulfilling the resonance condition quickly drops until reaching $f_{NS}(m_a)\approx 0$ at $m_a \geq 10^{-4}\,$eV.
Since we assume a total number $N_{NS}=10^9$ in this work, dropping below $f_{NS}\sim 10^{-8}$ effectively excludes NS-collisions from occurring in our galaxy.
Accordingly, the high-$m_a$ cutoffs in figure \ref{fig:Int_Coll_NS-AS} amount to ALP parameters, which yield no suitable NSs for the production of radio signals.
\noindent
As a consequence of the scaling $\Gamma_{\star-NS}(m_a)\propto m_a$ in figure \ref{fig:Int_Coll_NS_AS0} for large $m_a$ and from the NS fraction $f_{NS}(m_a)$, the signal rates of NS-AS encounters in figure \ref{fig:Int_Coll_NS-AS} peak around $m_a\approx 10^{-5}\,$eV.
Overall, we find that galactic radio signals from NS-AS collisions are expected to be extremely rare - for any of the ALP parameters, but especially for ALPs with $n=3.34$ and the QCD axion.\\
The same remains true for a modified MCMF slope of $\alpha=-0.7$, which boosts the signal rates in figure \ref{fig:Int_Coll_NS-AS} by $n$-dependent factors of order $\sim 10$ to $\sim100$ giving $\Gamma_{\star-NS} < 10^{-1}\,$yr$^{-1}$ for $n=0$ and even smaller rates for $n>0$ and the QCD axion.
\\
Unless a majority of the galactic dark matter is contained in ALP stars rather than in miniclusters (thus yielding $f_\star\sim 1$), this issue will persist.
Currently there is no evidence for such a scenario, which leaves us with the conclusion, that galactic NS-AS collisions are far less promising than anticipated.

\subsection{Neutron-Star-Minicluster Collisions} \label{subsec:NS-MC}
With the encounter rates for NS-AS signatures being insufficiently frequent, a reasonable next step is to explore the same scenario of ALP-photon conversion in the NS magnetosphere but with miniclusters instead.
This idea is especially appealing because the size of miniclusters with $\mathcal{R}\sim 10^7\,$km is typically much larger than that of their AS cores, thus enhancing the cross-section of their interactions.
Another advantage is the fact that the size of miniclusters in the spherical model \eqref{eq:R_mc} scales with their mass as $\mathcal{R}\propto\mathcal{M}^{1/3}$ as opposed to the inverse scaling $R_\star\propto 1/M_\star$ of stable ALP stars.
This means that the collision rates of heavier objects, which can potentially yield stronger signals due to the larger abundance of ALP particles for conversion, are less suppressed by the cross-section \eqref{eq:Cross-Section}.\\
On the other hand, the weaker gravitational binding and large size of miniclusters makes them more prone to tidal disruption, especially in dense environments such as the galactic bulge.
A more detailed study of minicluster survival in the context of tidal disruption from stars can be found in \citep{edwards_transient_2021, kavanagh_stellar_2021, Dandoy_2022, Shen_2024}.
For the purpose of this work we adapt a simplified approach based on the results of \citep{edwards_transient_2021} by applying a minimum galactocentric radius $R_\mathrm{surv}=1\,$kpc to our integration in equation \eqref{eq:Gamma_Integrated_Cylindrical}.
We hence assume that any minicluster in the region $R< R_\mathrm{surv}$ will be tidally disrupted, while essentially all of the MCs outside of the MW central region have survived.
\begin{figure}[t]
\centering
\includegraphics[width=\columnwidth]{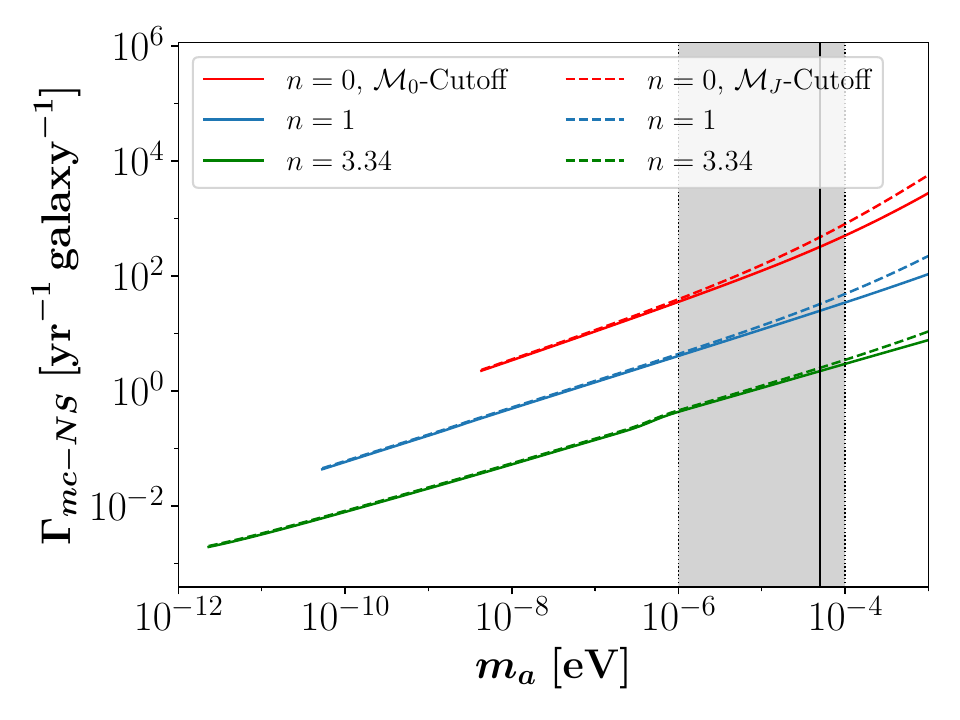}
\caption{Mass-integrated collision rates per year and galaxy between ALP miniclusters and neutron stars as a function of ALP mass $m_a$.
Colored lines indicate the temperature dependence of the ALP mass, solid and dashed lines represent the two different low-$\mathcal{M}$-cutoffs from subsection \ref{subsec:MCMF_Parametr}, both for $\alpha=-1/2$.
\label{fig:Int_Coll_NS_MC}}
\end{figure}
\begin{figure}[t]
\centering
\includegraphics[width=\columnwidth]{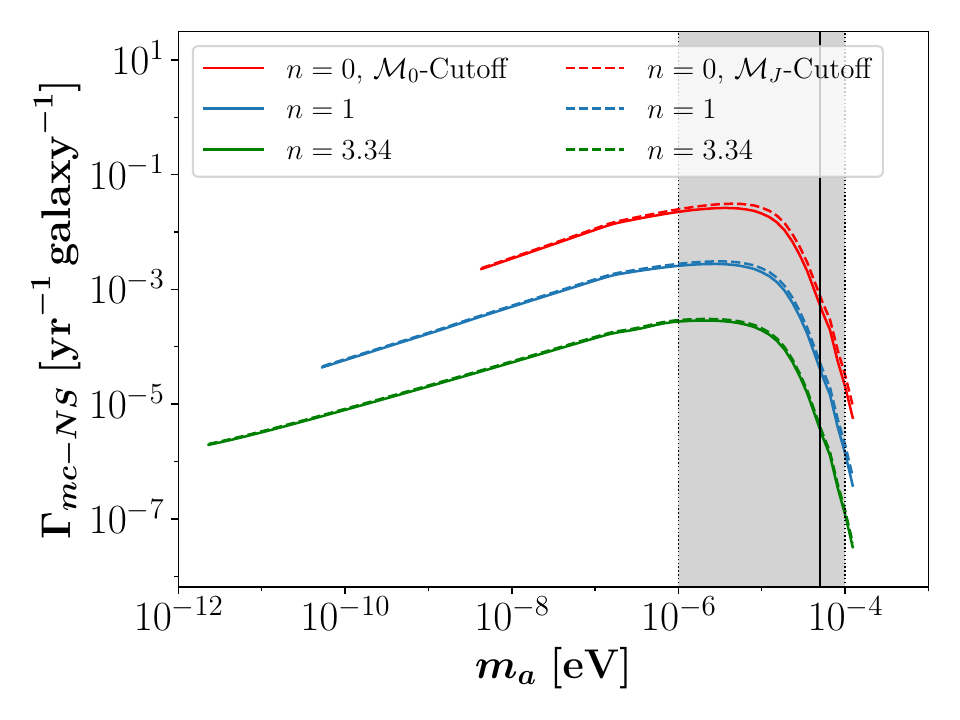}
\caption{Mass-integrated signal rates per year and galaxy between ALP miniclusters and neutron stars after applying the resonance condition $f_{NS}(m_a)$, as a function of ALP mass $m_a$ for $\alpha=-1/2$.
\label{fig:Int_Coll_NS_MC_f_NS}}
\end{figure}
\noindent
The resulting collision rates before and after applying the NS requirements $f_{NS}(m_a)$ from subsection \ref{subsec:NS-AS} are shown for an MCMF slope of $\alpha=-1/2$ in figures \ref{fig:Int_Coll_NS_MC} and \ref{fig:Int_Coll_NS_MC_f_NS}.\\
As before, the $n=0$ scaling of the NS-MC encounter rates in red in figures \ref{fig:Int_Coll_NS_MC} and \ref{fig:Int_Coll_NS_MC2} may be divided into two regimes using $\mathcal{M}_0(n=0) \ll M_{NS}$, $\mathcal{N}_\mathrm{tot}\propto \mathcal{M}_0^{-1}$ and $\mathcal{R}\propto \mathcal{M}^{1/3}$ from equation \eqref{eq:R_mc}
\begin{align}
\Gamma_{NS-\mathrm{mc}}\Big|_{n=0} & \propto
\begin{cases}
\mathcal{N}_\mathrm{tot} \mathcal{R}^2
\propto m_a^{1/2}, \quad &\eta < 1\\
\mathcal{N}_\mathrm{tot} \mathcal{R}
\propto m_a, \quad &\eta > 1
\end{cases}\,,
\end{align}
where the turnaround occurs roughly at the QCD axion mass $m_a\approx 50\,\mu$eV and the cases $n>0$ in blue and green show a similar trend with slightly different, $n$-dependent scalings with $m_a$.
Note that the results in figure \ref{fig:Int_Coll_NS_MC} are nearly independent of the low-mass cutoffs $\mathcal{M}_{J,\min}$ and $\mathcal{M}_0/25$ because the major contribution to the mass-integrated collision rates is given by the high-mass tail with $\mathcal{M}\gtrsim \mathcal{M}_0$ and $\mathcal{R}(\mathcal{M}) \gtrsim \mathcal{R}(\mathcal{M}_0)$.\\
Without considering the resonance condition, i.e. in figure \ref{fig:Int_Coll_NS_MC}, NS-MC collisions appear rather frequently, reaching $\simeq 4\,$yr$^{-1}$ galaxy$^{-1}$ for the QCD axion with both MCMF cutoffs and up to $\sim 10^3\,$yr$^{-1}$ galaxy$^{-1}$ for $n=0$ and $m_a=50\,\mu$eV.
Coincidentally, the regions where $\Gamma_{mc-NS}(m_a)$ becomes large are strongly suppressed by $f_{NS}(m_a)$ so that the effective rates for producing astrophysical signatures are typically well below $1$ per decade for most $m_a,n$ in figure \ref{fig:Int_Coll_NS_MC_f_NS}.\\
We emphasize that this result strongly depends on the power-law index $\alpha=-0.5$ of the MCMF $\rm{d}n/\rm{d}\ln\mathcal{M}\propto \mathcal{M}^{\alpha}$ which we have assumed until now.
For comparison, the authors of \citep{edwards_transient_2021} used a steeper power-law with $\alpha=-0.7$, mainly motivated by observations in numerical simulations of minicluster formation and their subsequent evolution \citep{eggemeier_first_2020}.
In this case, the contribution of the low-$\mathcal{M}$ components is significantly boosted, yielding enhancements by a factor of $\sim10$ to $\sim100$ for the $n$-dependent encounter rates of ASs and MCs, respectively.
Neglecting the NS resonance and for the MC masses $3.3\cdot 10^{-19}\,M_\odot \leq \mathcal{M} \leq 5.1\cdot 10^{-5}\,M_\odot$ used in \citep{edwards_transient_2021} with $\alpha=-0.7$, we confirm their prediction of $\Gamma_{mc-NS}\simeq4\,/$day.
We also note that different to \citep{edwards_transient_2021}, we use the phenomenological NS distribution fit by \citep{taani_modeling_2012} instead of the stellar distribution used by \citep{edwards_transient_2021}.
The results of our calculations with power-law index $\alpha=-0.7$ are shown in figures \ref{fig:Int_Coll_NS_MC2} and \ref{fig:Int_Coll_NS_MC2_f_NS}.\\
For the slope index $\alpha=-0.7$ in figure \ref{fig:Int_Coll_NS_MC2}, the low-mass cutoff dependence becomes stronger due to the larger contribution of light miniclusters with $\mathcal{M} < \mathcal{M}_0$, however the general trend remains valid.
\begin{figure}[t]
\centering
\includegraphics[width=\columnwidth]{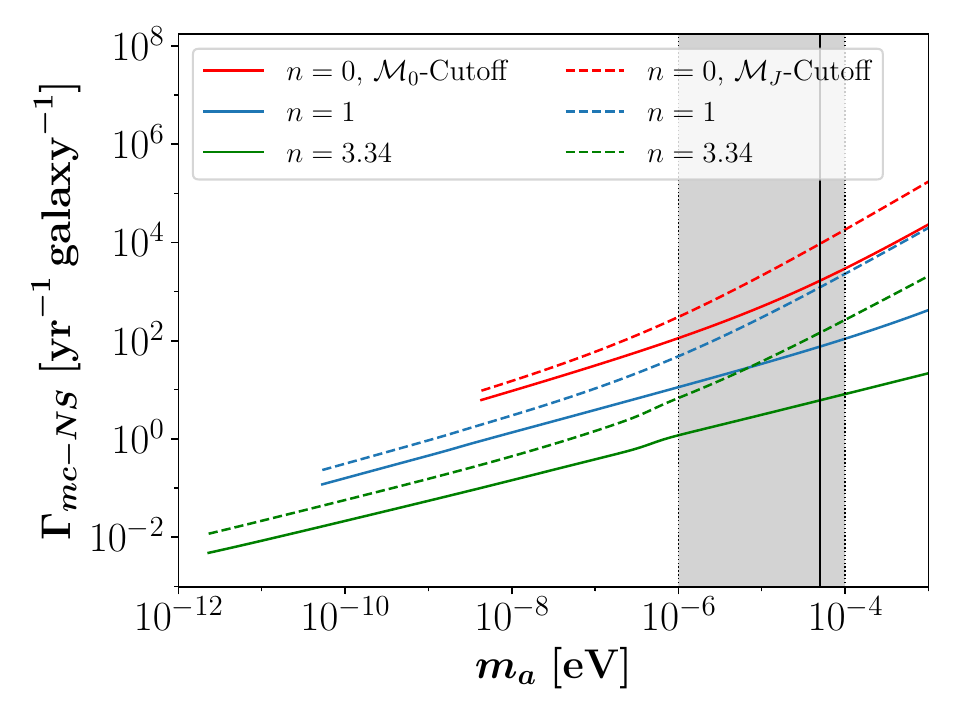}
\caption{Mass-integrated collision rates per year and galaxy between ALP miniclusters and neutron stars as a function of ALP mass $m_a$.
Same as in figure \ref{fig:Int_Coll_NS_MC} but for an MCMF slope of $\alpha=-0.7$ instead.
\label{fig:Int_Coll_NS_MC2}}
\end{figure}
\begin{figure}[t]
\centering
\includegraphics[width=\columnwidth]{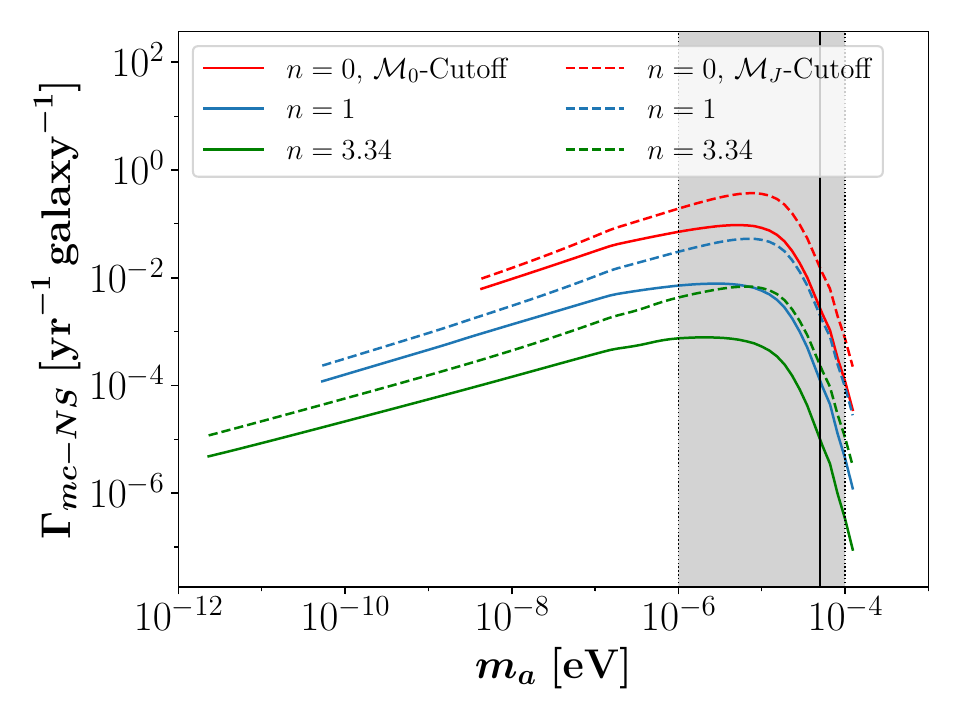}
\caption{Mass-integrated signal rates per year and galaxy between ALP miniclusters and neutron stars and after applying the resonance condition $f_{NS}(m_a)$, as a function of ALP mass $m_a$.
Same as in figure \ref{fig:Int_Coll_NS_MC_f_NS} but for an MCMF slope of $\alpha=-0.7$ instead.
\label{fig:Int_Coll_NS_MC2_f_NS}}
\end{figure}
\noindent
The conclusion to draw from our analysis of NS collisions is that despite the very promising encounter rates with AS-MC systems, the occurrence of actual signatures from ALP-photon conversions is much less common than previously expected.
This is especially true for NS-AS collisions which are basically undetectable in the galactic minicluster scenario.
For the NS-MC case, detection might still be possible, especially for $n=0$, $\alpha=-0.7$ and ALP masses around $m_a\simeq 10\,\mu$eV shown in figure \ref{fig:Int_Coll_NS_MC2_f_NS}.
Taking into account the large uncertainties in the NS properties and in the detailed evolution of the minicluster population, the occurrence of radio signals from NS-MC collisions cannot be ultimately ruled out.
Our results suggest however, that future research on ALP miniclusters should aim to explore new detection mechanisms due to the small expected rates of NS-AS/MC signals in our galaxy.
\\
We also mention that assuming constant NS magnetic fields as in the first model in \cite{Safdi_2019}, the survival factor becomes $f_{surv} = 1$, yielding an overall boost of order $10^3$ to our predictions in figures \ref{fig:Int_Coll_NS-AS}, \ref{fig:Int_Coll_NS_MC_f_NS} and \ref{fig:Int_Coll_NS_MC2_f_NS}.
This would raise the signal rates of NS encounters above detection threshold for $n=0,1$
.
However since the decay of magnetic fields is expected from Ohmic dissipation and other processes \cite{Safdi_2019}, and since we have been very optimistic on the NS abundance $N_{NS}=10^9$, the observation of NS signals should still be unlikely under realistic circumstances.

\subsection{ALP-Star Collisions and relativistic Bursts} \label{subsec:bosenova}
\begin{figure}[t]
\centering
\includegraphics[width=\columnwidth]{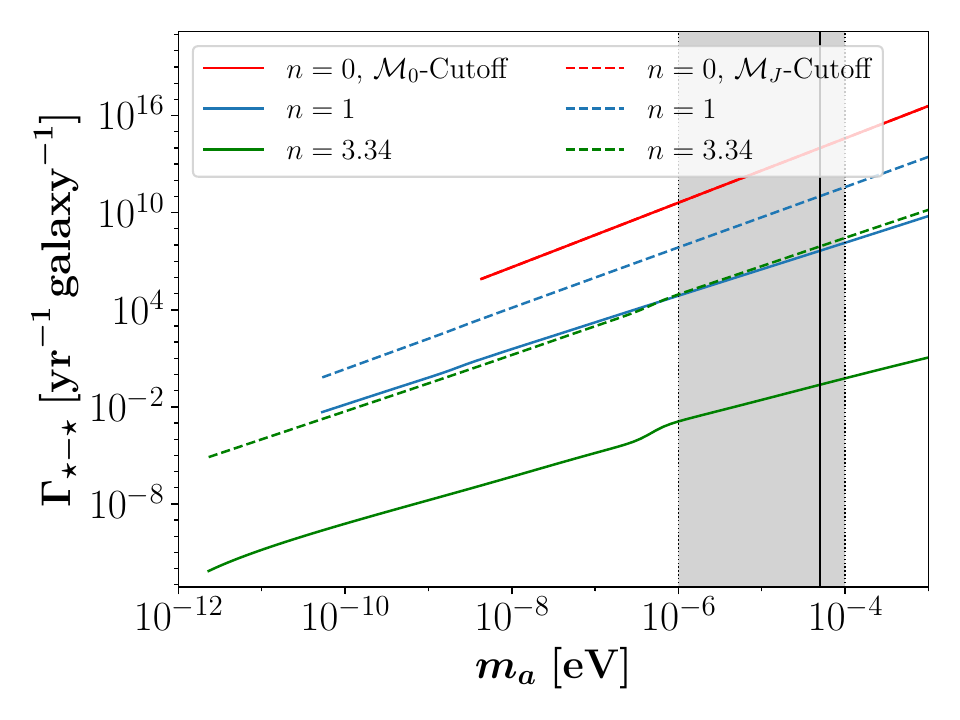}
\caption{Mass-integrated collision rates of ALP stars per year and galaxy as a function of ALP mass $m_a$. 
Colored lines indicate the temperature dependence of the ALP mass, solid and dashed lines represent the two different low-$\mathcal{M}$ cutoffs for $\alpha=-1/2$.
\label{fig:Int_Coll_AS}}
\end{figure}

\begin{figure}[t]
\centering
\includegraphics[width=\columnwidth]{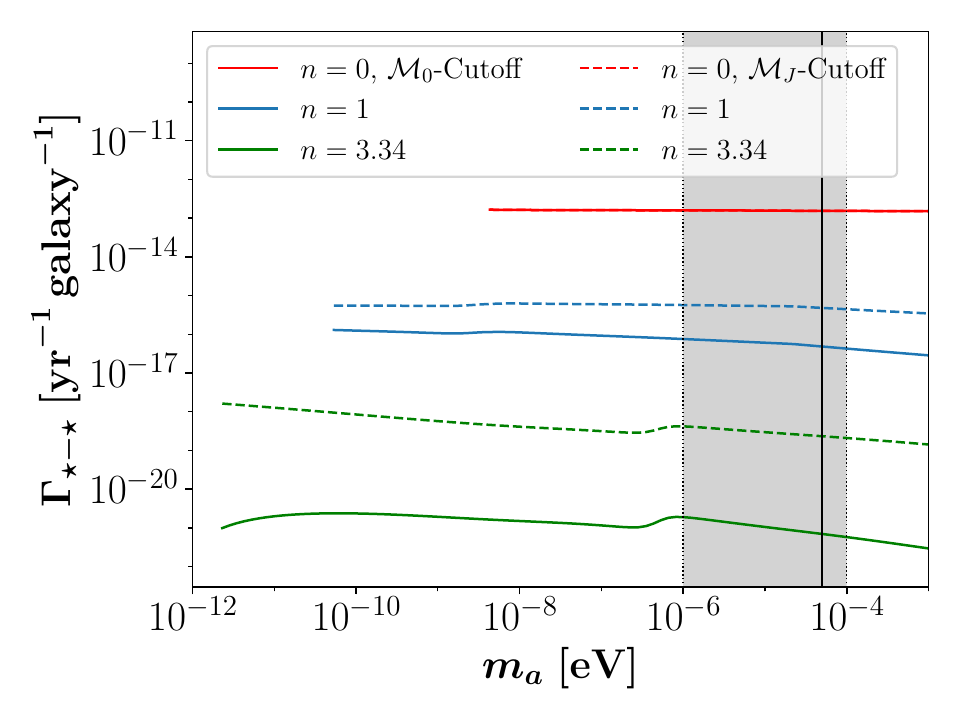}
\caption{Mass-integrated total merger rates of ALP stars per year and galaxy as a function of ALP mass $m_a$ for $\alpha=-1/2$.
\label{fig:Int_Coll_AS_vesc}}
\end{figure}

\begin{figure}[t]
\centering
\includegraphics[width=\columnwidth]{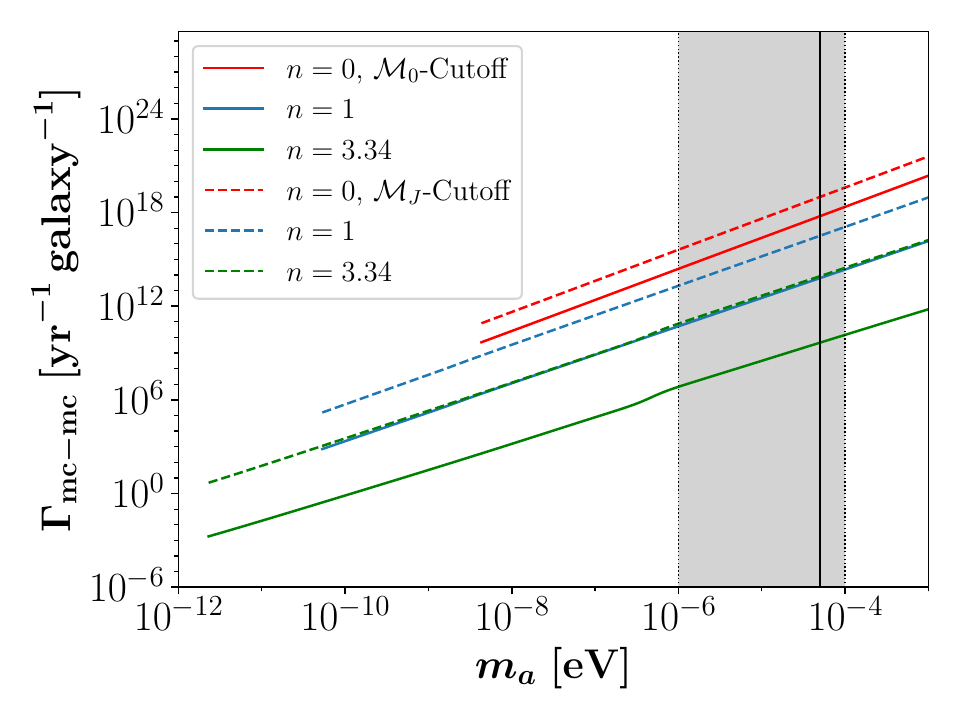}
\caption{Mass-integrated MC collision rates per year and galaxy as a function of ALP mass $m_a$. 
Colored lines indicate the temperature dependence of the ALP mass, solid and dashed lines represent the two different low-$\mathcal{M}$ cutoffs for $\alpha=-1/2$.
\label{fig:Int_Coll_MC_0.5}}
\end{figure}

\begin{figure}[t]
\centering
\includegraphics[width=\columnwidth]{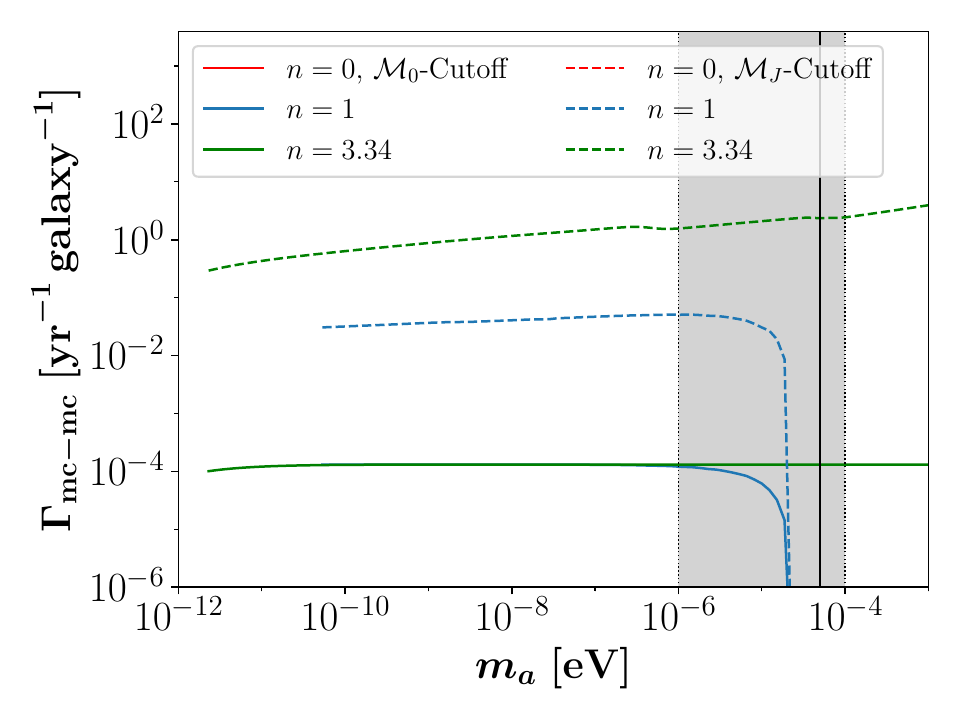}
\caption{Mass-integrated Bosenova merger rates with $\mathcal{M}_1 + \mathcal{M}_2 \geq \mathcal{M}(M_{\star,\lambda})$ per year and galaxy as a function of ALP mass $m_a$ for $\alpha=-1/2$. 
MC mergers could also produce radio bursts when $M_{\star,\gamma}<M_{\star,\lambda}$.
\label{fig:Int_Coll_Bosenova_0.5}}
\end{figure}

\begin{figure}[t]
\centering
\includegraphics[width=\columnwidth]{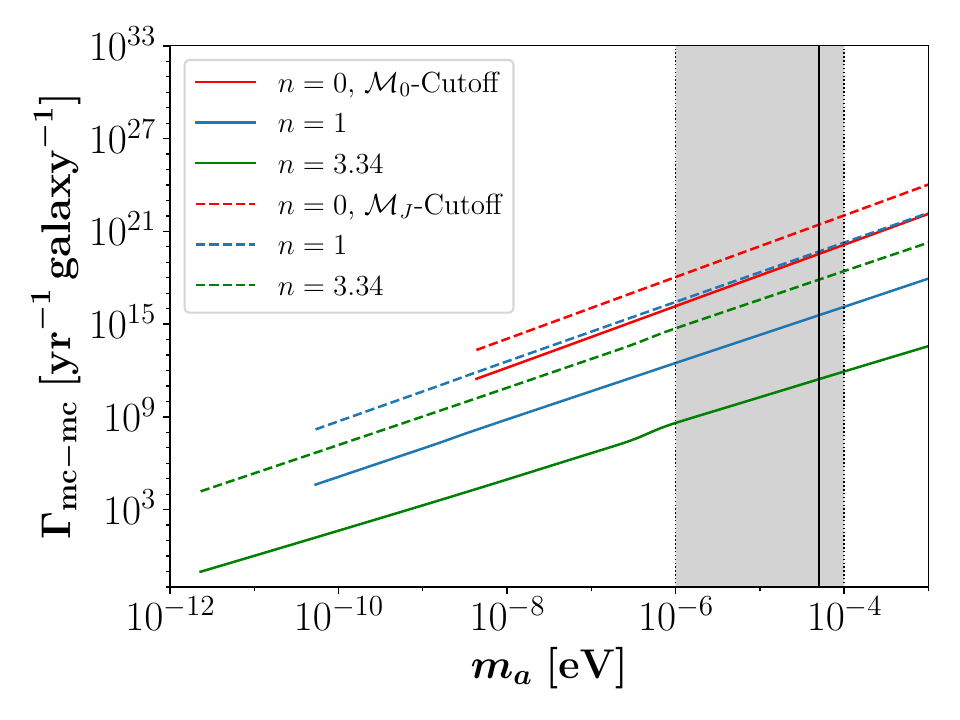}
\caption{Mass-integrated MC collision rates per year and galaxy as a function of ALP mass $m_a$. 
Colored lines indicate the temperature dependence of the ALP mass, solid and dashed lines represent the two different low-$\mathcal{M}$ cutoffs for $\alpha=-0.7$.
\label{fig:Int_Coll_MC_0.7}}
\end{figure}

\begin{figure}[t]
\centering
\includegraphics[width=\columnwidth]{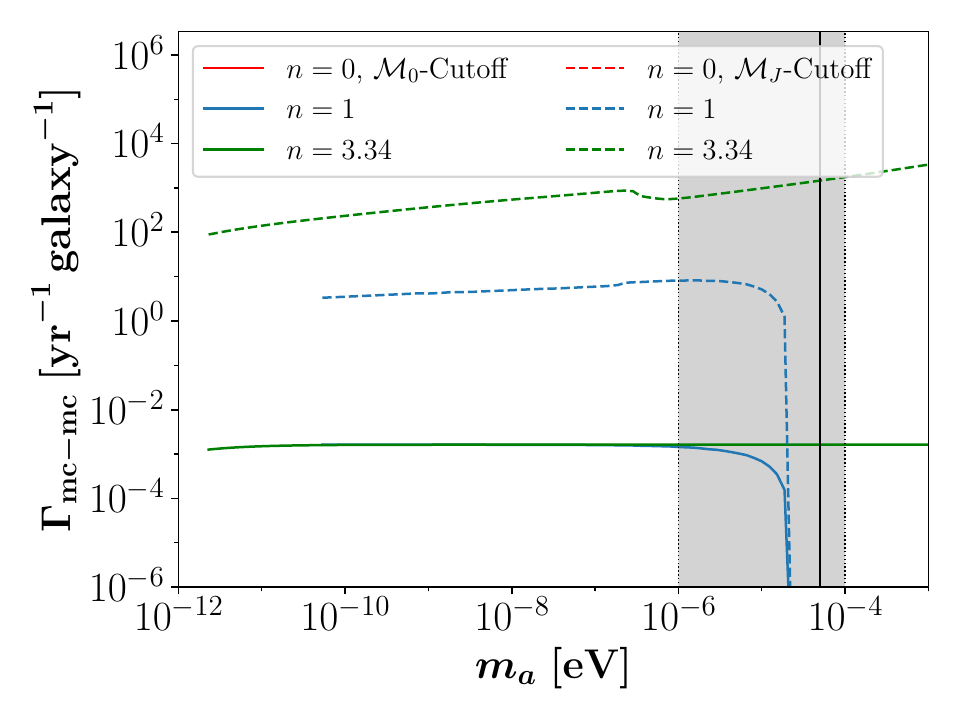}
\caption{Mass-integrated Bosenova merger rates with $\mathcal{M}_1 + \mathcal{M}_2 \geq \mathcal{M}(M_{\star,\lambda})$ per year and galaxy as a function of ALP mass $m_a$ and for $\alpha=-0.7$.
MC mergers could also produce radio bursts when $M_{\star,\gamma}<M_{\star,\lambda}$.
\label{fig:Int_Coll_Bosenova_0.7}}
\end{figure}

Another important mechanism for detection of ALP dark matter occurs when a soliton exceeds the critical mass $M_\star \geq M_{\star,\lambda}$ hence triggering the self-interaction instability \cite{levkov_relativistic_2017, arakawa2024bosenovae, eby_probing_2022}. 
The resulting collapse of a super-critical ALP star induces relativistic multi-particle interactions that lead to strong emission of weakly relativistic ALPs from the collapsing soliton.
Levkov et al. \cite{levkov_relativistic_2017} argued that the cycled ALP bursts observed in their simulations could repeat until the star eventually relaxes back to a sub-critical state.
Another more recent work on the detection of Bosenovae with quantum sensors was published in \citep{arakawa2024bosenovae}
.\\
In this subsection we will ignore the details of the Bosenova evolution and assume that the bursts emerging from it could eventually be detected by earth based experiments \citep{eby_probing_2022, arakawa2024bosenovae}.
To estimate how common the occurrence of such ALP bursts in our galaxy is, we start by computing the total collision- and merger rates of galactic ALP stars considering the full range of AS masses $M_\star$ in the ASMF in figures \ref{fig:Int_Coll_AS} and \ref{fig:Int_Coll_AS_vesc}.
In this case we integrate over the mass distributions of both of the ASs/MCs involved in a single collision by writing
\begin{align}
    \Gamma_{i-i} =\,& 4 \pi S \int_{0}^{R}d r\, r^2
    n_i^2(r)
    \int\,dM_i\, p_i(M_i) \nonumber \\
     & \times  \int\,dM_i'\, p_i(M_i')\left\langle \sigma_{\text{eff}}(v, M_i, M_i') \,v \right\rangle_v \,.\label{eq:Gamma_Integrated_Spherical_ii}
\end{align}
For both ALP stars and miniclusters, the gravitational enhancement is negligible, $\eta \ll 1$, so that the encounter rates for $n=0$ in red lines in figures \ref{fig:Int_Coll_AS} with $N_{\star,\mathrm{tot}}\propto \mathcal{M}_0^{-1}$ and $\langle R_\star \rangle \propto m^{-1/2}$ (see eq. \eqref{eq:NS-AS_Scaling}) simply scale as
\begin{align}
\Gamma_{\star-\star}\Big|_{n=0} & \propto
N_{\star,\mathrm{tot}}^2 \langle R_\star \rangle^2
\propto m_a^2
\end{align}
and similarly for the minicluster rates in red in figures \ref{fig:Int_Coll_MC_0.5} and \ref{fig:Int_Coll_MC_0.7}.
For the remaining cases $n>0$, the scalings of the binary collision rates with $m_a$ will be marginally different but qualitatively similar as argued before.
We can furthermore calculate the total number of AS-/MC-mergers from eq. \eqref{eq:Gamma_Integrated_Spherical_ii} by replacing the velocity cutoff $v_\mathrm{esc}$ in equation \eqref{eq:SigmaVAverage} with the escape velocity $v_{\star,\mathrm{esc}}(M)\simeq \sqrt{2GM_\star / R_\star}$ of the binary ALP star system.
The corresponding fraction of collisions which can lead to a merger
\begin{align}
f_\mathrm{esc}(M_\star, n=0) 
&\lesssim\left[ \frac{v_{\star,\mathrm{esc}}(M_{\star,\lambda})}{v_\mathrm{esc}} \right]^4 \nonumber \\
&\hspace*{-0.7em} \stackrel{50\,\mu\mathrm{eV}}{\sim}  \left( \frac{10\,\mathrm{m}\,\mathrm{s}^{-1}}{100\,\mathrm{km}\,\mathrm{s}^{-1}} \right)^4
\sim 10^{-16}  \,, \label{eq:Merger_suppresion_AS}
\end{align}
depends on the different star masses $M_\star \leq M_{\star,\lambda}$ in the ASMF, where we have taken the maximum AS properties as an upper bound for $n=0$ and neglected the impact of the reduced number density at large $M_\star$ for simplicity.
Integrating the $M_\star$-dependent suppression \eqref{eq:Merger_suppresion_AS} over the whole range of AS masses and taking into account the reduced number densities at large values of $M_\star$, the effective suppression factor can become orders of magnitude smaller than $f_\mathrm{esc}(M_{\star,\lambda})\sim 10^{-16}$ - depending on $n$ and the low-mass cutoffs.
Accordingly, the AS merger results obtained from our results with an MCMF slope of $\alpha=-1/2$ in figure \ref{fig:Int_Coll_AS_vesc} are generally many orders of magnitude lower than the corresponding collision rates in figure \ref{fig:Int_Coll_AS}.
\noindent
While galactic ALP star encounters in figure \ref{fig:Int_Coll_AS} are very common for every case except $n=3.34$ with the $\mathcal{M}_0$-cutoff, they are extremely unlikely to merge.
Both of the above rates are enhanced in the case of the $\mathcal{M}_J$-cutoff due to the larger abundance of ASs/MCs in the galaxy.
Equivalent predictions were already made by the authors of \citep{hertzberg_merger_2020}.
The simple explanation for the strong suppression are the small binding energy of ALP stars and their large typical velocities in the Milky Way halo with velocity dispersion $v_0=239\,$km\,s$^{-1}$.\\
In the above scenario, we have so far neglected the merger probability of the host miniclusters, which can be many orders of magnitude larger than that of the ALP star cores.
Replacing the AS parameters in equation \eqref{eq:Gamma_Integrated_Spherical_ii} by the corrresponding MC properties and using $v_{mc,\mathrm{esc}}(\mathcal{M})\simeq \sqrt{2G \mathcal{M} / \mathcal{R}}$ as a cutoff instead, we can similarly compute the collision- and merger rates of miniclusters shown in figures \ref{fig:Int_Coll_MC_0.5} and \ref{fig:Int_Coll_Bosenova_0.5}, again for $\alpha=-1/2$.
\noindent
It is important to note that only MC-MC mergers with a total mass $\mathcal{M}_1 + \mathcal{M}_2 \geq \mathcal{M}(M_{\star,\lambda})$ will safely lead to the production of relativistic bursts (where we have inverted the core-halo relation \eqref{eq:CoreHalo} to find the MC mass corresponding to $M_{\star,\lambda}$).
For this reason we plot the \textit{Bosenova merger rates} in figures \ref{fig:Int_Coll_Bosenova_0.5} and \ref{fig:Int_Coll_Bosenova_0.7} by only counting collisions which pass the velocity cutoff $v_{\mathrm{mc,esc}}$ and fulfill the requirement $\mathcal{M}_1 + \mathcal{M}_2 \geq \mathcal{M}(M_{\star,\lambda})$.
We emphasize that the dynamics and timescale of the merger evolution, especially for the two soliton cores, are beyond the scope of this work.
Instead, we will argue in subsection \ref{subsec:extra-galactic} that the typical timescale between two MC merger events is much larger than the timescale of the AS merger and that we can thus neglect the effects of the merger dynamics in our estimations.\\
For $n=1$ in blue lines in figure \ref{fig:Int_Coll_Bosenova_0.5}, the Bosenova merger rates from MC mergers quickly drops to zero beyond $m_a \gtrsim 10^{-5}\,$eV where $\mathcal{M}_{\min} + \mathcal{M}_{\max} < \mathcal{M}(M_{\star,\lambda})$.
As expected, the overall rate of collisions in figure \ref{fig:Int_Coll_MC_0.5} is significantly boosted by the larger mass, radius and number of MCs compared to the AS case in figure \ref{fig:Int_Coll_AS}.
More importantly, their merger rates can be significantly enhanced in the case of the $\mathcal{M}_J$-cutoff due to the large total number of MCs, reaching $\Gamma_{mc-mc}\gtrsim 1\,$yr$^{-1}$ galaxy$^{-1}$ for ALPs with $n=3.34$ and at $m_a\approx50\,\mu$eV, $\alpha=-1/2$.
The weak dependence of the merger rates on $m_a$ indicates that for larger ALP masses and hence smaller $\mathcal{M}_0(m_a)$, the boost from having an increased number of objects $\mathcal{N}_\mathrm{tot}\propto 1/\mathcal{M}_0$ roughly cancels with their decreased merger rates due to the smaller typical size $\mathcal{R}\propto\mathcal{M}_0^{1/3}$.
The corresponding MC suppression factor 
\begin{align}
f_\mathrm{esc}(\mathcal{M}, n=0) 
\lesssim & \,\left[ \frac{v_{\star,\mathrm{esc}}(\mathcal{M}_{\max})}{v_\mathrm{esc}} \right]^4 \nonumber \\
& \hspace*{-1.7em} \stackrel{50\,\mu\mathrm{eV}}{\sim} \left( \frac{100\,\mathrm{m}\,\mathrm{s}^{-1}}{100\,\mathrm{km}\,\mathrm{s}^{-1}} \right)^4
\sim 10^{-12}  \,, \label{eq:Merger_suppresion_MC}
\end{align}
is orders of magnitude larger than in the AS case in equation \eqref{eq:Merger_suppresion_AS}, as expected.
\noindent
We have again evaluated $f_\mathrm{esc}(\mathcal{M})$ at the maximum mass $\mathcal{M}_{\max}$ to obtain an upper bound on $f_\mathrm{esc}$ while again neglecting the reduced number density d$n/\text{d}\mathcal{M}$ at $\mathcal{M}=\mathcal{M}_{\max}$.\\
Similar arguments hold for the case $\alpha=-0.7$ in figure \ref{fig:Int_Coll_MC_0.7}, where the total number of miniclusters $\mathcal{N}_\mathrm{tot}$ is significantly boosted due to the smaller fraction of heavy MCs and hence larger number of light MCs in the Milky Way with total mass $\mathcal{M}_\mathrm{tot} = f_\mathrm{mc}\,M_\mathrm{MW}$.
For the same reason, the total number of encounters $\Gamma_{mc-mc}\propto\mathcal{N}_\mathrm{tot}^2$ and mergers in figures \ref{fig:Int_Coll_MC_0.7} and \ref{fig:Int_Coll_Bosenova_0.7} is strongly enhanced compared to the $\alpha=-1/2$ case in figures \ref{fig:Int_Coll_MC_0.5} and \ref{fig:Int_Coll_Bosenova_0.5}.\\
We conclude that for the detection of Bosenovae from galactic AS-MC systems, the low-$\mathcal{M}$ cutoff and the slope index $\alpha=-1/2,-0.7$ of the MCMF have a strong impact on the expected event rates.
Our results suggest that for the QCD axion case $m_a\approx 50\,\mu$eV, $n=3.34$ and ALPs with similar temperature evolution, Bosenovae can occur as often as $\sim 1$ per year for $\alpha=-1/2$ and as often as $\sim 3$ per day for $\alpha=-0.7$, both with the $\mathcal{M}_{J}$-cutoff.
Conversely, for the $\mathcal{M}_{0}$-cutoff the expected merger rates in figures \ref{fig:Int_Coll_Bosenova_0.5} and \ref{fig:Int_Coll_Bosenova_0.7} are well below one per year - independent of $m_a$ and $n$.
Bosenovae from AS-MC mergers thus require large numbers of miniclusters and benefit from a larger maximum mass $\mathcal{M}_{\max}$ as seen for $n>0$ in dashed lines in figures \ref{fig:Int_Coll_Bosenova_0.5} and \ref{fig:Int_Coll_Bosenova_0.7}. 
Note that other mechanisms such as AS accretion, which we have neglected here, could still trigger large numbers of Bosenovae even for the $\mathcal{M}_0$-cutoff.

\subsection{Parametric Resonance and ALP Star Accretion} \label{subsec:Glowing_AS}
\begin{figure}[t]
\centering
\includegraphics[width=\columnwidth]{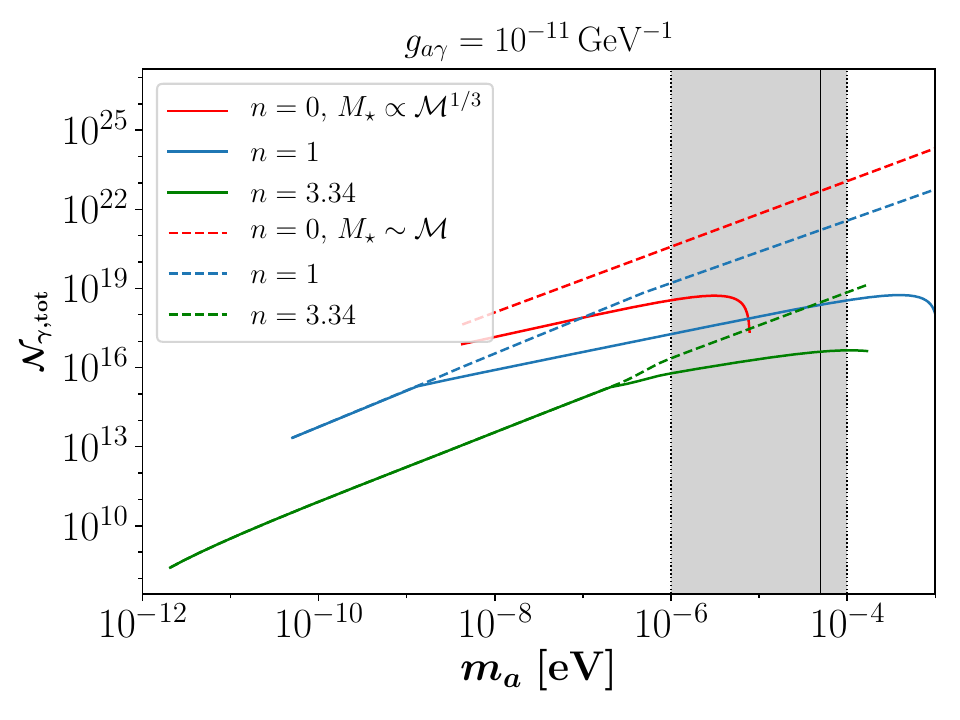}
\caption{Total number of parametrically resonant AS-MC systems in the Milky Way as a function of ALP mass $m_a$ at $g_{a\gamma}=10^{-11}\,$GeV and with $\alpha=-1/2$. 
Both using the $\mathcal{M}_0$-cutoff of the MCMF, for $M_\star$ given by the core-halo relation \eqref{eq:CoreHalo} with $M_\star \propto \mathcal{M}^{1/3}$ in solid lines and with $M_\star \sim \mathcal{M}$ suggested from the accretion model in \citep{dmitriev_self-similar_2024} in dashed colored lines.
\label{fig:Decay_gagamma-11}}
\end{figure}

\begin{figure}[t]
\centering
\includegraphics[width=\columnwidth]{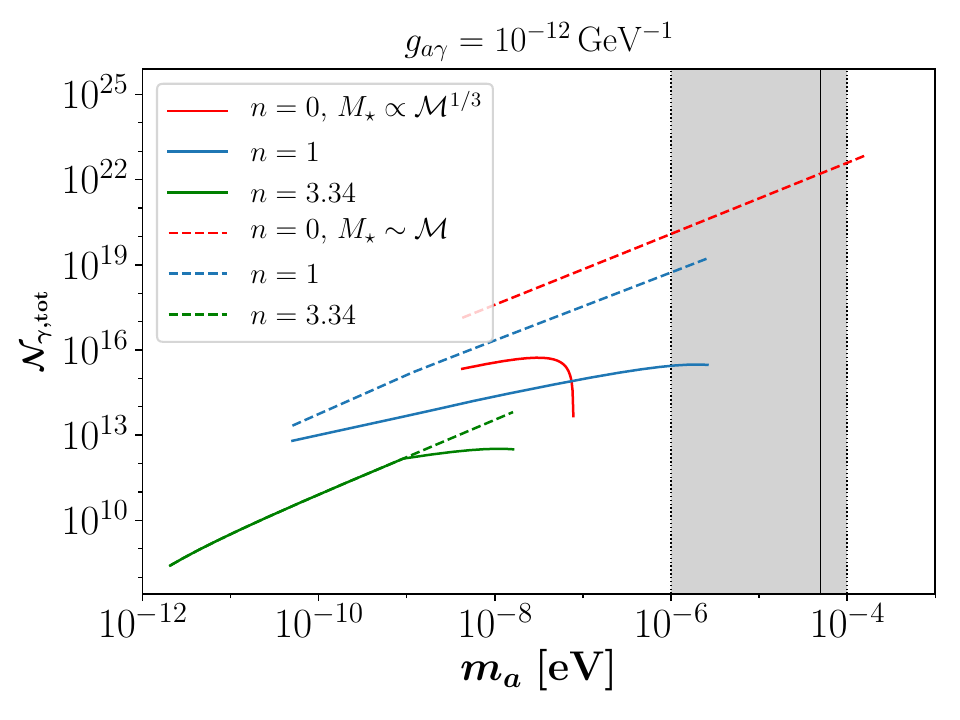}
\caption{Total number of parametrically resonant AS-MC systems in the Milky Way as a function of ALP mass $m_a$ at $g_{a\gamma}=10^{-12}\,$GeV and with $\alpha=-1/2$. Same as in figure \ref{fig:Decay_gagamma-11}.
\label{fig:Decay_gagamma-12}}
\end{figure}

A fundamental property of ALP dark matter is its weak coupling to the electromagnetic field.
Despite the resulting low probability of ALP-photon interactions in cosmological background fields, ALP stars can serve as highly efficient radio converters when the coherent soliton condensate undergoes a process called parametric resonance \cite{hertzberg_dark_2018, levkov_radio-emission_2020, Du_2024}.\\
In this scenario, which is mainly constrained by the size and density of the soliton solution, the stimulated emission from a single photon can stimulate further $a\rightarrow 2\gamma$ conversions thus creating a cascade with exponentially growing photon number $n_\gamma \propto \exp(2\mu t)$ \citep{levkov_radio-emission_2020}.
The growth exponent $\mu$ of the ALP star is generally derived from the growth exponent $\mu_{\infty}$ of a homogeneous ALP field \citep{hertzberg_dark_2018}.
Using this approximation, the condition for parametric resonance can be formulated by comparing the homogeneous growth rate 
\begin{align}
\mu_{\infty} &= \frac{1}{4} g_{a\gamma\gamma} m_a \phi_0 \geq \mu_\mathrm{esc}  \label{eq:Growth_Rate_Cond}
\end{align}
to the escape rate $\mu_\mathrm{esc} \approx 1/(2 R_\star)$ of the ALP star with radius $R_\star$ and amplitude $\phi_0\equiv \phi(\vec{x}=0)$.
Further combining equation \eqref{eq:Growth_Rate_Cond} with the mass-radius relation for the Gaussian profile, one can derive the decay mass of ALP stars with attractive self-interactions \cite{maseizik_radio_2024, hertzberg_merger_2020}
\begin{align}
    M_{*,\gamma} \simeq &\,\, 3.2 \times 10^{-14} M_{\odot}\left(\frac{50\,\mu{\rm eV}}{m_a}\right)\left(\frac{10^{-11} {\rm GeV}^{-1}}{g_{a\gamma\gamma}}\right)^2 \nonumber \\
    & \times \left(\frac{10^{11} {\rm GeV}}{f_a}\right)
    \sqrt{\left(\frac{g_{a\gamma\gamma}f_a}{0.23}\right)^2-\frac{5}{3}} \,.    \label{eq:M_Decay_Gaussian}
\end{align}
Stars with $M_\star \geq M_{\star,\gamma}$ will develop parametric resonance leading to an exponentially growing emission of photons with frequency $\omega_\gamma \approx m_a/2$.
To this date, Levkov et al. \cite{levkov_radio-emission_2020} presented the most detailed analysis of the process, including effects of exponential growth and the on-switch of photon backreactions.
We will use some of their results to argue in the following that the decay mass \eqref{eq:M_Decay_Gaussian} can lead to interesting phenomenological consequences in the context of (galactic) AS-MC systems:
In the regime, where $M_\star$ exceeds the critical mass only gradually, i.e. $M_\star \gtrsim M_{\star,\gamma}$, the growth exponent may be approximated as
\begin{align}
\mu&= 0.197 \,\frac{m_a^2}{m_p^2} (M_\star - M_{\star,\gamma})  \,, \label{eq:mu_Glowing}
\end{align}
where self-interactions have been neglected \citep{levkov_radio-emission_2020}.
The latter assumption is essentially ensured by the gravitational limit of ALP stars $M_\star \ll M_{\star,\lambda}$.
Equation \eqref{eq:mu_Glowing} demonstrates that the growth exponent and resulting on-set of resonant emission will initially be very small.
As the resonance develops, the star loses an exponentially increasing fraction of its mass to the conversion of ALPs into photons.
The mass-loss is expected to continue until the parametric resonance shuts off once the ALP star becomes sub-critical again at $M_\star < M_{\star,\gamma}$.\\
Applying this scenario to our galactic AS-MC population has profound consequences due to two different mechanisms.
First, we have assumed the core-halo relation \eqref{eq:CoreHalo} to describe the equilibrium state of virialization between the star and its host minicluster.
As a direct consequence, we predict a large number of stars residing in heavier miniclusters to have $M_{\star,\gamma} \leq M_\star < M_{\star,\lambda}$.
These stars however can not reach a virialized state due to the conversion process described above.
The outcome of this scenario is an AS-MC system that continuosly feeds ALP dark matter into its soliton trying to reach an equilibrium state that is prohibited by the exponential decay into radio photons.\\
Similarly, in a second scenario numerical simulations suggest that the accretion from the minicluster onto the ALP star does continue even at late times \citep{eggemeier_formation_2019, chen_new_2021} with a recent semi-analytical study by \citep{dmitriev_self-similar_2024} suggesting that up to an order one fraction of the MC mass could be absorbed by the ALP star over time.
Incorporating the effects of long-time accretion in AS-MC systems can induce continuous growth of the soliton mass until reaching either $M_\star=M_{\star,\gamma}$ or $M_\star\simeq \mathcal{M}$ (where we are assuming that $M_{\star,\gamma}\leq M_{\star,\lambda}$).\\
In both of the above scenarios, a considerable fraction of the galactic DM halo can be converted into radio photons in the narrow frequency band $\nu_\gamma\approx m_a/4\pi$, where $\Delta\nu \sim 10^{-3} \nu_\gamma$ is set by the galactic Doppler-shift.
Estimating the time- and mass-dependent rate of accretion onto the soliton is non-trivial and depends on the chosen AS-MC model, which is why we dedicate a follow-up paper to the possibility of constraining ALP properties from AS accretion and galactic radio backgrounds \citep{maseizik_radio_2024}.
\begin{figure}[t]
\centering
\includegraphics[width=\columnwidth]{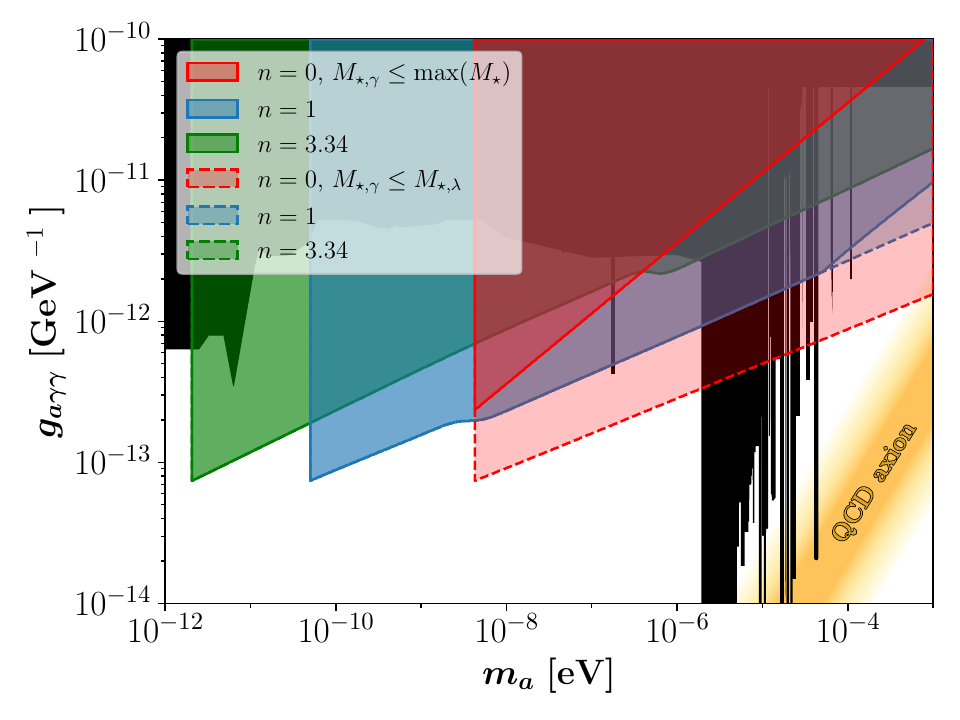}
\caption{\textit{Solid colored lines and shades}:
Regions of parameter space with ALP-photon coupling $g_{a\gamma\gamma}$ in GeV$^{-1}$ and ALP mass $m_a$ in eV, where the core-halo relation \eqref{eq:CoreHalo} predicts the existence of photon-critical ALP stars below the maximum AS mass in the ASMF, i.e. where $M_{\star,\gamma} \leq \max(M_\star)|_{n, m_a}$.
\textit{Dashed colored lines and light shades:}
ALP-photon couplings, with $M_{\star,\gamma}\leq M_{\star,\lambda}$, where parametric resonance can occur 
before the self-interaction instability develops at $M_{\star,\lambda}$.
This part of parameter space could be explored when including effects of AS-MC accretion.
Current constraints are shown in black and the QCD axion band is indicated by the yellow-shaded region \cite{AxionLimits}.
\label{fig:g_m_a}}
\end{figure}
\noindent
For the scope of this paper, we only plot the parameter space in $m_a$ and $g_{a\gamma\gamma}$ for which radio emission from parametric resonance can occur in the galactic ASMFs for different $n$ in figure \ref{fig:g_m_a}.
The shaded regions in figure \ref{fig:g_m_a} indicate the two scenarios introduced above: 
First in darker shades and solid lines, we plot the requirement that the core-halo relation and MCMF parametrization by \cite{fairbairn_structure_2018} predict the existence of parametrically resonant ALP stars, $M_{\star,\gamma}(n, m_a, g_{a\gamma\gamma}) \leq \max(M_\star)|_{n, m_a}$, where $\max(M_\star)$ indicates the maximum predicted AS mass in the ASMF.
This condition amounts to our conservative approach of using only the core-halo relation \eqref{eq:CoreHalo} to determine the soliton mass while neglecting the (currently uncertain) long-time effects of AS accretion.\\
On the other hand, the second case in light shades and dashed colored lines indicates the weaker constraint that $M_{\star,\gamma}(n, m_a, g_{a\gamma\gamma}) \leq M_{\star,\lambda}(n,m_a)$, which basically shows where ALP stars with $m_a,n,f_a$ following the procedure in subsection \ref{subsec:M0} can experience parametric resonance before suffering the self-interaction instability given by \eqref{eq:M_Star_Max_R_Star_min}.
This case is especially relevant when including the effects of long-time accretion from the MC onto its AS core, similar to what was suggested in \cite{dmitriev_self-similar_2024}.
In the most optimistic case \cite{dmitriev_self-similar_2024}, the accreting solitons absorb an order one fraction of the mass of their host MCs - unless prevented by the critical masses $M_{\star,\gamma}$ and $M_{\star,\lambda}$.
As a consequence, the population of parametrically resonant ALP stars in the Milky Way would be significantly boosted and every MC with $\mathcal{M} \gtrsim M_{\star,\gamma}$ could serve as a site of radio conversion.
Note also, that both of the above conditions require the existence of miniclusters, which is why the low-$m_a$ regions in figure \ref{fig:g_m_a} are excluded by the constraint $f_a < 8.2 \cdot 10^{12}\,$GeV from subsection \ref{subsec:M0} and figure \ref{fig:f_a}.\\
We show the predicted total number of resonant AS-MC systems $\mathcal{N}_{\gamma,\mathrm{tot}}$ in the galaxy using the above two conditions for representative $g_{a\gamma}=10^{-11}\,$GeV in figure \ref{fig:Decay_gagamma-11} and $g_{a\gamma}=10^{-12}\,$GeV in figure \ref{fig:Decay_gagamma-12}.
\noindent
The solid lines in figures \ref{fig:Decay_gagamma-11}, \ref{fig:Decay_gagamma-12} show our results using the core-halo relation \eqref{eq:CoreHalo} with $M_\star \propto \mathcal{M}^{1/3}$ (i.e. the first scenario and solid lines in figure \ref{fig:g_m_a}), while the dashed lines show the results from the more optimistic second accretion scenario from \cite{dmitriev_self-similar_2024} yielding $M_\star \sim \mathcal{M}$ and hence larger numbers of resonant AS-MC systems.
For the conservative case $M_\star \propto \mathcal{M}^{1/3}$, $\mathcal{N}_{\gamma,\mathrm{tot}}$ drops to zero at the point where $\max(M_\star)|_{n,m_a} = M_{\star,\gamma}$.
The detailed shape of the curves depends on the temperature-dependence $n$ of the axion mass and on the interplay of the different cutoffs of the ASMF in subsection \ref{subsec:ASMF_Cutoff} with the decay mass $M_{\star,\gamma}$.
Conversely, in the case $M_\star \sim \mathcal{M}$, the number of resonant systems vanishes at higher ALP mass $m_a$, when $M_{\star,\gamma} \geq M_{\star,\lambda}$ again depending on $n$.
For both of the discussed scenarios, we predict large numbers of potentially resonant AS-MC systems in the Milky Way.
The resulting diffuse radio background signal from these objects would be peaked around a narrow frequency range $\nu_\gamma\approx m_a/4\pi$ and we will use it to constrain ALP- and QCD axion models in our follow-up paper \cite{maseizik_radio_2024}.
We conclude that there is significant potential in exploiting the combined effects of ALP star accretion and parametric resonance in the context of AS-MC systems.\\
Note that in theory, the AS decay mass \eqref{eq:M_Decay_Gaussian} may also be reached through AS merger events (see figure \ref{fig:Int_Coll_AS_vesc}), which are extremely rare in our galaxy due to the large relative velocities in a typical encounter.
\noindent
We will briefly discuss a related third mechanism of resonance involving the more common host MC-MC mergers and subsequent AS core mergers in subsection \ref{subsec:extra-galactic}.
Let us emphasize that in this third scenario, the parameter spaces $M_{\star,\gamma}  \leq M_{\star,\lambda}$ in figure \ref{fig:Decay_gagamma-11} could also be probed using the MC-MC mergers from figures \ref{fig:Int_Coll_Bosenova_0.5} and \ref{fig:Int_Coll_Bosenova_0.7} since for these values of $m_a$, $g_{a\gamma}$, the AS resonance occurs before the self-interaction instability. 
In this case, the host MC mergers would induce subsequent AS core mergers with a final mass $M_\star \geq M_{\star,\gamma}$ where the AS stable-mass excess $M_\star - M_{\star,\gamma}$ gets converted into radio emission.
The multi-messenger signals emerging from an ALP star merger were recently investigated in \cite{chungjukko_2024_multimessenger}.\\
For the remaining parameter space with $M_{\star,\lambda} \leq M_{\star,\gamma}$, a recent study performed in \cite{di2023stimulated} suggests that ALP stars can also trigger stimulated decay of ALPs into radio photons rather than producing bursts of relativistic ALPs during collapse.
This scenario could potentially yield observable radio bursts even from super-critical ALP stars with $M_\star \geq M_{\star,\lambda}$ and for small photon couplings where $M_{\star,\lambda} \leq M_{\star,\gamma}$, which we have both neglected in our work.

\subsection{Cosmological Event Rates} \label{subsec:extra-galactic}
The different event rates calculated in this paper need to be combined with the signal strength of the corresponding single events in order to infer observational predictions.
While the galactic rates of AS/MC collisions with NSs are generally too low for detection $\Gamma < 1\,$yr$^{-1}$ galaxy$^{-1}$, we can additionally consider the cosmological scenario, assuming a galaxy density of $n_\mathrm{gal}\sim 10^{-2} \,$Mpc$^{-3}$ \cite{Weaver_2023}, to estimate the isotropic emission from extra-galactic AS/MC-NS and MC-MC collisions.
Neglecting redshift defects, we define the duty cycle
\begin{align}
    \mathcal{D} \sim \gamma \frac{4\pi}{3} d_\mathrm{obs}^3 t_s  \simeq
     \,,\quad \gamma = n_{gal} \Gamma_s 
    \label{eq:dutycycle}
\end{align}
 of the different AS/MC encounters with galactic event rates $\Gamma_s$ given in terms of the extra-galactic event rate $\gamma = n_{gal}\Gamma_s$ in units of s$^{-1}$Mpc$^{-3}$.
 Here, the typical observation distance is taken to be $d_\mathrm{obs}=2\,$Gpc,  and the signal duration $t_s$ depends on the specific encounter under consideration.
The duty cycle within the beam size $\Delta\Omega$ of a given radio telescope is then given by $\mathcal{D}\Delta\Omega/(4\pi)$. 
For a typical beam size of $\simeq 1^{\circ}$ one has $\Delta \Omega / (4\pi)\simeq 3\cdot 10^{-4}\,$sr.
We require the duty cycle per beam $ \mathcal{D} \Delta \Omega / (4\pi)$ to not be much smaller than one in order to have individual events, which are not too infrequent.
A rough estimate for the observed flux of a single event with total emitted energy $E_s$ and at cosmological distance $d_{\rm obs}$ is given by 
\begin{align}
    j_s & \sim \frac{E_s }{4 \pi d_\mathrm{obs}^2 \Delta m_a t_s}
    \simeq 0.5 \, \text{Jy} \, 
    \left( \frac{E_s}{10^{52}\,\text{eV}} \right) 
    \left( \frac{m_a}{50\,\mu\text{eV}} \right)^{-1} \nonumber \\
    & \times 
    \left( \frac{t_s}{0.1\,\mathrm{s}} \right)^{-1}
    \left( \frac{d_\mathrm{obs}}{2\,\mathrm{Gpc}} \right)^{-2}
    \label{eq:Extragal_flux} \,,
\end{align}
where we assume the relative bandwidth $\Delta\simeq 10^{-3}$ from the galactic velocity dispersion.
In the following, we insert $\Gamma_s$ based on the results from subsections \ref{subsec:NS-AS} to \ref{subsec:bosenova} for $m_a=50\,\mu$eV combined with $E_s$, $t_s$ from different literature about single event signals.\\
~\\
\textbf{NS-AS Collisions}\\
Starting with the AS-NS collisions from figure \ref{fig:Int_Coll_NS-AS}, we get $\gamma \sim 10^{-16}\,$Mpc$^{-3}$s$^{-1}$ for the $\mathcal{M}_J$-cutoff at $\alpha=-1/2$.
The duration $t_s$ can be inferred from the results in \cite{witte_transient_2023} by considering the signal from an AS-NS transient event with non-zero impact parameter $b=10^8\,$km.
The resulting signal duration $t_s\sim 50\,$s leads to a duty cycle of roughly $\mathcal{D} \simeq 2\cdot 10^{-4} \ll 1$.
Using the same event with $b=10^8\,$km in \cite{witte_transient_2023}, we obtain the total emitted energy $E_s \sim 4\pi \,\text{kpc}^2 t_s m_a \Delta S_T \sim 10^{44}\,$eV from the flux density $S_T\sim 10^{5}\,$mJy at a distance of $1\,$kpc, leading to an observed flux of $j_s \simeq 10^{-11}\,$Jy way below the sensitivity of current radio telescopes.\\
~\\
\textbf{NS-MC Collisions}\\
For the more common NS-MC collisions we predicted larger signal rates $\gamma \sim 10^{-12}\,$Mpc$^{-3}$s$^{-1}$ in figure \ref{fig:Int_Coll_NS_MC2_f_NS} for the $\mathcal{M}_J$-cutoff at $\alpha=-0.7$.
From \cite{witte_transient_2023} and for an impact parameter of $b=10^8\,$km, we find $t_s\sim 150\,$d leading to a large duty cycle $\mathcal{D} \simeq 4\cdot 10^5 \gg 1 $.
To further estimate the observed flux of a single NS-MC collision according to equation \eqref{eq:Extragal_flux}, we can take the results from \cite{witte_transient_2023} and find that $E_s \sim 4\pi \,\text{kpc}^2 t_s m_a \Delta S_T \sim 10^{39}\,$eV from $S_T\sim 10^{-6}\,$mJy at $1\,$kpc distance, which yields an essentially undetectable signal with flux $j_s\simeq 4\cdot 10^{-22}\,$Jy.\\
~\\
\textbf{Parametric Resonance and MC-MC Merger}\\
The last and most relevant scenario in the cosmological context is the occurrence of parametric resonance in AS-AS mergers following a successful MC-MC merger as calculated in subsection \ref{subsec:bosenova} for Bosenovae.
For sufficiently large ALP-photon coupling $g_{a\gamma}$, the parametric resonance can be triggered before the self-interaction instability develops, $M_{\star,\gamma}<M_{\star,\lambda}$, leading to strong radio emission following a MC merger.
An important detail to this scenario is the question how long it takes for the ALP star cores to merge after their host miniclusters have merged.
For the scope of this work, we can estimate the typical time between two MC mergers with final mass $\mathcal{M}_1 + \mathcal{M}_2 \geq \mathcal{M}(M_{\star,\lambda})$ by dividing the corresponding rate $\Gamma_{mc-mc} / N_{\star,\mathrm{tot}} \sim 10^{3}\,\text{yr}^{-1} / 10^{23}$ from figure \ref{fig:Int_Coll_Bosenova_0.7} by the total number of MCs for $m_a=50\,\mu$eV, $n=3.34$, $\alpha=-0.7$ in figure \ref{fig:N_tot_MC}, which gives $t_\mathrm{merg} \sim 10^{20}\,$yr.
This time should be compared to the intrinsic timescale of the AS-MC system.
Note that the condensation time from \cite{levkov_gravitational_2018, chen_relaxation_2022}, which measures the required time for soliton formation starting from random initial conditions, does not apply here since the merged MC system provides a pre-defined potential well to the merging AS cores.
Instead we use the free-fall time of the merged miniclusters as an estimate for the timescale of the AS core merger and find $\tau_\mathrm{ff} = \pi \mathcal{R}^{3/2} / [4 \sqrt{G\mathcal{M}(M_{\star,\lambda})}] \simeq 0.2\,$yrs for the QCD axion with $\mathcal{M}(M_{\star,\lambda})\simeq 4\cdot 10^{-7}\,M_\odot$ and $\mathcal{R}\simeq 2\cdot 10^9\,$km from equation \eqref{eq:R_mc}.
With the timescale of MC merger interactions being much larger than the free-fall time, $t_\mathrm{merg} \gg t_\mathrm{ff}$, we can assume that AS mergers happen quasi-instantaneously in the following (see equation \eqref{eq:dutycycle}).\\
The resulting energy emitted in a single radio burst can roughly be estimated from $M_{\star,\gamma}$ in equation \eqref{eq:M_Decay_Gaussian} for the QCD axion with $m_a=50\,\mu$eV, $f_a\simeq 10^{11}\,$GeV at $g_{a\gamma}=10^{-11}\,$/GeV.
For these values we find $M_{\star,\gamma}\approx 1.3\cdot10^{-13}\,M_\odot$ and we assume that an order $0.1$ fraction of the resonant star mass will be converted into photons, i.e. $E_s\sim M_{\star,\gamma}/10 \sim 10^{-14}\,M_\odot \sim 10^{52}\,$eV, which is roughly consistent with the total emitted energy calculated in \cite{di2023stimulated}.
Combining these numbers with $t_s\simeq0.1$ \cite{di2023stimulated} and $\gamma \sim 10^{-6}\,$Mpc$^{-3}$s$^{-1}$ from figure \ref{fig:Int_Coll_Bosenova_0.7} for the $\mathcal{M}_J$-cutoff with $\alpha=-0.7$, equation \eqref{eq:Extragal_flux} gives $j_s \simeq 0.5\,$Jy\,sr$^{-1}$ and $\mathcal{D} \simeq 3 \cdot 10^{4}$.
For a beam size of $\simeq 1^\circ$ with $\Delta \Omega / (4\pi) \simeq 3 \cdot 10^{-4}$, we obtain a beam duty cycle of order unity for the resonant MC mergers in equation \eqref{eq:dutycycle}. 
This means that within one beam we would expect a popcorn like signal that should be easy to distinguish from backgrounds as long as the time integrated intensity is above the sensitivity of the radio telescope considered.
Note that our above estimates predict lines which are as narrow as $\Delta\simeq10^{-3}$, but at different redshifted central frequencies.\\
~\\
To summarize, the isotropic cosmological backgrounds of NS-AS/MC collisions are expected to be negligible even from our most optimistic estimates for QCD axion parameters.
We use the SKA-mid sensitivity $S \sim 10 \, \mu\text{Jy}\,\text{hr}^{-1/2}$ \cite{ska_web} which, integrated over a signal duration $t_s=0.1$ \cite{di2023stimulated}, gives $S\sim 2\,\text{mJy}\,(0.1\,{\rm s}/t_s)^{1/2}$. 
This is smaller than the estimate in equation \eqref{eq:Extragal_flux} for the flux from parametric resonance of a single MC merger at cosmological distances which should thus indeed be detectable at the representative value of $g_{a\gamma}=10^{-11}\,$GeV$^{-1}$.
We conclude that parametric resonance and MC mergers are the most promising mechanism in the context of extra-galactic background signals.
Note that a similar study involving soliton mergers rates of cosmological DM halos (opposed to our approach for ALP miniclusters with $z=z_\mathrm{eq}$) was already performed in \cite{Du_2024}.

\section{Summary and Conclusion} \label{sec:conclusions}
In this paper, we have established a full formalism for inferring soliton properties from their host minicluster mass distributions for both QCD axions and more generally for ALPs.
We suggest that the core-halo relation \eqref{eq:CoreHalo} from \cite{schive_understanding_2014} can be applied to stable ALP stars on the dilute branch as an estimate, but emphasize the need for an extended core-halo relation including the effects of self-interactions in the condensate.
Using this assumption, we improved previous predictions for collision rates of ASs and MCs in the literature \cite{eby_collapse_2016, bai_diluted_2022, edwards_transient_2021} by inferring the full ASMF from the MCMF taking into account different ALP star cutoffs in subsection \ref{subsec:ASMF_Cutoff}.\\
We find that for the collapse redshifts $z\simeq z_\mathrm{eq}$ that we use, the minimum halo mass $\mathcal{M}_{h,\min}$, \eqref{eq:M_h_min_CoreHalo} provides the strongest low-$M_\star$ cutoff.
This is opposed to some previous works, who used a collapse redshift of $z=0$ and found the radius-cutoff $\mathcal{M}_{R,\min}$ as the predominant cutoff to the ASMF.
More generally, we also calculated the different scalings of the $M_\star$-cutoffs with $z$ analytically in subsection \ref{subsec:ASMF_Cutoff}.\\
After normalization of the total AS-MC mass to the mass of the Milky Way DM halo, the above approach allowed us to directly determine the ALP star properties and -abundance in our galaxy.
We compared the resulting fraction $f_\star$ of dark matter contained in ALP stars to the estimates $10^{-4} \leq f_\star \lesssim 1$ by previous authors \citep{hertzberg_merger_2020, eby_collisions_2017, bai_diluted_2022} and predict much smaller values, reaching down to $f_\star \sim 10^{-7}$ for the QCD axion and ALP masses with a similar temperature evolution.
The reason for this is the fact that heavy miniclusters still contain at most a single, relatively light ALP star and thus make up more of the total mass of the AS-MC system.
Similarly, we showed that the typical mass of ALP stars characterized by $\varepsilon = \langle M_\star \rangle / M_{\star,\lambda}$ can indeed reach values close to the maximum stable AS mass $M_{\star,\lambda}$, as favored in the literature.
Specifically for QCD axions and ALPs with similar temperature evolution, we find $\varepsilon \lesssim 1$, albeit at significantly reduced abundance $f_\star \lesssim 10^{-6}$.
On the other hand, our results suggest that $\varepsilon$ can be significantly smaller $\varepsilon \sim 10^{-4}$, especially for temperature-independent ALPs, which in turn have the largest DM abundance $f_\star\sim 10^{-3}$.\\
Our mock population of neutron stars in subsection \ref{subsec:NS-AS} indicates that the previously neglected plasma resonance criterion $\omega_p \gtrsim m_a$ imposes a strong suppression on the signal rates received from NS-AS and NS-MC encounters.
We have re-evaluated the rate of radio signals from NS-AS encounters in our galaxy showing that such events are generally rare $\Gamma_{\star-NS}<10^{-3}\,$yr$^{-1}$ galaxy$^{-1}$, especially for the QCD axion with $\Gamma_{\star-NS}<10^{-7}\,$yr$^{-1}$ galaxy$^{-1}$.
For the more common NS-MC encounters we predict signal rates on the order of $\Gamma_{mc-NS}\sim 10^{-5}\,$yr$^{-1}$ galaxy$^{-1}$ at $\alpha=-1/2$ and $\Gamma_{mc-NS}\sim 10^{-3}\,$yr$^{-1}$ galaxy$^{-1}$ at $\alpha=-0.7$ for the QCD axion, depending on the low-$\mathcal{M}$ cutoff of the MCMF and its slope index $\alpha$.
For most of the ALP models and both values of $\alpha=-1/2$ and $\alpha=-0.7$, the resonance suppression $f_{NS}$ renders radio signals from NS-AS/MC collisions essentially undetectable.\\
In the context of Bosenovae, discussed in subsection \ref{subsec:bosenova}, and for $n=3.34$, our results suggest that MC-MC mergers can appear as often as $\sim 10\,$yr$^{-1}$ galaxy$^{-1}$ for $\alpha=-1/2$ and $\sim 10^3\,$yr$^{-1}$ galaxy$^{-1}$ for $\alpha=-0.7$, using the $\mathcal{M}_J$-cutoff.
This prediction has important consequences for the future detection of AS signals from both parametric resonance and Bosenovae triggered by AS core mergers \cite{chungjukko_2024_multimessenger, di2023stimulated, hertzberg_merger_2020, Amin_2021}.
For the $\mathcal{M}_0$-cutoff, the total number of miniclusters is significantly lower which is why AS mergers can not be efficiently triggered by MC-MC collisions using this cutoff.
We emphasize that the $\mathcal{M}_0$-cutoff does not exclude Bosenovae in general since the long-time effects of accretion can still play a vital role in the AS evolution.\\
In subsection \ref{subsec:Glowing_AS} we have used the galactic AS distribution to argue that the most promising mechanism of dark matter detection with AS-MC systems in our galaxy is given by solitons in, or close to the state of parametric resonance.
Depending on the ALP-photon coupling and ALP model, we find strong evidence for the numerous existence of heavy MCs hosting a resonant AS core.
This prediction can be combined with considerations of AS accretion and MC-MC mergers to yield additional observable signatures such as isotropic background emission and radio burst signals.\\
Lastly in subsection \ref{subsec:extra-galactic}, we have briefly discussed the potential of extra-galactic NS-AS/MC encounters and MC-MC mergers with a parametrically resonant AS core.
Our rough estimates suggest that NS-AS/MC signals are too faint for individual detection but that the extra-galactic radio bursts from resonant AS mergers can have large fluxes of $\sim 0.5\,$Jy even at cosmological distances of $d_\mathrm{obs}\simeq 2\,$Gpc, with a duty cycle that can reach order one within a typical radio telescope beam with degree-scale opening angle.\\
Altogether, our results highlight fundamental difficulties in the detection of ALP substructure using NS collisions and they strongly suggest that the search for ALP dark matter using these structures should be directed towards signatures of parametric resonance in AS-MC systems or Bosenovae.
We conclude by listing the uncertainties in our ASMF determination scheme for future authors to improve on (see also figure \ref{fig:Scheme}).
First and mainly, we have used the core-halo relation from \cite{schive_understanding_2014} and the linear growth Press-Schechter theory predictions from \citep{fairbairn_structure_2018} for the present-day minicluster distributions.
The most relevant uncertainties in the Press-Schechter model include the low-$\mathcal{M}$ cutoff of the MCMF and its connection to non-Gaussianities of the MC density field, the initial power spectrum and non-linear effects of structure formation.
There is also active research on the MCMF slope $\alpha$, the derivation of an extended core-halo relation, the $\mathcal{M}$-scaling of the $\lambda=0$ core-halo relation and on the survival rate of miniclusters in the galactic environment.
For all of these uncertainties we have used the currently favoured assumptions, but we emphasize that our approach can be easily updated by using modified versions of the above relations without loss of generality.

\section*{Acknowledgments}
This work is funded by the Deutsche Forschungsgemeinschaft (DFG, German Research Foundation) under Germany’s Excellence Strategy -- EXC 2121 ``Quantum Universe'' -- 390833306. This article is based upon work from the COST Action COSMIC WISPers CA21106, supported by COST (European Cooperation in Science and Technology).
We thank Samuel Witte for very useful suggestions on the neutron star properties and Benedikt Eggemeier for detailed discussions on the evolution of miniclusters and axion stars.
We also thank Hyeonseok Seong and Virgile Dandoy for various discussions about miniclusters as well as Giuseppe Lucente, Pierluca Carenza and Alessandro Mirizzi for general discussions on neutron star encounters with miniclusters and ALP stars.

\appendix

\section{ALP Star Properties} \label{app:AS-Params}
\begin{figure}[t]
\centering
\includegraphics[width=\columnwidth]{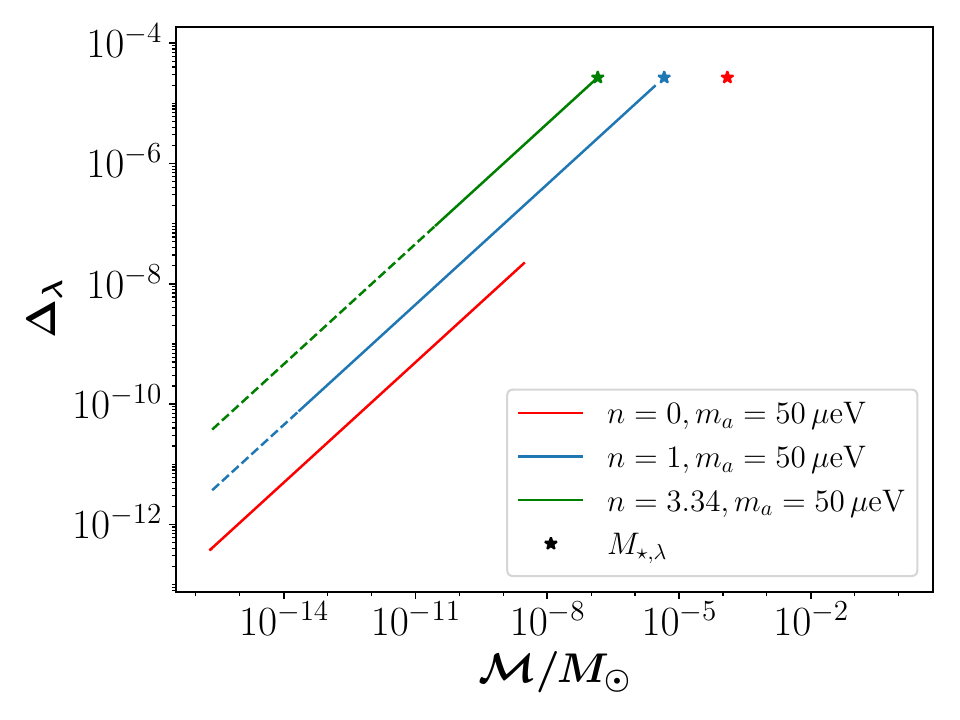}
\caption{Perturbation term $\Delta\lambda(\mathcal{M},f_a)$ from equation \eqref{eq:DeltaLambda} measuring the modification of $M_\star(\mathcal{M})$ at $|\lambda|>0$ compared to the $\lambda=0$ relation \eqref{eq:CoreHalo}.
Both with colors indicating different values for the ALP mass $m_a$ and its temperature dependence $n$.
The ranges of $M_\star$ obtained with the $\mathcal{M}_0$- and $\mathcal{M}_J$-cutoffs from subsection \ref{subsec:MCMF_Parametr} are shown in solid and dashed lines.
Stars correspond to the maximum stable AS mass $M_{\star,\lambda}$ from equation \eqref{eq:M_Star_Max_R_Star_min}.} \label{fig:DeltaLambda}
\end{figure}

\begin{figure}[t]
\centering
\includegraphics[width=\columnwidth]{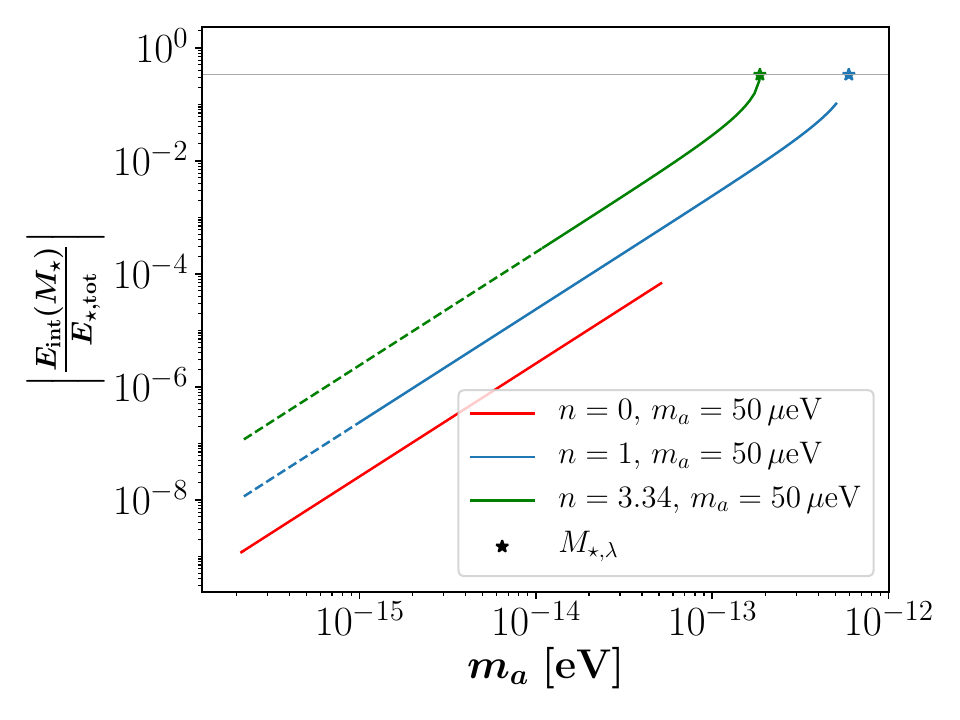}
\caption{Relative core energy fractions $|\Delta E_{\star} / E_{\mathrm{grav}}|$ as a function of AS mass $M_\star$ reaching up to $|\Delta E_{\star} / E_{\mathrm{grav}}| = 1/3$ indicated by the grey line.
Stars correspond to the maximum stable AS mass $M_{\star,\lambda}$ from equation \eqref{eq:M_Star_Max_R_Star_min}.} \label{fig:E_SI}
\end{figure}

\begin{figure}[t]
\centering
\includegraphics[width=\columnwidth]{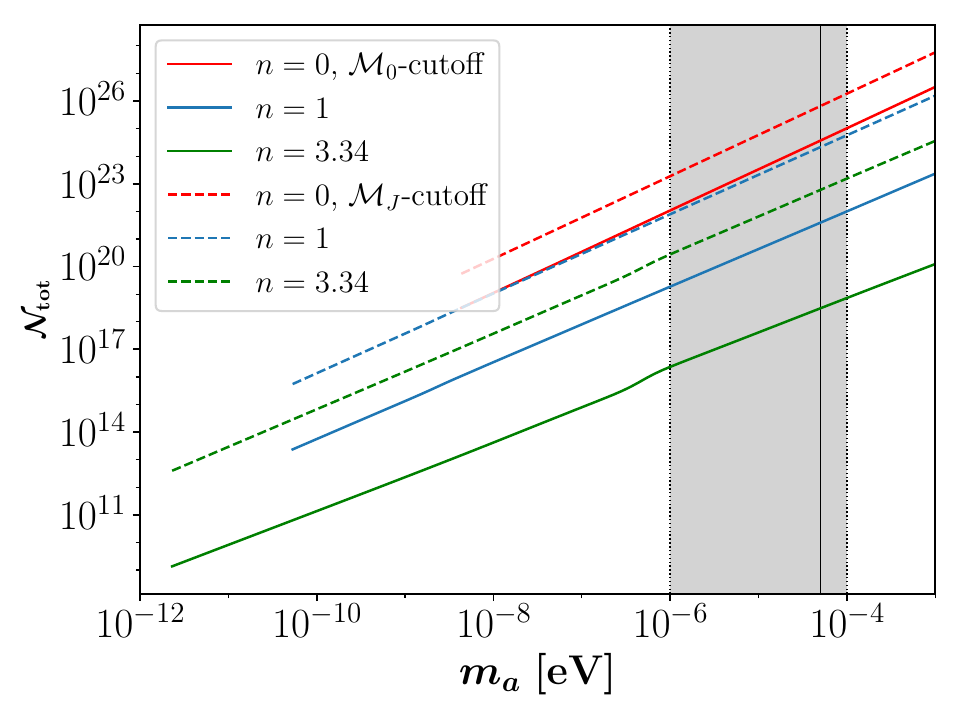}
\caption{Total number of MCs as a function of ALP mass $m_a$ and its temperature index $n$, obtained from the MCMF of ALP miniclusters at $m_a$.
\label{fig:N_tot_MC}}
\end{figure}

As argued in \cite{bar_galactic_2018}, the virialization condition $v_\star \simeq v_\mathrm{mc}$ is equivalent to the requirement
\begin{align}
\frac{|E_{\star,\mathrm{tot}}|}{M_\star} & \simeq \frac{|\mathcal{E}|}{\mathcal{M}} 
\quad \stackrel{E_{\star,\mathrm{tot}} \sim E_\mathrm{grav}}{\Longrightarrow}
\frac{G M_\star}{R_\star} \simeq \frac{G\mathcal{M}}{\mathcal{R}}\,. \label{eq:virial_Ansatz_E}
\end{align}
for the specific energies of the AS-MC system and where the total star energy is typically assumed to be on the order of the gravitational AS binding energy $E_{\star,\mathrm{tot}} \sim E_\mathrm{grav}$ \cite{schive_understanding_2014, padilla_core-halo_2021}.
Instead of the modified virialization approach from \cite{padilla_core-halo_2021} in equation \eqref{eq:virial_Ansatz_v2} one could thus consider the change in $E_{\star,\mathrm{tot}}$ and $\mathcal{E}$ respectively.\\
Starting with the change $\Delta\mathcal{E}$ in MC energy, we can argue that for typical AS-MC systems with overdensity parameter $\delta\sim 1$, the minicluster density \eqref{eq:rho_mc} will be much lower than that of their ALP star cores $\rho_\star \lesssim M_{\star,\lambda} / R_\star^3 \sim 10^{23}\,$GeV/cm$^3$.
In these dilute systems, the short-range self-interaction will be negligible compared to the long-range gravitational force (see also \cite{padilla_core-halo_2021} for a detailed calculation).
Thus assuming $\Delta \mathcal{E} \ll \mathcal{E}$, the relevant shift in the equilibrium state described by equation \eqref{eq:virial_Ansatz_E} is 
\begin{align}
\Delta E_\star \equiv | E_{\star,\mathrm{tot}} - E_{\star,\mathrm{tot}}(\lambda=0) | 
= | E_\mathrm{int} | \label{eq:Delta_E_star} \,,
\end{align}
where $E_{\star,\mathrm{tot}}(\lambda=0)$ is the star energy evaluated at $\lambda=0$.
We show the relative energy shifts $|\Delta E_{\star} / E_{\mathrm{grav}}| = E_\mathrm{int} / E_{\mathrm{grav}}$ in figure \ref{fig:E_SI} and the perturbation term $\Delta \lambda$ from equation \eqref{eq:DeltaLambda} in figure \ref{fig:DeltaLambda} with $m_a=50\,\mu$eV.
The ranges of AS masses in figures \ref{fig:DeltaLambda} and \ref{fig:E_SI} correspond to the core-masses derived from the MCMF using equation \eqref{eq:CoreHalo} with the $\mathcal{M}_J$-cutoff in dashed lines and for the $\mathcal{M}_0$-cutoff in solid colored lines.
The colored curves demonstrate that the main factor increasing the prediction for $M_\star$ and thus $\Delta \lambda$, $\Delta E_\star$ is the temperature dependence $n$ of the ALP mass.
Accordingly, the range of AS masses for $n=1,3.34$ in blue and green extends to larger $M_\star$ compared to the temperature-independent cases in red.\\
Figure \ref{fig:DeltaLambda} shows that the predicted perturbation $\Delta \lambda$ is well below one for any $m_a,n$ considered in this work as claimed in the main text.
Figure \ref{fig:E_SI} depicts the energy perturbation $\Delta E_\star$ from the modified ansatz in \eqref{eq:virial_Ansatz_E} for the same AS distributions.
\noindent
Figure \ref{fig:E_SI} thus demonstrates that the condition \eqref{eq:virial_Ansatz_E} is more stringent and that it yields qualitatively similar results by predicting $|\Delta E_\star / E_\mathrm{grav}| \ll 1$ for the majority of the mass range compared to $\Delta \lambda$ and equation \eqref{eq:virial_Ansatz_v2}.\\
In the small region, where $M_\star$ becomes similar to the maximum stable AS mass $M_{\star,\lambda}$ (i.e. close to the grey line and colored stars), $|\Delta E_\star / E_\mathrm{grav}| \approx 1/3$ becomes relevant.
In this range, we expect any extended core-halo relation to be modified compared to equation \eqref{eq:CoreHalo} by \cite{schive_understanding_2014}.
We emphasize that even at $M_\star=M_{\star,\lambda}$ the expected energy shift in equation \eqref{eq:virial_Ansatz_E} is of order one, so that our $\lambda=0$ approach should still yield an estimate that is within the large uncertainties of the MCMF in section \ref{sec:MCMF}.
We also note, that in a more general sense, the soliton solutions on the dense, unstable branch in figure \ref{fig:Mass-Radius_QCD_minimal} with $R_\star \leq R_{\star,\lambda}$ would be subject to much larger modifications $|\Delta E_\star / E_\mathrm{grav}| \gg 1$ due to their significantly higher densities.\\
~\\
For future works, we also attach some relevant AS- and MC parameters obtained from our approach in figures \ref{fig:N_tot_MC} to \ref{fig:R_AS(m_a, n)}.
\begin{figure}[t]
\centering
\includegraphics[width=\columnwidth]{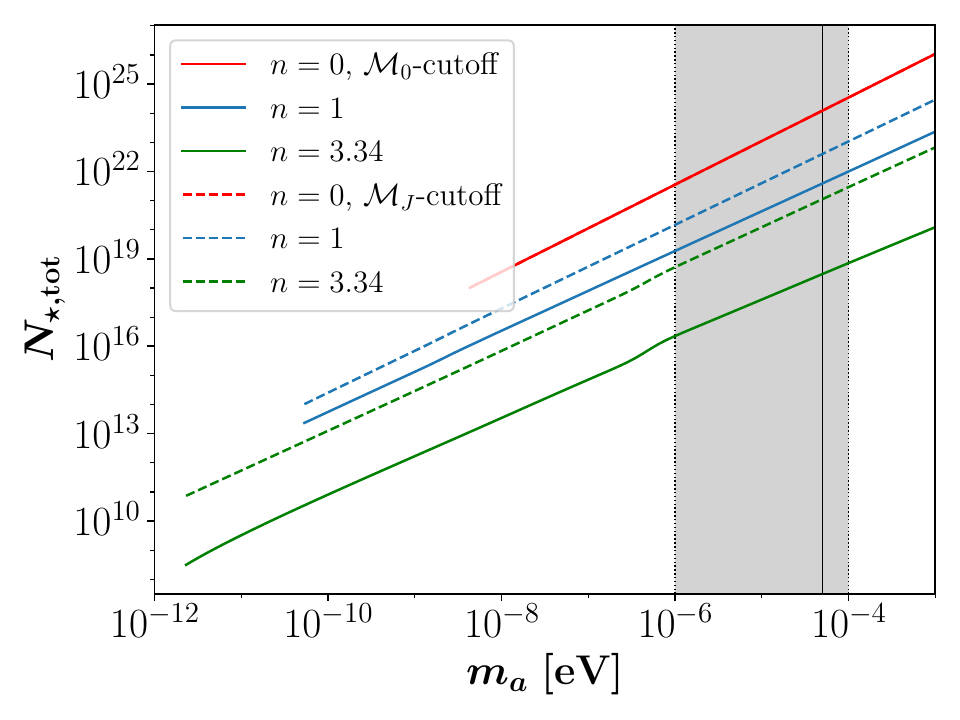}
\caption{Total number of ASs as a function of ALP mass $m_a$ and its temperature index $n$, obtained from the MCMF of ALP miniclusters at $m_a$.
For $n=0$, the AS number is independent of the low-$\mathcal{M}$ cutoffs.
\label{fig:N_tot_AS}}
\end{figure}
Figure \ref{fig:N_tot_MC} shows the total number of miniclusters obtained from the MCMF cutoffs from subsection \ref{subsec:MCMF_Parametr} and the corresponding total number of ALP stars obtained from the additional ASMF cutoffs from section \ref{subsec:ASMF_Cutoff} is given in figure \ref{fig:N_tot_AS}.
The average mass of ALP stars used in equation \eqref{eq:f_AS_alpha} is determined by 
\begin{align}
\langle M_\star \rangle &= \frac{\int_{\min(M_\star) }^{\max(M_\star)} d M_\star \, M_\star \frac{\mathrm{d}n_\star}{\mathrm{d}M_\star}}{\int_{\min(M_\star) }^{\max(M_\star)} d M_\star \, \frac{\mathrm{d}n_\star}{\mathrm{d} M_\star}} \,, \label{eq:M_average}
\end{align} 
where $\min(M_\star)$, $\max(M_\star)$ are the respective low- and high-$M_\star$ cutoffs from subsection \ref{subsec:ASMF_Cutoff}.
The average AS radius $\langle R_\star \rangle$ is defined as the radius of the average AS mass, i.e. $\langle R_\star \rangle \equiv R_\star(\langle M_\star \rangle)$ according to the mass-radius relation \eqref{eq:Radius-Mass-Rel_Physical}.
Figures \ref{fig:M_AS(m_a, n)} and \ref{fig:R_AS(m_a, n)} shows the average mass and radius of ALP stars using both low-$\mathcal{M}$ cutoffs.
\begin{figure}[h]
\includegraphics[width=\columnwidth]{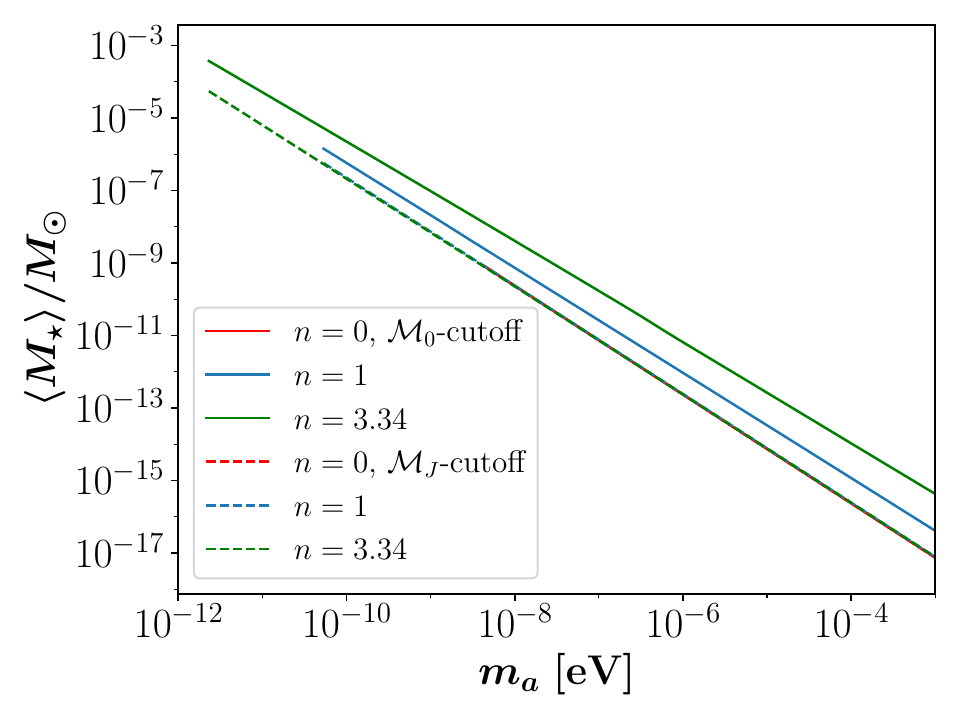}
\caption{Average mass of ALP stars as a function of ALP mass $m_a$ and $n$, obtained from the MCMF of ALP miniclusters at $m_a$.
The different $\mathcal{M}_J$-cutoffs in dashed lines and the $n=0$ $\mathcal{M}_0$-cutoff in red solid lines yield almost identical results for both $\langle M_\star \rangle$ and $\langle R_\star \rangle$.
\label{fig:M_AS(m_a, n)}}
\end{figure}
\begin{figure}[t]
\includegraphics[width=\columnwidth]{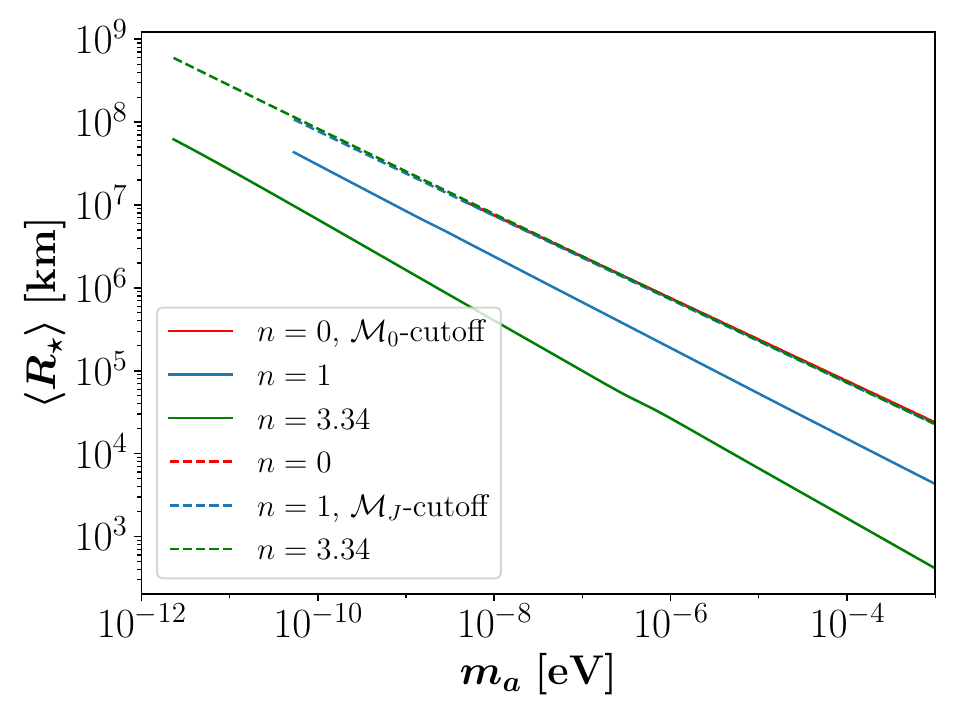}
\caption{Average radius of ALP stars as a function of ALP mass $m_a$ and $n$, obtained from the MCMF of ALP miniclusters at $m_a$.
The different $\mathcal{M}_J$-cutoffs in dashed lines and the $n=0$ $\mathcal{M}_0$-cutoff in red solid lines yield almost identical results for both $\langle M_\star \rangle$ and $\langle R_\star \rangle$.
\label{fig:R_AS(m_a, n)}}
\end{figure}

\section{AS-MC Quantities}

For comprehensibility, we provide a list of all important MC- (top) and AS (bottom) observables in table \ref{tab:Params}.
The left and right columns show the definitions of each quantity, while the center column explains its meaning.
\begin{table*}[t]
    \centering
    \begin{tabular}{|c|c|c|}
    \hline
        Quantity & Explanation & Definition \\
        \hline
        $\rho_\mathrm{mc}$   & Characteristic MC density & \eqref{eq:rho_mc}   \\
        $\delta$   & MC overdensity parameter & $\delta = \delta\rho_a / \rho_a$   \\
        $\mathcal{M}_{0}$   & Characteristic MC mass & \eqref{eq:M0}   \\
        $\mathcal{R}$   & Spherically homogeneous MC radius & \eqref{eq:R_mc}   \\
        $\mathcal{M}_{J,\min}$   & Low-mass MCMF cutoff from the Jeans mass $\mathcal{M}_J$ & \eqref{eq:M_h_min_J} \\
        $\mathcal{M}_{\min}$   & Applied low-mass MCMF cutoff at $z=0$ & $\mathcal{M}_0/25$ or $\mathcal{M}_{J,\min}$ \\
        $\mathcal{M}_{\max}$   & High-mass MCMF cutoff at $z=0$  & \eqref{eq:M_h_max}   \\
        $\mathcal{M}_{h,\min}$   & MC mass of ASMF cutoff from core-halo relation& \eqref{eq:M_h_min_CoreHalo} \\
        $\mathcal{M}_{R,\min}$   & MC mass of ASMF radius cutoff where $\mathcal{R} = R_\star$ &  \eqref{eq:M_h_min_RadiusCutoff}\\
        $\mathcal{M}_{\mathrm{tot}}$   & Total mass of MCs in the MW & \eqref{eq:M_tot_mc}\\
        $\mathcal{N}_{\mathrm{tot}}$   & Total number of MCs in the MW &  \eqref{eq:N_tot_mc}\\
        $\mathcal{N}_{\gamma,\mathrm{tot}}$   & Number of MCs hosting a resonant AS with $M_\star\geq M_{\star,\gamma}$ &  see \eqref{eq:M_Decay_Gaussian}\\
        \hline
        $M_{\star,\lambda}$   & Maximum stable AS mass imposed by self-interactions & \eqref{eq:M_Star_Max_R_Star_min}\\
        $R_{\star,\lambda}$   & Minimum stable AS radius imposed by self-interactions & \eqref{eq:M_Star_Max_R_Star_min}\\
        $M_{\star,h}$   & Low-mass ASMF cutoff from core-halo relation & see \eqref{eq:M_h_min_CoreHalo} \\        
        $M_{\star,R}$   & Low-mass ASMF radius cutoff where $\mathcal{R} = R_\star$ & \eqref{eq:M_h_min_RadiusCutoff_AS}\\ 
        $M_{\star,\gamma}$   & Decay mass of ASs triggering parametric resonance & \eqref{eq:M_Decay_Gaussian}\\
        $M_{\star,\mathrm{tot}}$   & Total mass of ASs in the MW & \eqref{eq:M_tot_star}\\
        $N_{\star,\mathrm{tot}}$   & Total number of ASs in the MW &  \eqref{eq:N_tot_star}\\
        $f_\star$   & Fraction of total MW mass contained in ASs &  \eqref{eq:f_AS_alpha}\\
        $\varepsilon$   & Parameter describing the typical AS mass &  \eqref{eq:f_AS_alpha}\\
        \hline
    \end{tabular}
    \caption{Different minicluster (top) and ALP star parameters (bottom) used in this paper}
    \label{tab:Params}
\end{table*}

\section{Milky Way Parameters} \label{app:MW}
In this section we will briefly summarize the physical parameters and observables used to calculate the collision rates in section \ref{sec:detection}.

\subsection{DM Density Distribution} \label{app_sub:DM}
For the profile of the galactic DM halo we use the Navarro-Frank-White (NFW) profile \cite{navarro_structure_1996}
\begin{align}
\rho_\mathrm{NFW}(r)=\frac{\rho_{0}}{\frac{r}{R_{s}}\left(1+\frac{r}{R_{s}}\right)^{2}}\,,\label{eq:NFW}
\end{align}
with characteristic density $\rho_0=\rho_\mathrm{DM}=0.32\,$GeV/$\mathrm{cm}^3$ and core radius $R_s=20.2\,$kpc \citep{mcmillan_mass_2011}.

\subsection{Neutron Star Distribution} \label{app:MW_NS}
We model the galactic neutron star distribution using the fit from \cite{taani_modeling_2012}
\begin{align}
n_{NS}(\rho,z) &= \frac{C_{NS} }{2 \pi \rho} p_\rho(\rho) p_z(\rho,z)\,,\\
p_{\rho}(\rho) &=A_{0, \rho}+A \frac{\rho^{\gamma-1}}{\lambda^\gamma} e^{-\rho / \lambda}\\
p_z(\rho, z) =&\,\,A_{0, z} \,\theta(z-0.1 \mathrm{kpc}) \nonumber \\
&+ A_{1, z} e^{-z / h_1(\rho)}+ A_{2, z} e^{-z / h_2(\rho)},
\end{align}
where $\theta(x)$ is a Heaviside function. 
The scale heights $h_{1,2}(\rho)$ are defined by
\begin{align}
h_1(\rho) &=k_1 \rho + b_1 \,, \\
h_2(\rho) &= \begin{cases}k_2^{<} \rho+b_2^{<}, & \rho \leq 4.5 \mathrm{kpc} \\
k_2^{>} \rho+b_2^{>}, & \rho \geq 4.5 \mathrm{kpc}\end{cases}\,.
\end{align}
with the relevant parameters summarized in table \ref{tab:NS_Params}.
\begin{table}[h]
    \centering
    \begin{minipage}[h]{0.5\columnwidth}
        \centering
        \begin{tabular}{|c|c|}
            \hline
            Param. & Value \\
            \hline
            $\gamma$ & 1.83 \\
            $A_{0, z}$ & $1.8 \cdot 10^{-5}$ kpc$^{-1}$ \\
            $A_{1, z}$ & 1.87 kpc$^{-1}$ \\
            $A_{2, z}$ & $35.6 \cdot 10^{-3}$ kpc$^{-1}$ \\
            $k_1$ & $13 \cdot 10^{-3}$ \\
            $k_2^{<}$ & $18.4 \cdot 10^{-3}$ \\
            $k_2^{>}$ & 0.05 \\
            \hline
        \end{tabular}
    \end{minipage}
    \hspace*{-1.5em}
    \begin{minipage}[h]{0.5\columnwidth}
        \centering
        \begin{tabular}{|c|c|}
            \hline
            Param. & Value \\
            \hline
            $A$ & $95.6 \cdot 10^{-3}$ \\
            $\lambda$ & 4.48 kpc \\
            $b_1$ & $12.8 \cdot10^{-3}$ kpc \\
            $b_2^{<}$ & 0.03 kpc \\
            $b_2^{>}$ & 0.65 kpc \\
            $A_{0, \rho}$ & $5 \cdot 10^{-3}$ kpc$^{-1}$ \\
            \hline
        \end{tabular}
    \end{minipage}
    \caption{Neutron star best-fit parameters obtained from \citep{taani_modeling_2012} and used in section \ref{sec:detection}.}
    \label{tab:NS_Params}
\end{table}

\bibliographystyle{apsrev4-2}
\bibliography{References}

\end{document}